\documentclass[structabstract]{aa}  
\usepackage{graphicx}
\usepackage{natbib}
\bibpunct{(}{)}{;}{a}{}{,}

\def\lya{Ly$\alpha$} 

\def\ha{H$\alpha$} 
\def\hb{H$\beta$}

\def\ebv{\ifmmode E_{B-V} \else $E_{B-V}$\fi}

\def\fesc{\ifmmode f_{\rm esc}(\lambda) \else $f_{\rm esc}(\lambda)$\fi}
\def\fescuv{\ifmmode f_{\rm esc}({\rm UV}) \else $f_{\rm esc}(\rm UV)$\fi}
\def\flya{\ifmmode f_{\rm esc}({\rm{Ly}\alpha}) \else $f_{\rm esc}({\rm{Ly}\alpha})$\fi}
\def\wlya{\ifmmode {\rm EW}_{\mathrm{Ly}\alpha}\else ${\rm EW}_{\mathrm{Ly}\alpha}$\fi}

\def\cm2{cm$^{-2}$}

\def\nh{\ifmmode N_{\mathrm{HI}}\else $N_{\mathrm{HI}}$\fi}
\def\vexp{\ifmmode {\rm v}_{\rm exp} \else v$_{\rm exp}$\fi}
\def\nhi{\ifmmode \overline{N_{\mathrm{HI}}}\else $\overline{N_{\mathrm{HI}}}$\fi}
\def\taua{\ifmmode \overline{\tau_{a}}\else $\overline{\tau_{a}}$\fi}
\def\tauc{\ifmmode \overline{\tau_{c}}\else $\overline{\tau_{c}}$\fi}
\def\nratio{\ifmmode n_{\rm IC}/n_{\rm C}\else  $n_{\rm IC}/n_{\rm C}$\fi}
\def\nc{\ifmmode n_{\rm C}\else  $n_{\rm C}$\fi}
\def\nic{\ifmmode n_{\rm IC}\else  $n_{\rm IC}$\fi}
\def\tratio{\ifmmode T_{\rm IC}/T_{\rm C}\else  $T_{\rm IC}/T_{\rm C}$\fi}

\usepackage{color}
\usepackage{amstext}

\begin{document}

   \title{The Lyman alpha Reference Sample VI: Lyman alpha escape from the edge-on disk galaxy Mrk1486}


   \author{Florent Duval
                 \inst{1,2}
                \,,
                G\"oran \"Ostlin
                \inst{2}
                \,,
                Matthew Hayes
                \inst{2}
                \,,
                Erik Zackrisson
                \inst{3}
                \,,
                 Anne Verhamme
                \inst{4}
                \,,
                Ivana Orlitova
                \inst{5}
                \,,
                Angela Adamo
                \inst{2}
                \,,
                Lucia Guaita
                \inst{6,2}
                \,,
                 Jens Melinder
                \inst{2}
                \,,
                John M. Cannon
                \inst{7}
                \,,       
                Peter Laursen   
                \inst{8}
                \,,         
                Thoger Rivera-Thorsen
                \inst{2}
                \,,
                        E. Christian Herenz
                \inst{9}
               \,,
                Pieter Gruyters
                \inst{2}
                \,,
                J.Miguel Mas-Hesse
                \inst{10}
                \,,   
                Daniel Kunth
                \inst{11}
                \,,       
                 Andreas Sandberg
                \inst{2}
                \,,            
                Daniel Schaerer
                \inst{4,12}
                \,,
                Tore M\r{a}nsson
                \inst{2}
}

   \institute{Institute for Cosmic Ray Research, The University of Tokyo, 5-1-5 Kashiwanoha, Kashiwa, Chiba 277-8582, Japan
              \email{fduval@icrr.u-tokyo.ac.jp }
         \and
         Department of Astronomy, Stockholm University, Oscar Klein Center, AlbaNova, Stockholm SE-106 91, Sweden
        \and
        Department of Physics and Astronomy, Division of Astronomy and Space Physics, Uppsala University, Box 516, 75120 Uppsala, Sweden
        \and
        Observatoire de Gen\`eve, Universit\'e de Gen\`eve, 51 Ch. des Maillettes, 1290 Versoix, Switzerland
         \and 
         Astronomical Institute, Czech Academy of Sciences, Bo\v cn{\'\i} II/1401, 141 00 Prague, Czech Republic
         \and
        INAF Observatorio Astronomico di Roma, Via Frascati 33,00040 Monteporzio (RM), Italy
        \and
        Department of Physics \&\ Astronomy, Macalester College, 1600 Grand Avenue, Saint Paul, MN 55105
        \and
        Dark Cosmology Centre, Niels Bohr Institute, University of Copenhagen, DK-2100, 
        \and
        AIP, Leibniz-Institut für Astrophysik Potsdam, An der Sternwarte 16, D-14482 Potsdam, Germany
        \and
        Centro de Astrobiologa (CSIC-INTA), Departamento de Astrofsica, POB 78, 28691, Villanueva de la Canada, Spain
        \and
        Institut d$'$Astrophysique Paris, 98bis Bd Arago, 75014 Paris
         \and
         CNRS, IRAP, 14 Avenue E. Belin, 31400 Toulouse, France
                     }

\authorrunning{F.Duval}
\titlerunning{Ly$\alpha$ escape from the edge-on disk galaxy Mrk1486}
   \date{Received 2 July 2015 ; accepted 19 October 2015  }

 
 \abstract
{Recent numerical simulations suggest that the strength of the Lyman alpha (\lya) line of star-forming disk galaxies 
strongly depends on the inclination at which they are observed: from edge-on to face-on, we expect to see a change from a strongly attenuated Ly$\alpha$ line to a strong Ly$\alpha$ emission line.
} 
{We aim to understand how a strong Ly$\alpha$ emission line is able to escape from the low-redshift highly inclined (edge-on) disk galaxy Mrk1486 (z $\sim$ 0.0338). To our knowledge, this work is the first observational study of Ly$\alpha$ transport inside an edge-on disk galaxy.  
}
{Using a large set of HST imaging and spectroscopic data, we investigated the interstellar medium (ISM) structure and the dominant source of Ly$\alpha$ radiation inside Mrk1486. Moreover, using a 3D Monte Carlo Ly$\alpha$ radiation transfer code, we studied the radiative transfer of Ly$\alpha$ and UV continuum photons inside a 3D geometry of neutral hydrogen (HI) and dust that models the ISM structure at the galaxy center. Our numerical simulations predicted the Ly$\alpha$ line profile that we then
compared to the one observed in the HST/COS spectrum of Mrk1486. 
}
{While a pronounced \lya\ absorption line emerges from the disk of Mrk1486, very extended Ly$\alpha$ structures are observed at large radii from the galaxy center: a large \lya-halo and two very bright Ly$\alpha$ regions located slightly above and below the disk plane. The analysis of IFU H$\alpha$ spectroscopic data of Mrk1486 indicates the presence of two bipolar outflowing halos of HI gas at the same location as these two bright Ly$\alpha$ regions. Comparing different diagnostic diagrams (such as [OIII]$_{5007}$/H$\beta$ versus [OI]$_{6300}$/H$\alpha$) to photo- and shock-ionization models, we find that the Ly$\alpha$ production of Mrk1486 is dominated by photoionization inside the galaxy disk. From this perspective, our numerical simulations succeed in reproducing the strength and shape of the observed Ly$\alpha$ emission line of Mrk1486 by assuming a scenario in which the Ly$\alpha$ photons are produced inside the galaxy disk, travel along the outflowing halos, and scatter on cool HI materials toward the observer.  
}
{Extended bipolar galactic winds are frequently observed from star-forming disk galaxies. Given the advantage Ly$\alpha$ photons take of such outflowing HI materials to easily escape from Mrk1486, this mechanism may explain the origin of strong Ly$\alpha$ emission lines frequently observed from highly inclined galaxies at high-redshift. This therefore challenges the robustness of the expected viewing-angle effect on the Ly$\alpha$ properties of star-forming disk galaxies. 
}

   \maketitle
%


\section{Introduction}

During the past two decades, the detection of high-redshift galaxies has experienced a great step forward by using the intrinsically brightest spectral signature of these objects: the Lyman alpha (\lya) recombination line of the hydrogen atom \citep{hu98, hu04, cowie98, rhoads00, taniguchi03, taniguchi05, shimasaku06, gronwall07, guaita10, ouchi03, Ouchi08, ouchi10, kashikawa11, hu10, hibon11, schenker14}. 
Thanks to its rest wavelength of 1216 \AA\ (which makes it accessible to optical or near-IR ground-based telescopes for redshifts $z > 2$) and its intrinsic brightness in the spectra of remote young galaxies \citep{patridge67,schaerer03,charlot_fall93}, the potential of the Ly$\alpha$ emission line for a redshift confirmation of distant galaxies (Chapman et al. 2005, Lehnert et al. 2010), for deriving the star formation rate (SFR), for investigating the circumgalactic medium of galaxies \citep{steidel10, steidel11, cantalupo14}, and for probing the reionization epoch \citep{ouchi10, fan02} is remarkable and unique in astrophysics and cosmology. 

As a result of its resonant nature, however, the Ly$\alpha$ line is subject to a very complex radiative transfer inside the interstellar medium (ISM) of star-forming galaxies. As revealed by different studies \citep{meier81,neufeld90,charlot_fall93,kunth98,tenorio99,mashesse03,ostlin09,ostlin14,hayes14}, many ISM quantities enter into the Ly$\alpha$ transport (dust, neutral hydrogen kinematics, geometry of the ISM, and hydrogen column density) and alter the most relevant features of the Ly$\alpha$ line (luminosity, Ly$\alpha$ equivalent width EW(\lya), and line profile). As long as no robust and reliable calibration technique of the Ly$\alpha$ line is established, these radiative transport problems will always prevent a proper interpretation of all very promising \lya-oriented studies in astrophysics and cosmology. 

To remedy this situation and to proceed in our understanding of the Ly$\alpha$ radiative transfer inside the ISM of galaxies, astronomers have taken an active interest in studying the Ly$\alpha$ line of low-redshift star-forming galaxies through the past three decades (see Hayes 2015 for a review). Although this strategy requires investing time on UV space telescopes, the high-resolution observations of those nearby analogs of remote young galaxies provide a more detailed scale in the ISM than observations of high-redshift targets. This leads to a more robust constraint of the relevant physical quantities that affect Ly$\alpha$ in the ISM of galaxies. 

One of those observational programs, the Lyman Alpha Reference Sample (LARS) of \cite{ostlin14} and \cite{hayes14}, has provided the deepest and most extensive data set for studying detailed Ly$\alpha$ astrophysics to date. LARS is a sample of 14 star-forming galaxies selected from the cross-correlated GALEX General Release 2 and SDSS DR6 catalogs at redshift between 0.028 and 0.18. All targets have been observed with the Hubble Space Telescope (HST; in cycles 18, 19, and 20) and with a large array of ground- and space-based telescopes (providing data of various electromagnetic fields, from X-rays to radio frequencies). Based on the first set of global results reported in \cite{hayes14}, the intensity maps derived from these observations show a large portion of \lya-emitting galaxies (12/14). Moreover, six galaxies might be detected by deep Ly$\alpha$ narrowband surveys that select objects with EW(\lya) above $20 \ {\rm \AA}$. 

Of all the very bright \lya-emitting galaxies of the LARS sample, Mrk1486 exhibits undoubtedly the most surprising and intriguing Ly$\alpha$ emission given its unfavorable geometry. Mrk1486 is a highly inclined (edge-on) disk-like star-forming galaxy \citep[the inclination is 85$^o$ from the line of sight;][]{Chisholm14} located at redshift ${\rm z} \sim 0.0338$. Based on a diameter of 4 kiloparsecs (kpc), \cite{hayes14} derived a stellar mass M$_*$ = 4.75x10$^9$ M$_{\odot}$ for Mrk1486, which means that this object is a low-mass galaxy. Its Ly$\alpha$ photometric properties characterize Mrk1486 as one of the highest Ly$\alpha$ luminosites (L$_{Ly\alpha}$ = 1.11$\times$10$^{42}$ erg s$^{-1}$), Ly$\alpha$ equivalent widths (${\rm EW(Ly\alpha)} = 45     \ {\rm \AA}$), and Ly$\alpha$ escape fractions of the LARS sample \citep[${\rm fesc(Ly\alpha)} = 0.174$;][]{hayes14}. While these properties indicate a very easy escape of Ly$\alpha$ photons from the galaxy, this result clearly contradicts recent numerical simulations that suggest the formation of either a very attenuated Ly$\alpha$ emission line or a pronounced Ly$\alpha$ absorption feature in the spectra of highly inclined low-mass disk galaxies \citep{laursen07, laursen09a, verhamme12, behrens14}. Based on the calculation of Ly$\alpha$   and UV continuum radiative transfer inside the ISM of high-resolution, dusty, isolated disk simulations, all numerical works have clearly highlighted a strong inclination effect on the Ly$\alpha$        properties of disk galaxies. 
Phenomenologically it is easy to understand that the observed inclination of a galaxy disk would affect the escape of \lya, since Ly$\alpha$ photons always tend to follow the path of least opacity in an ISM. Therefore, Ly$\alpha$ photons should escape more easily from a disk galaxy seen face-on than edge-on.

In the edge-on disk galaxy Mrk1486, it seems that a particular mechanism helps Ly$\alpha$ photons to easily escape from the core of the galaxy. Investigating the Ly$\alpha$        radiative transport and understanding the observed Ly$\alpha$     properties of the galaxy (i.e., the strength and line shape) is necessary and very challenging given our current knowledge of the Ly$\alpha$ transport in galactic environments. This is precisely the aim of this paper: to figure out the path that is followed by the Ly$\alpha$ photons to escape from Mrk1486.

The remainder of this paper is structured as follows. In Sect. 2 we describe the main observations of Mrk1486 from HST and other telescopes, as well as their reduction. In Sect. 3 we present the images of Mrk1486 in Ly$\alpha$ and other relevant wavelengths (H$\alpha$ and UV continuum wavelengths). Sects. 4, 5, and 6 present our results for the Ly$\alpha$ radiative transport inside the galaxy. We study the structure of the ISM in Sect. 4 and investigate the dominant source and strength of the intrinsic Ly$\alpha$ line in Sect. 5. In Sect. 6 we model the radiative transfer of the Ly$\alpha$ and UV continuum photons inside Mrk1486 and compare our predicted Ly$\alpha$ line profile to the one observed in our HST spectroscopic data. Section 7 is dedicated to a discussion of these results, and our main conclusions are summarized in Sect. 8.

Throughout we assume a cosmology of (H$_0$, $\Omega_0$, $\Omega_\lambda$) = (70 km s$^{-1}$ Mpc$^{-1}$, 0.3, 0.7).



\section{Data sample and image processing}

\begin{table*}
\caption{HST imaging observations that are used in this present analysis of Mrk1486. From the left column, we list the HST filters, the instrument or camera, the bandpass width ($\lambda_{min}$ - $\lambda_{max}$), the pivot wavelength, and the exposure time for each filter in question. (a) Archival observations from the program \# GO 11110 (PI McCandliss). (b) New imaging observations from the program \# GO 12310 (PI G. Ostlin).}        
\label{HSTfilter}      
\centering          
\begin{tabular}{c c c c c}     
\hline\hline
Filter & Instrument & Bandpass & Pivot wavelength & Exposure time \\
 & & (\AA) & (\AA) & (sec) \\
\hline
Archive data & &\\
\hline
F125LP (FUV)$^a$ & ACS/SBC & 1218 - 1987 & 1453 & 2669 \\
F140LP (FUV)$^a$ & ACS/SBC & 1313 - 1989 & 1538 & 2669 \\
F150LP (FUV)$^a$ & ACS/SBC & 1420 - 1992 & 1618 & 2669 \\
\hline
New observations & &\\
\hline
F336W (U)$^b$ & WFC3/UVIS & 3014 - 3707 & 3355 & 1020 \\
F438W (B)$^b$ & WFC3/UVIS & 3895 - 4710 & 4324 & 850 \\
F775W (I)$^b$ & WFC3/UVIS & 6869 - 8571 & 7648 & 620 \\
F502N (\hb)$^b$ & WFC3/UVIS & 4963 - 5059 & 5010 & 2000 \\
F673N (\ha)$^b$ & WFC3/UVIS & 6681 - 6860 & 6766 & 871 \\
\hline  
\hline                  
\end{tabular}
\end{table*}

\subsection{Data sample and reduction}

A large set of imaging and spectroscopic data were used in our analysis of Mrk1486. We describe the different data and their reduction here in detail. 

Mrk1486 was imaged with the HST from far-ultraviolet to optical ($\sim$ 8000 \AA) wavelengths. A summary of the HST data is given in Table\ \ref{HSTfilter}. In optical wavelengths, we used new Wide Field Camera 3 (WFC3) optical broadband imaging in U (F336W), B (F438W), and I (F775W) from the program \# GO 12310 (PI G. Ostlin). We supplemented these broadband data with narrowband imaging that isolate the H$\alpha$ and H$\beta$ lines of the galaxy (WFC3/F673N and WFC3/F502N, respectively). In UV wavelengths, we used archival ACS/SBC broadband imaging from the program \#GO 11110 (PI McCandliss; ACS/SBC F125LP, F140LP, and F150LP filters). The Ly$\alpha$ emission line of the galaxy was isolated from a UV long-pass pair-subtraction method that used all available UV and optical HST data of Mrk1486 \citep{hayes09}.

The spectroscopic observations of Mrk1486 were carried out with several instruments. First of all, we obtained new high-resolution ultraviolet spectra of Mrk1486 with the Cosmic Origins Spectrograph (COS) onboard the HST. These observations were associated with the program \# GO 12583 (PI. M.Hayes). To target both the H$\alpha$ line and the neighboring  N[II]$\lambda$$\lambda$6548.1,6583.6\AA\ emission lines, we then obtained two intermediate-resolution optical spectra with PMAS at the Calar Alto 3.5m telescope and with ALFOSC at the Nordic Optical Telescope (NOT).

\subsubsection{HST data reduction}

The individual reduced frames were obtained from the Mikulski Archive for Space Telescopes. We then processed the individual data frames and drizzled them to create stacked images in each filter with a pixel scale of 0.04\arcsec/px using DrizzlePac (version 1.1.16)\footnote{Gonzaga, S., Hack, W., Fruchter, A., Mack, J., eds. 2012, The DrizzlePac Handbook. (Baltimore, STScI)}. For a description of the general data reduction methodology we
refer to \cite{ostlin14} or \cite{hayes14}, although some improvements to the method were implemented for our work here. These improvements included point spread function (PSF) matching of the different filters, charge transfer inefficiency corrections for the WFC3 filters, and more efficient cosmic-ray masking. The steps taken to improve these are briefly summarized below (more details will be given in an forthcoming paper, Melinder et al. in prep).

To maximize the time spent on target, all of the optical imaging was observed using long exposures (two exposures per HST orbit). Unfortunately, this means that removing cosmic rays was not as efficient as it would have been with more exposures. We therefore used LAcosmic \citep{vandokkum01} on the individual frames to mask cosmic rays before drizzling. Given the fairly small amount of cosmic rays in the frames (on the order of 10-100), the effect on the stacked images is mainly cosmetic, but when applying the PSF matching convolution kernels to images with remaining image defects, the faulty information spreads over a larger region and becomes problematic. The charge transfer inefficiency (CTI) of the optical detectors on HST is a problem for our study. We corrected the individual data frames for this effect by using the pixel-based empirical CTI correction software \citep{anderson10} supplied by STScI in all optical filters\footnote{http://www.stsci.edu/hst/wfc3/tools/cte$\_$tools}. All results presented in earlier papers were using CTI corrections only for the ACS/WFC filters, while we here also include corrections of the WFC3/UVIS filters. A proper CTI correction in the WFC3/UVIS filters successfully removes almost all of the medium-scale noise patterns associated with the charge traps and thus substantially
improves the image fidelity. 

The final drizzled and registered images in each filter were then convolved with a delta-function-based kernel to match the PSFs (similar to the technique described in \cite{becker12} using the filter with the widest PSF (F775W) as the reference. This is especially important when comparing the UV filters in ACS/SBC to the optical filters because the SBC PSF profile is significantly different from the UVIS PSF. Because the field of view in these observations is small and we study UV wavelengths, there were no stars in common to all filters that could be used to match
the PSF. Instead, we created models of the PSFs using the techniques described in \cite{krist11}. Details of the PSF modeling and computing optimum convolution kernels will be presented in Melinder et al. (in prep.). Using PSF-matched data is important for our work here because we study emission on small scales ($\sim$0.1 arcseconds) in the center of the galaxy where flux gradients are high. Colors determined from non-matched data will suffer from unknown systematic errors that depend on the PSF differences and the flux gradient at the location.

We binned all HST scientific frames and weighted maps of Mrk1486 using the Voronoi tesselation algorithm. As a reference bandpass, we used the UV broadband image obtained through the SBC/F125LP fillter. A threshold signal-to-noise ratio S/N = 10 per bin was required, but we allowed the bin sizes to exceed 1", that is, about 625 pixels. 

\subsubsection{PMAS spectroscopy data}

  \begin{figure}
   \centering
   \includegraphics[width=90mm]{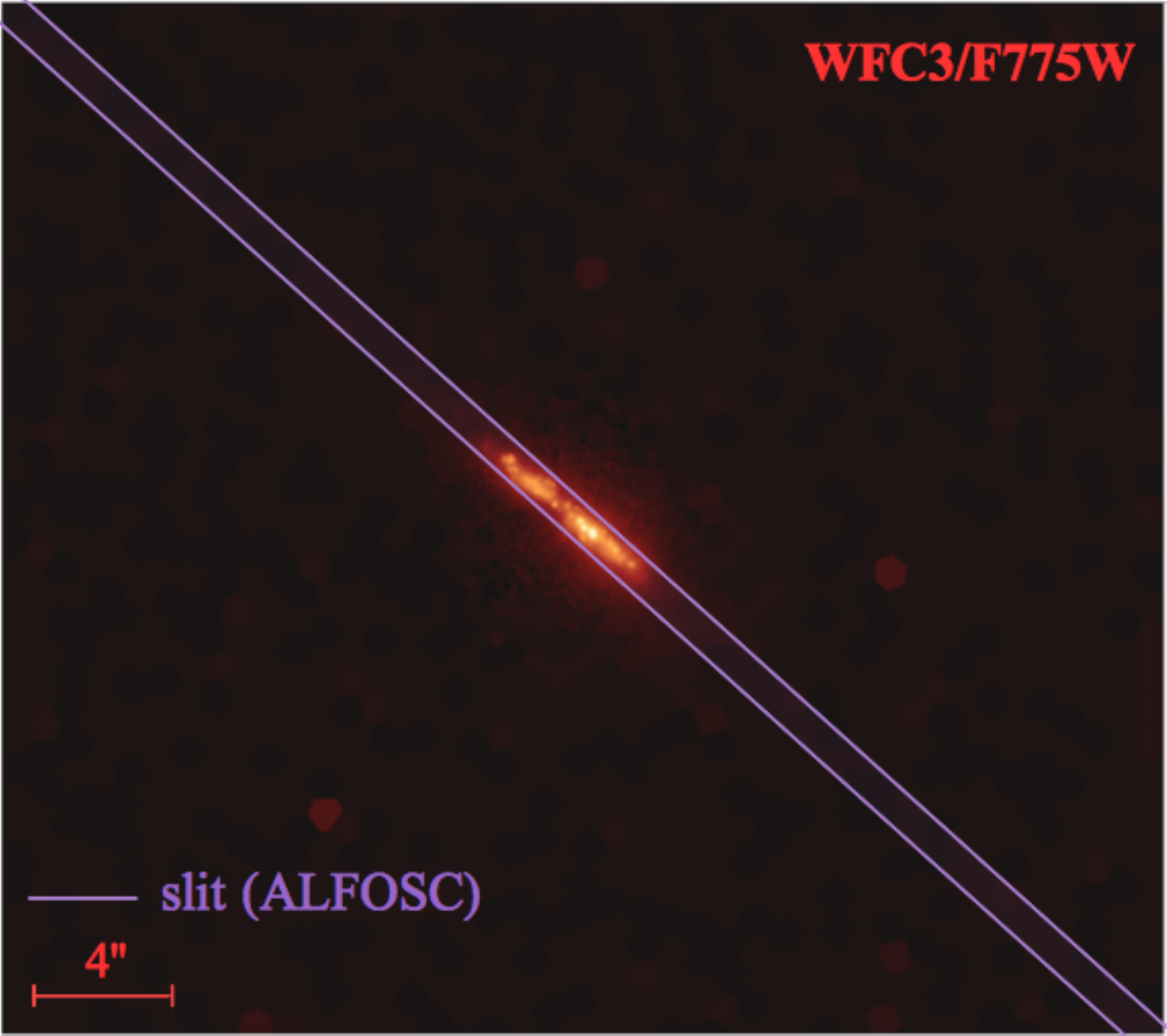}
      \caption{Image of Mrk1486 as obtained from the WFC3/F775W exposure with HST. We show the exact position of the slit of the ALFOSC spectrograph. It was placed along the major axis of the galaxy disk.  
              }
       \label{slit}
   \end{figure}


Integral field spectroscopy observations targeting the H$\alpha$ emission of all LARS galaxies were carried out with the Potsdam Multi Aperture Spectrophotometer in Lens-Array mode \citep[PAMS;][]{roth05} at the 3.5m telescope of Calar Alto Observatory. We briefly summarize the observations, data reduction, and analysis steps relevant for the present study. A more detailed description will be given in a subsequent publication that will present an analysis of the H$\alpha$   kinematics of the full sample \citep{herenz15}. 

Mrk1486 was observed on March 12, 2014. To cover the whole extent of the galaxy, we used the 2$\times$magnification mode, giving us a field of view of 16\arcsec{}$\times$16\arcsec{} sampled at 1\arcsec{}$\times$1\arcsec{}. The R1200 grating was mounted on the spectrograph, and the CCD was read out unbinned along the dispersion axis \citep{Roth10}. We took three on-target exposures with $t_\mathrm{exp.}=1800\,$s and exposed neighboring blank sky with $t_\mathrm{exp.}=400\,$s before the first and after the second exposure. At the beginning of the night, the \cite{oke90} spectrophotometric standard star BD+75d325 was observed. 

To reduced the raw data we employed the p3d\footnote{http://p3d.sourceforge.net}-pipeline     \citep{sandin10,sandin11}, which covers all the important steps in the reduction of fiber-fed integral field spectroscopic data: 
bias subtraction, flat fielding, cosmic-ray removal, tracing, and extraction of spectra. The final data products resulting from \texttt{p3d} are flux-calibrated data cubes and the associated pipeline-propagated variances for all on-target and sky exposures. The data cubes cover a wavelength range from 5970\AA{} to 7700\AA{}. To remove the telluric emission, we subtracted from the on-target exposure cubes the blank sky cubes nearest in time, and finally we co-added all three on-target exposure cubes using the variance-weighted mean; variances accordingly in the last two steps.       


\subsubsection{COS spectroscopy data reduction}

Mrk1486 was observed with COS onboard HST using the primary science aperture and grating G130M. We first used the ACS/SBC/F140LP and WFC3/UVIS/F336W, obtained as part of the imaging campaign \citep{ostlin14} to identify the regions of peak UV surface brightness.  Knowing the target coordinates for the COS observations precisely, we acquired the target with {\tt ACQ/IMG} with the NUV
imager. We performed spectral observations with the G130M grating; we chose
the {\tt CENWAVE} setting  to include Ly$\alpha$ and the largest possible number of low-ionization absorption lines. We performed spectral dithers using three {\tt FP-POS} settings. The total integration time used was 2166.56 seconds.

Data reduction was performed with {\tt CALCOS} version 2.15.6 (2011-11-03) pipeline tools.  Because our absorption features do not fall close to geocoronal emission lines, we made no attempt to correct for OI lines near 1302 \AA\ as was done in \citet{james14}, instead we collected continuum photons obtained at all Earth-limb angles.  Finally, data were resampled onto a uniform wavelength grid using the $\Delta \lambda$ corresponding to the coarsest value, and were stacked, for which we rejected any pixel for which the data quality flag was $>0$.

More details on the COS observations and reductions for the LARS sample are described in \cite{rivera15}

\subsubsection{ALFOSC spectroscopy data reduction}

Long-slit spectra of Mrk1486 were obtained on May 21, 2013, using the ALFOSC spectrograph at the Nordic Optical Telescope (NOT), La Palma, Spain. The ALFOSC observations were made at intermediate resolution (R=1000) and the spectral range covered 5825-8350 \AA. A slit width of 1'' was used for the spectroscopy. As shown in Fig.\ \ref{slit}, the latter was placed along the major axis of the galaxy disk to measure the spectrum variations at different distances from the galaxy center  (in particular, the H$\alpha$ and the N[II]$\lambda$$\lambda$6548.1,6583.6\AA\ emission lines). Three frames were exposed during the observations, with an exposure time of 1200 sec each. All 2D-spectra were bias-subtracted, flat-field
corrected, and wavelength- and flux-calibrated following usual IRAF procedures.

\begin{figure*}
   \centering
   \includegraphics[width=82mm]{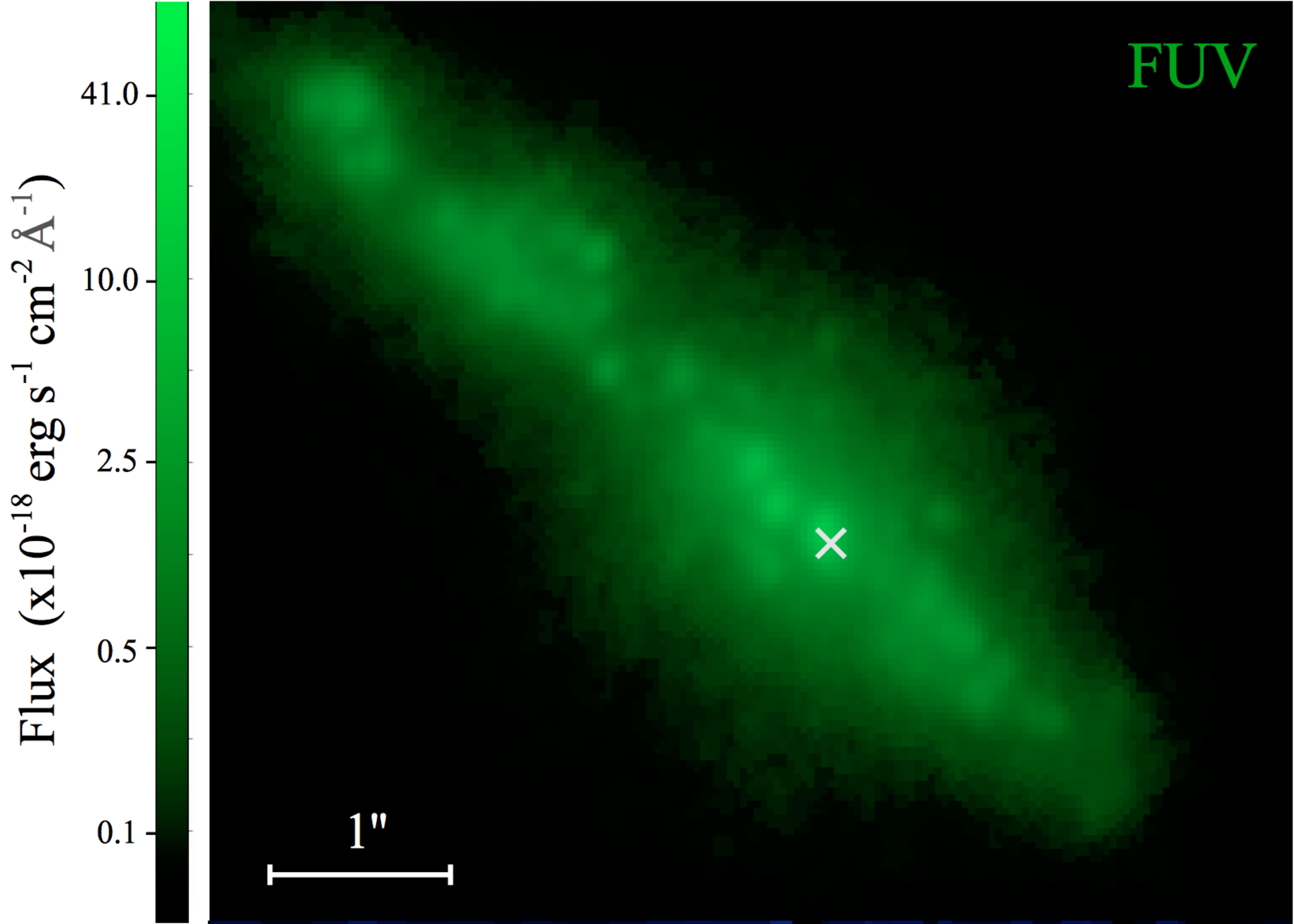}
   \includegraphics[width=82mm]{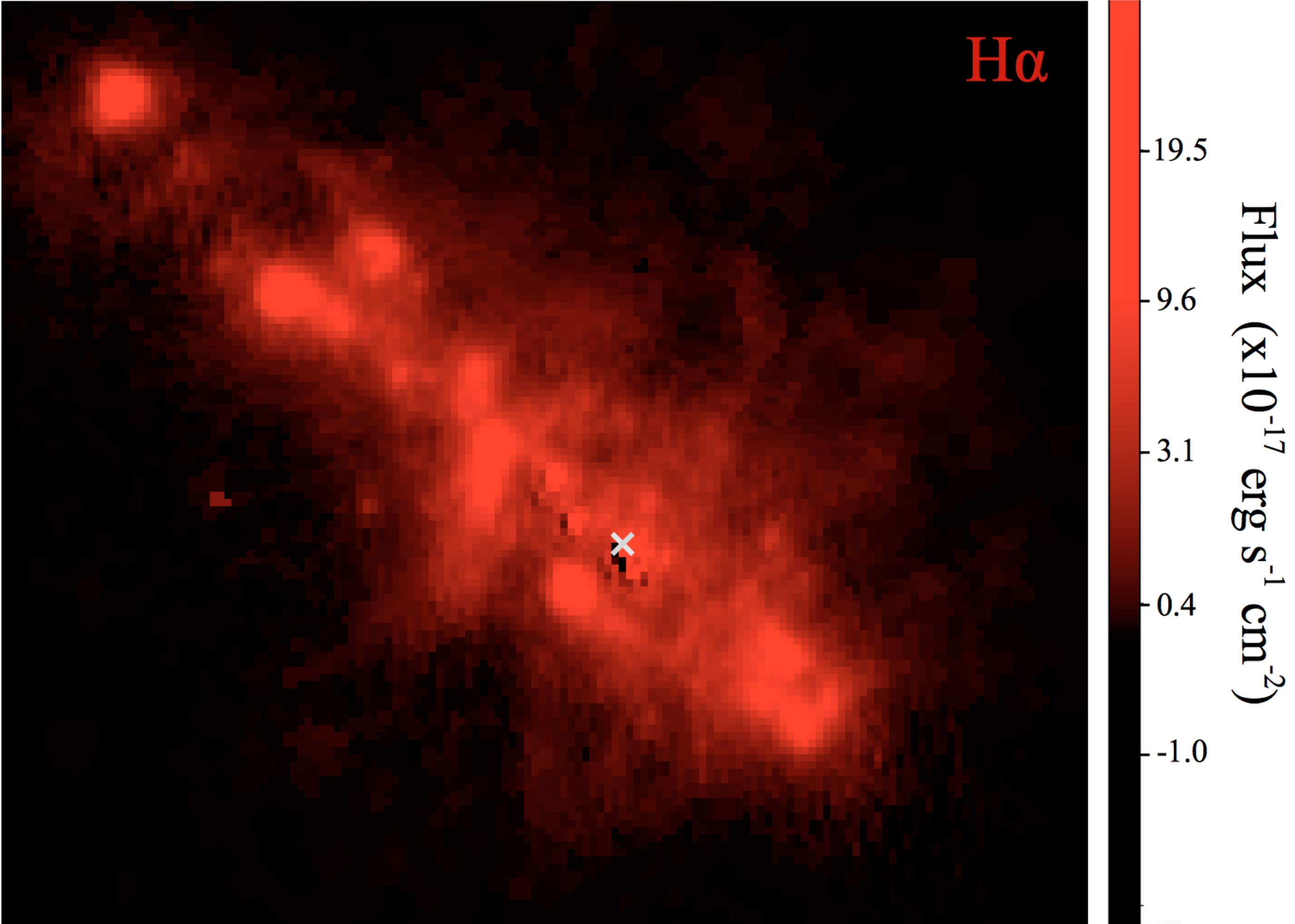}
   \includegraphics[width=82mm]{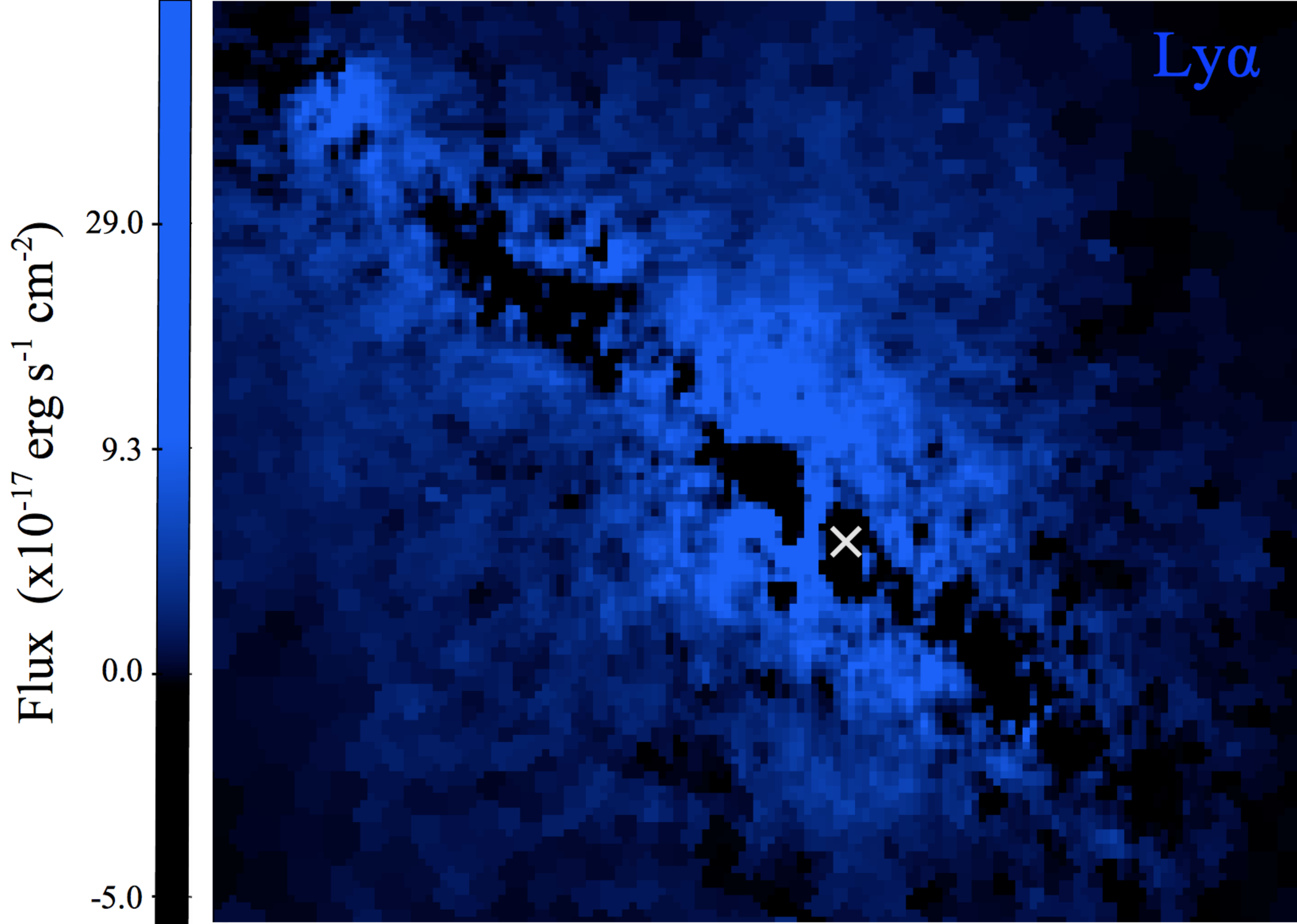}
   \includegraphics[width=81.8mm]{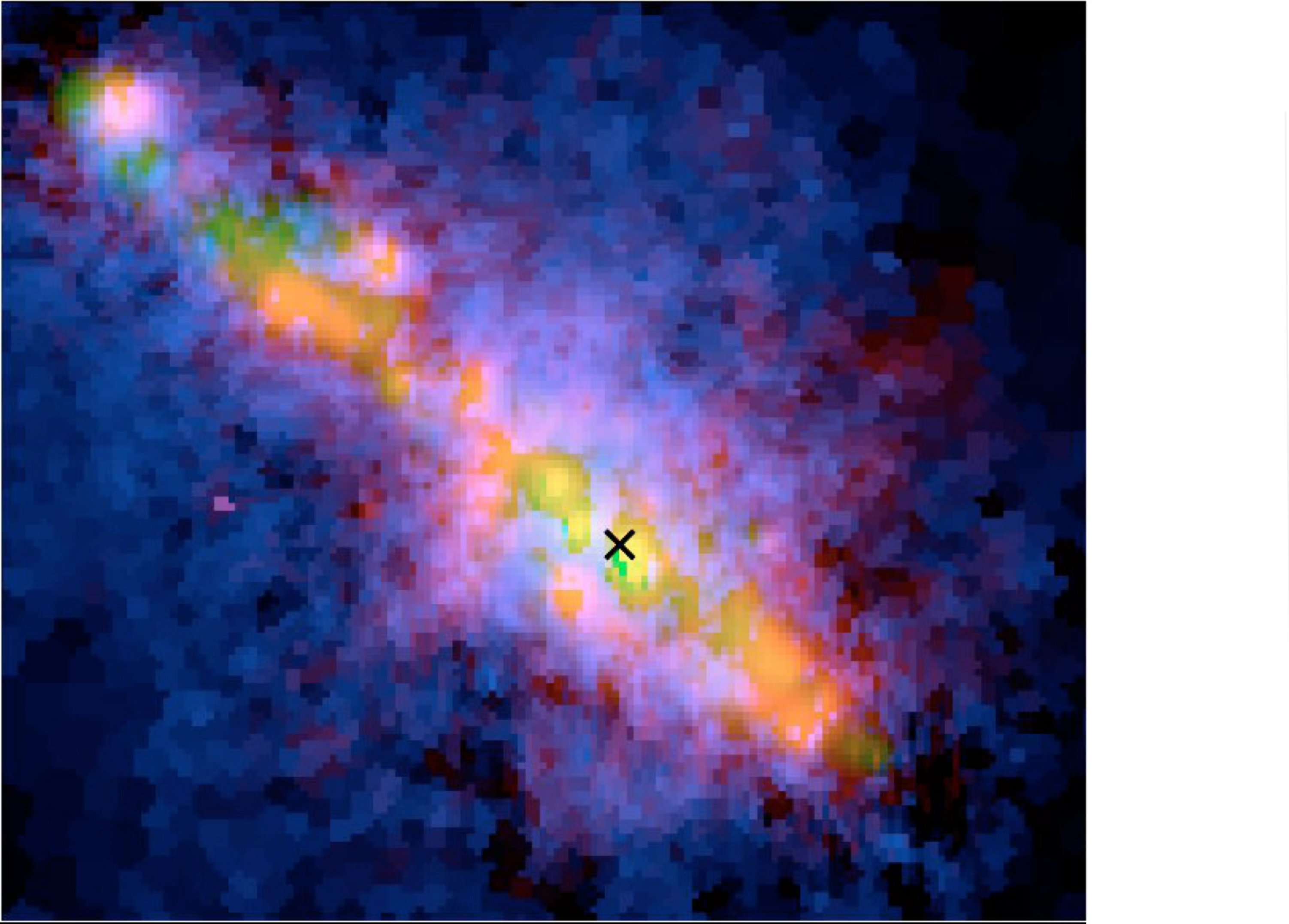}
      \caption{Multiwavelength images of Mrk1486. The \textit{upper left panel} shows the UV continuum at a rest-frame wavelength of 1600 \AA\ (image through the SBC/F150LP filter). This emission traces the most massive stars that suffer low dust extinction. The \textit{upper right panel} shows the continuum-subtracted H$\alpha$ map of the galaxy. It traces the ionized hydrogen nebulae (as the result of star formation) from which the \ha, H$\beta$ and Ly$\alpha$ photons are produced. The \textit{bottom left panel} shows the continuum-subtracted Ly$\alpha$ map of Mrk1486, and the \textit{bottom right panel} is made by superimposing all images on top of each other. Given the redshift of the galaxy (z $\sim$ 0.0338), 1 arcsec corresponds to 0.67 kpc here. Finally, the cross shown in each panel indicates the position of the brightest UV continuum pixel used as the center of the isophotal area in the right panel of Fig.\ \ref{SB_LARS5}.  
              }
       \label{maps}
   \end{figure*}

\subsection{Derivation of the Ly$\alpha$ line and continuum maps of Mrk1486}

Creating the continuum-subtracted Ly$\alpha$ map of nearby star-forming galaxies in the SBC-only data set is nontrivial and requires a good knowledge of many parameters: the stellar UV continuum emission near \lya, the stellar absorption below \lya, and the presence of UV interstellar absorption lines (LIS) in the spectral
energy distribution (SED) of the galaxy \citep{hayes05, hayes06, hayes09}. We derived the Ly$\alpha$ map of Mrk1486 throughout
from a continuum-subtraction method using the \textit{Lyman-alpha eXtraction Software} (LaXs; Hayes et al. 2009) and our full set of HST data (see Table 1). 

While the combination of the ACS/F125LP and ACS/F140LP filters allowed us to generate a synthetic narrow bandpass that covers the Ly$\alpha$ line of Mrk1486 (i.e., the combination F125LP-F140LP; \"Ostlin et al. 2014), all information regarding the UV continuum of the galaxy are predicted by SED fitting on a pixel-by-pixel scale using LaXs. From a photometric point of view, what we here
refer to as Ly$\alpha$  is any deviation from the predicted stellar UV continuum, integrated over the synthetic narrow Ly$\alpha$ bandpass. 

A more detailed description of our continuum-subtraction method can be found in \cite{hayes14} and \cite{ostlin14}.

\section{Images and line maps of Mrk1486}

\subsection{\lya, H$\alpha,$ and UV maps}

  \begin{figure*}
   \centering
   \includegraphics[width=91mm]{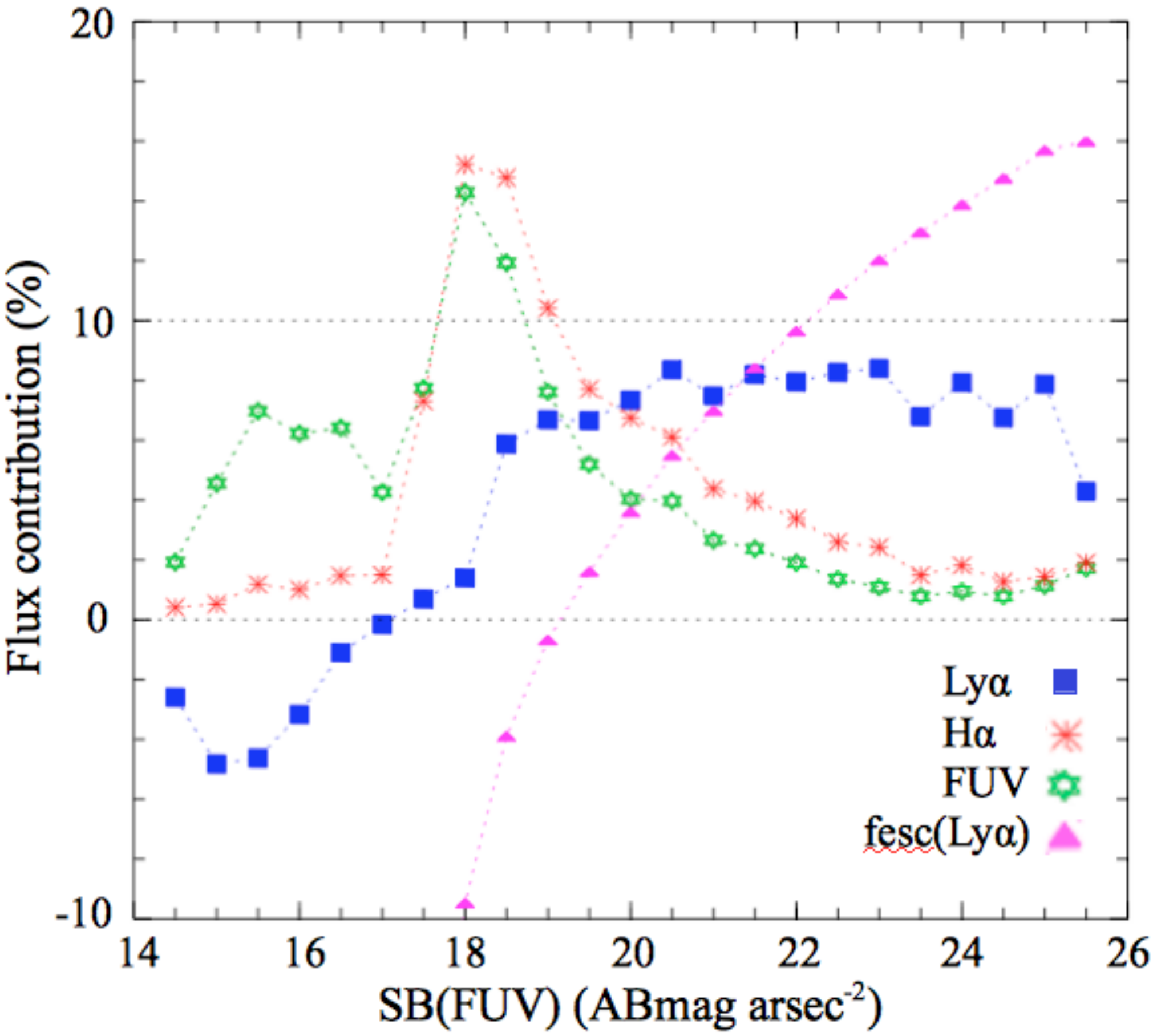}
   \includegraphics[width=91mm]{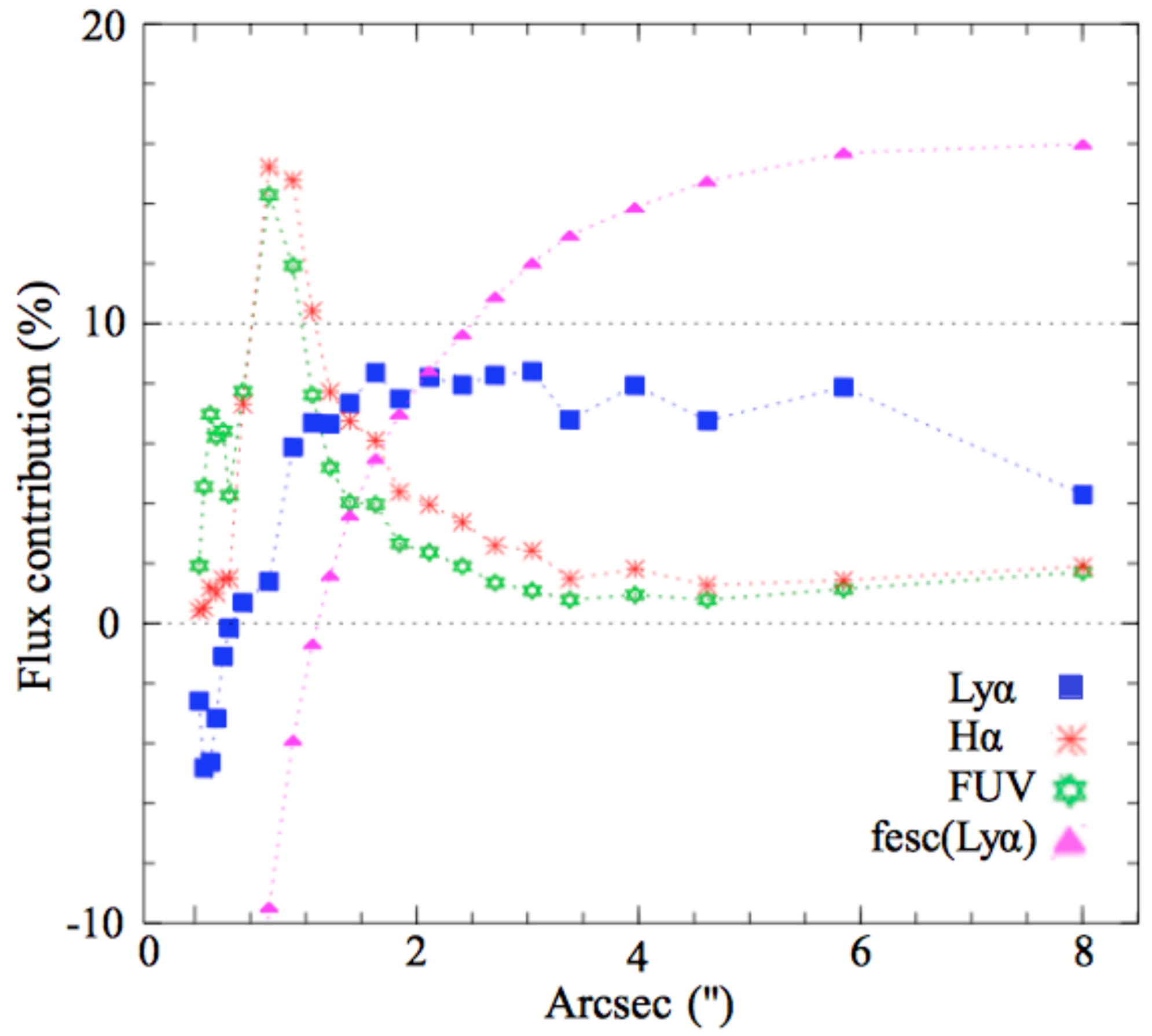}
      \caption{Escape fraction fesc(\lya) and spatial distribution of \lya, H$\alpha,$ and UV continuum radiation. The two panels show the fractional contribution to the integrated flux for UV (green), H$\alpha$ (red stars), and Ly$\alpha$ (blue squares) as a function of the far-UV surface brightness (ABmag arcsec$^{-2}$; left panel) or equivalent radius (i.e., the isophotal radius; right panel). We express here the radius in arcsec ('') (1'' corresponds to 0.67 kpc at z = 0.0338). The cumulative Ly$\alpha$ escape fraction is shown with magenta triangles. It starts out as negative and increases monotonically to finally reach 16\% at large radii from the center of the galaxy.
              }
       \label{SB_LARS5}
   \end{figure*}
   
We present our results in Fig.\ \ref{maps}. The upper left panel shows the observed intensity in the far-ultraviolet (at the rest-frame wavelength $\sim$1600 \AA, as observed though the WFC3/F150LP exposure with the HST). This radiation traces the most massive and hottest stars of the galaxy that suffer low dust extinction (O, B, and A type stars). In particular, the brightest UV knots might correspond to the sites with a copious production of ionizing photons (i.e., Lyman continuum photons, LyC). The upper right panel shows the continuum-subtracted H$\alpha$ map of the galaxy, as derived from LaXs. This hydrogen recombination line traces the ionized nebulae in which the ionizing photons are absorbed by the neutral gas of the ISM and are converted into nebular emission lines. 

The UV and H$\alpha$ maps show that both radiations are mostly concentrated in the galaxy disk. From the UV continuum image we can easily distinguish the structure of the elongated disk of Mrk1486 (seen edge-on in Fig.\ \ref{maps}). For the H$\alpha$ emission, we note smooth and extended H$\alpha$ filamentary-like structures at large scale from the disk plane (up to 3 kpc away). 

The discrepancies that appear between the H$\alpha$ and UV emission along the galaxy disk may have several explanations. This can result from dust extinction (which affects the UV continuum radiation more than H$\alpha$) or the presence of old and non-ionizing stellar populations (such as A-type stars, which are known to be bright in UV, but are not massive enough to produce ionizing photons). Moreover, the relative difference between the H$\alpha$ and UV intensity from individual star-forming knots may be explained by the fact that ionized gas can be located at a significant distance from the ionizing clusters because of ISM wiping by the effect of stellar winds and supernova explosions.  


Because both the H$\alpha$ and Ly$\alpha$ photons are produced in the same locations, the H$\alpha$ emission line can be used as a probe of the underlying production of the Ly$\alpha$ line in Mrk1486. We show in the bottom left panel of Fig.\ \ref{maps} and in Fig.      \ref{Lya_COS_ape} the continuum-subtracted Ly$\alpha$ map of Mrk1486. Interestingly, the Ly$\alpha$ morphology of Mrk1486 is clearly different from its counterparts in UV continuum and \ha. While the UV and H$\alpha$ radiation are mostly concentrated in the galaxy disk, this region of the galaxy exhibits large Ly$\alpha$ -absorbing regions. Bright or smooth extended Ly$\alpha$
structures are instead visible outside the galaxy disk. In particular, we note an extended Ly$\alpha$ halo that surrounds Mrk1486 at a radius of up to 4 kpc from the galaxy center \citep{hayes13}. 

The intensity scaling of all images shown in Fig.\ \ref{maps} is logarithmic. We here adopt a cut level that shows the maximum of the structures in each image and reduces the faintest features to background noise. However, because the absorption Ly$\alpha$ regions are cut at the same level, this image prevents any qualitative analysis of the Ly$\alpha$ strength below zero. We therefore provide a more quantitative comparison of the images in absolute surface intensity in Fig.\ \ref{SB_LARS5}. 
     
\subsection{Surface photometry}

\begin{table*}
\caption{Photometric properties and dust extinction of each ISM region shown in Fig.\ \ref{ISM_structure}. While the contribution in Ly$\alpha$ emission is shown in Col. 4, the UV contribution is shown in Col. 5 (both are derived from the analysis of the Ly$\alpha$ and UV maps of Mrk1486; see Fig.\ \ref{maps}).  
 The nebular color excess E$_{B-V}^{{\rm Neb}}$ (7) is measured from the Balmer decrement \ha/\hb       (6), assuming an intrinsic ratio of 2.86 for a 10,000 K gas (Case B, Osterbrock 1989) and the ratio determined using the SMC extinction law.}.      
\label{table:obs_proper}      
\centering          
\begin{tabular}{c c c c c c c}  
\hline\hline
Region & SB(\lya) & SB(UV) & Ly$\alpha$ contribution & UV contribution & H$\alpha$/H$\beta$ & E$_{B-V}^{{\rm Neb}}$ \\
  & 10$^{-14}$ erg s$^{-1}$ cm$^{-2}$ & 10$^{-15}$ erg s$^{-1}$ cm$^{-2}$ \AA$^{-1}$ & & & & \\ 
(1) & (2) & (3) & (4) & (5) & (6) & (7) \\
\hline
Galaxy disk & -7.96$^{+0.49}_{-0.31}$ & 5.05$^{+0.29}_{-0.36}$ & 0\%$^{a}$ & 76$^{+4}_{-5}$ \%\ & 3.62$^{+0.04}_{-0.02}$ & 0.34$^{+0.02}_{-0.02}$ \\
Halo 1 & 4.40$^{+0.02}_{-0.12}$ & 0.76$^{+0.15}_{-0.13}$ & 56$^{+2}_{-3}$ \% &  11$^{+3}_{-2}$\% & 3.09$^{+0.16}_{-0.15}$ & 0.11$^{+0.07}_{-0.07}$ \\
Halo 2 & 3.49$^{+0.34}_{-0.33}$ & 0.83$^{+0.13}_{-0.11}$ & 44$^{+3}_{-2}$ \% &  12$^{+3}_{-1}$ \% & 3.26$^{+0.16}_{-0.16}$ & 0.18$^{+0.07}_{-0.07}$ \\
\hline  
\hline                     
\end{tabular}
a) instead of adopting a negative value for the Ly$\alpha$ contribution of the galaxy disk, we arbitrarily set this parameter to 0\%.
\label{obs_brightness}
\end{table*} 

In Fig.\ \ref{SB_LARS5} we quantitatively compare the distribution of Ly$\alpha$ to that of the H$\alpha$ and UV continua. 
More precisely, we show the fractional contribution to the integrated flux for UV, H$\alpha,$ and Ly$\alpha$ as a function of UV surface brightness (integrated down to SB(UV) = 26 AB magnitudes per square arcsec; left panel) and radius (right panel). The center of the isophotal area is the brightest UV continuum pixel in the UV image (see Fig.\ \ref{maps}).   

As mentioned above, we note that the H$\alpha$ surface brightness
closely follows the one of the UV continuum. Although a relatively faint H$\alpha$ emission is detected from the brightest regions in UV continuum (SB(UV) $<$ 17 ABmag arcsec$^{-2}$, left panel in Fig.\ \ref{SB_LARS5}), most of the UV and H$\alpha$ luminosities come from regions showing a SB(UV) between 17 and 20 ABmag arcsec$^{-2}$. The situation for Ly$\alpha$ photons is different because most of the Ly$\alpha$ luminosity escapes from the galaxy at lower UV surface brightness. While the brightest regions in UV continuum exhibit a strong \lya\ absorption line, a net increase of the local Ly$\alpha$ luminosity starts to be observed for SB(UV) $>$ 18 ABmag.arcsec$^{-2}$. 

The luminosity distribution as a function of the UV isophotal radius ($\sqrt{{Area}/\pi}$, right panel) shows that most of the H$\alpha$ and UV emission escapes from the galaxy close to the disk (i.e., inside a radius R = 1.5 arcsec). The radial profile of the Ly$\alpha$ luminosity is clearly different. On the one hand, we note a clear absorption feature in the center of Mrk1486 (with negative contribution for ${\rm R}$ $<$ 0.5 arcsec, corresponding to the thickness of the disk). On the other hand, the Ly$\alpha$ radiation largely escapes from regions located outside the core of the galaxy. 

From the two panels we can derive the cumulative Ly$\alpha$ escape fraction fesc(\lya) (shown with magenta triangles in both panels). This parameter is derived as 

\begin{equation}
fesc(Ly\alpha) = \frac{F_{obs}(Ly\alpha)}{F_{int}(Ly\alpha)} = \frac{F_{obs}(Ly\alpha)}{8.7 \times F_{int}(H\alpha)} 
,\end{equation} 
where F$_{obs}$(\lya) corresponds to the integrated Ly$\alpha$ flux inside the isophotal area and F$_{int}$(\ha) is the integrated intrinsic H$\alpha$ one. We used the SMC dust extinction law  of \cite{prevot84} to correct the H$\alpha$ flux for dust attenuation. As expected, fesc(\lya) is negative at high UV surface brightness/small radius\footnote{Nonetheless, a negative fesc(\lya) has no physical meaning. It just indicates that the integrated Ly$\alpha$ line appears in absorption and that no escape of \lya\ photon is observed from this region of the galaxy.} (this is due to the negative Ly$\alpha$ flux we get from this region of the galaxy). Conversely, fesc(\lya) rapidly increases by up to 16\% above 4.5 arsec (i.e., above R = 3 kpc)  from the center of the galaxy disk
at larger radii. This asymptotic value is one of the highest Ly$\alpha$ escape fractions measured in the LARS sample \citep{hayes14}. 

\subsection{Reliability of the Ly$\alpha$ map of Mrk1486}  

As explained in Sect. 2.2, we derived the continuum and the line maps of Mrk1486 from a SED fitting approach \citep{hayes09, hayes14}. Throughout this process we used two instantaneous-burst stellar population models to fit the HST observations on a pixel-by-pixel scale. We considered both a very young stellar population (which
is responsible for the UV radiation captured through each pixel) and an old one (which is formed by previous star-forming events and responsible for a large portion of the optical radiation). The continuum subtraction of the Ly$\alpha$ line was then made from the best SED model. 

Overall, four physical quantities assigned to the stellar populations were considered as free-parameters to fit the HST data: the \textit{stellar dust extinction} E$_{B-V}^{{\rm Stel.}}$, the \textit{mass} of the old stellar population, and the \textit{age} and \textit{mass} of the young stellar population. However, for simplicity, we kept two important physical parameters in our SED fits constant: 1) the \textit{age of the old stellar population} (set to 4 Gyr
because this age provides the best chi-square when fitting the HST observations) and 2) the \textit{NII/H$\alpha$ emission line ratio} (NII/H$\alpha$ = 0.0378, as derived from the Sloan Digital Sky Survey spectrum of Mrk1486). Our approach thus assumes no spatial variation of these two quantities in Mrk1486, which cannot be true in reality and might challenge the reliability of our continuum-subtracted Ly$\alpha$ map shown in Fig.\ \ref{maps}. 

Before studying the Ly$\alpha$ radiative transfer inside the ISM of Mrk1486 (Sect. 6), we therefore investigated how strongly the Ly$\alpha$ map of Mrk1486 might be affected by even weak variations in these two input parameters. For this purpose, we computed different Ly$\alpha$ maps of Mrk1486 using various ages for the old stellar population (from 2 to 6 Gyr) and NII/H$\alpha$ line ratios (from 0.035 to 0.042, as measured along the galaxy disk from our ALFOSC spectroscopic data). In conclusion, no noticeable change was observed in the output continuum-subtracted Ly$\alpha$ map when any of these parameters were changed. The cumulative Ly$\alpha$ flux only varies by 0.01\% and all changes are exclusively observed from pixels located on the edge of the Ly$\alpha$ map. This test thus ensures the reliability of the Ly$\alpha$ map shown in Fig. \ref{maps}. 

The only consequences of varying the stellar age and the NII/H$\alpha$ line ratio are seen in 1) the total mass of the old stellar population when fitting the HST data points, 2) the continuum subtraction of the H$\alpha$ line, and 3) the derived nebular extinction E$_{B-V}^{{\rm Neb}}$ map of the galaxy (derived from the Balmer decrement \ha/\hb). We obtain a variation of 57\%\ for the old stellar mass, 1\%\ for the integrated \ha\ flux, and 2\% for the nebular extinction. However, none of these changes has a noticeable affect on any of the results presented below.

\section{ISM structure and possible scenarios for Ly$\alpha$ escape inside the COS aperture}

Constraining the ISM structure of Mrk1486 and identifying the bright regions in Ly$\alpha$ is the key to determine how Ly$\alpha$ photons easily escape from the ISM of the galaxy. In this section we identify the main ISM regions inside the COS aperture of Mrk1486. This will allow us to propose different scenarios regarding the path Ly$\alpha$ photons follow to escape from the galaxy. 

\subsection{Structure of the ISM inside the COS aperture}
   
\begin{figure*}
   \centering
   \includegraphics[width=78mm]{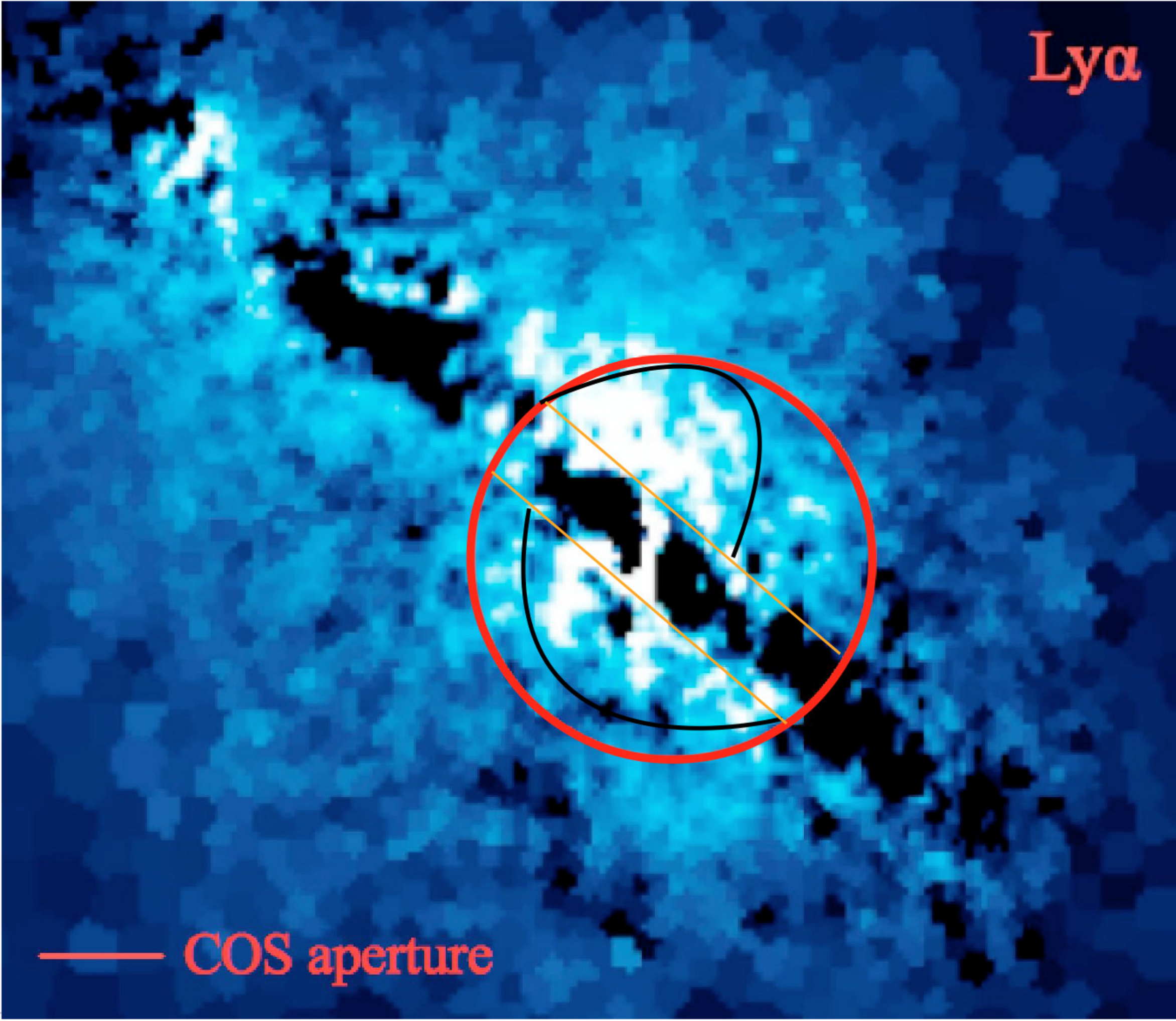}
   \includegraphics[width=104mm]{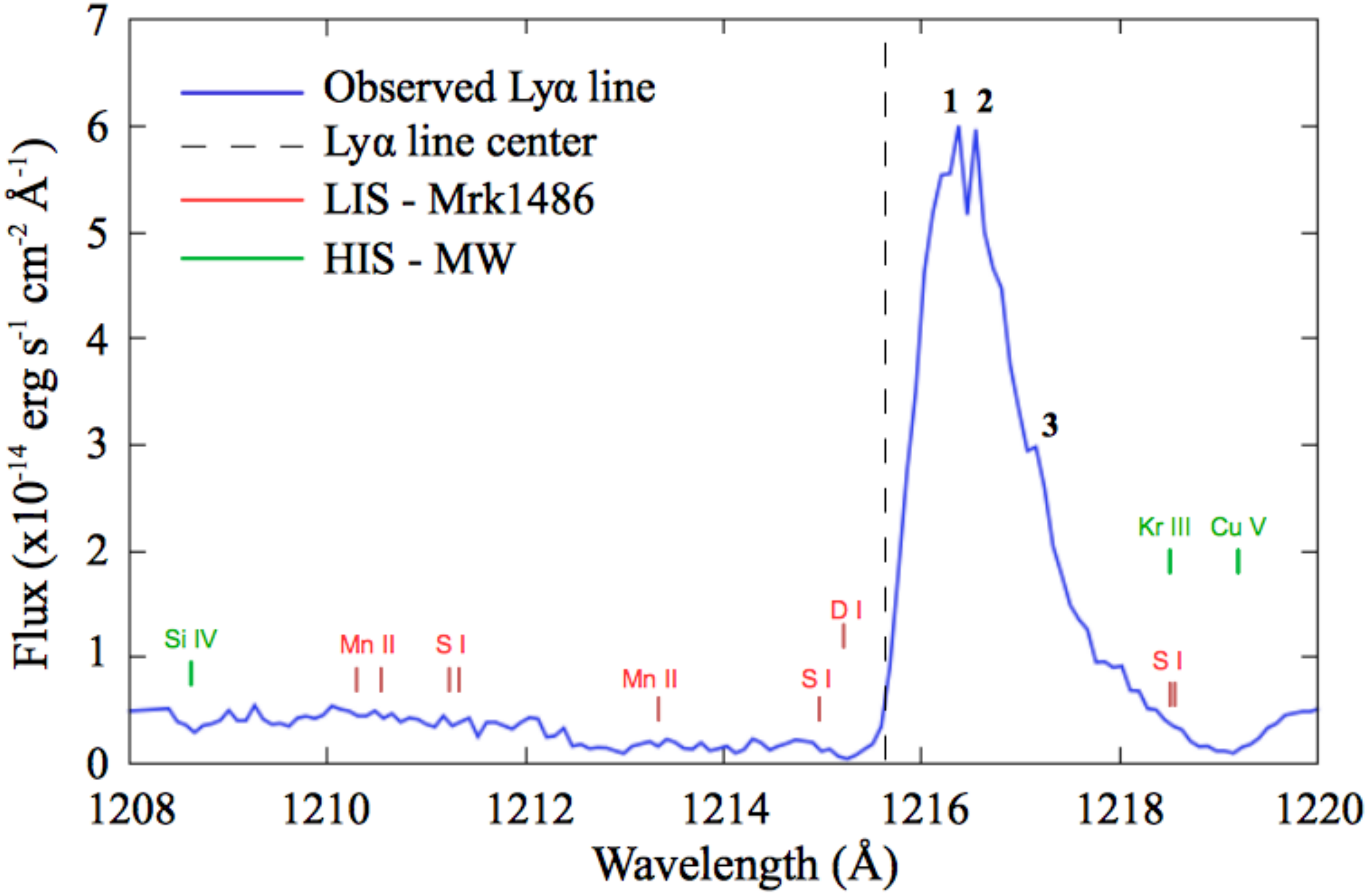}
      \caption{\textit{Left panel}: Ly$\alpha$ map of Mrk1486 and the position of the COS aperture (thick red circle). We also add the location of each ISM region identified in Fig. \ref{ISM_structure}. \textit{Right panel}: The Ly$\alpha$ line profile of Mrk1486 within the COS aperture. The x-axis shows the rest-frame wavelength. The Ly$\alpha$ line profile exhibits a P-cygni profile in which three significant peaks can be identified (listed by numbers on top of the Ly$\alpha$ line profile).  Finally, different LIS and HIS absorption lines from the ISM of Mrk1486 (red lines) and the Milky Way (green lines) are identified in the COS spectrum. 
              }
       \label{Lya_COS_ape}
   \end{figure*}

\begin{figure}
   \centering
   \includegraphics[width=85mm]{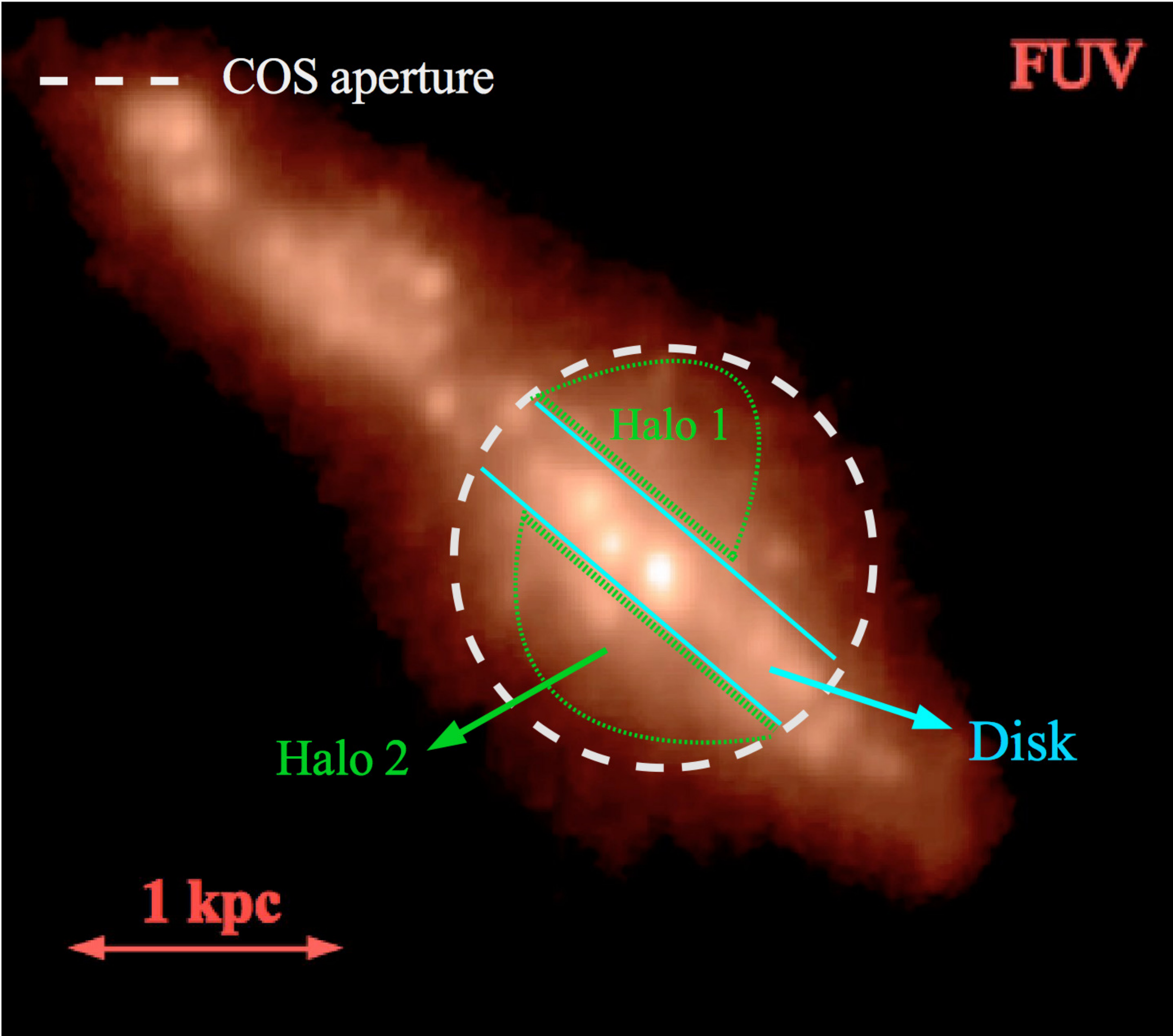}
      \caption{ISM structure of Mrk1486 inside the COS aperture. Based on the Ly$\alpha$ map of Mrk1486, three components are identified inside the COS aperture: the galaxy disk and two outflowing halos (halos 1 and 2).
              }
       \label{ISM_structure}
   \end{figure}

The apparent structure of the ISM is relatively complex inside the COS aperture. We show in the left panel of Fig.\ \ref{Lya_COS_ape} the continuum-subtracted Ly$\alpha$ map of Mrk1486 and the region covered by the COS aperture. As revealed by our imaging and spectroscopy data, the ISM is composed of three distinct regions inside this aperture. We illustrate the ISM structure in Fig.       \ref{ISM_structure} and explain below the exact nature of these different regions. 

     \begin{figure}
   \centering
   \includegraphics[width=93mm]{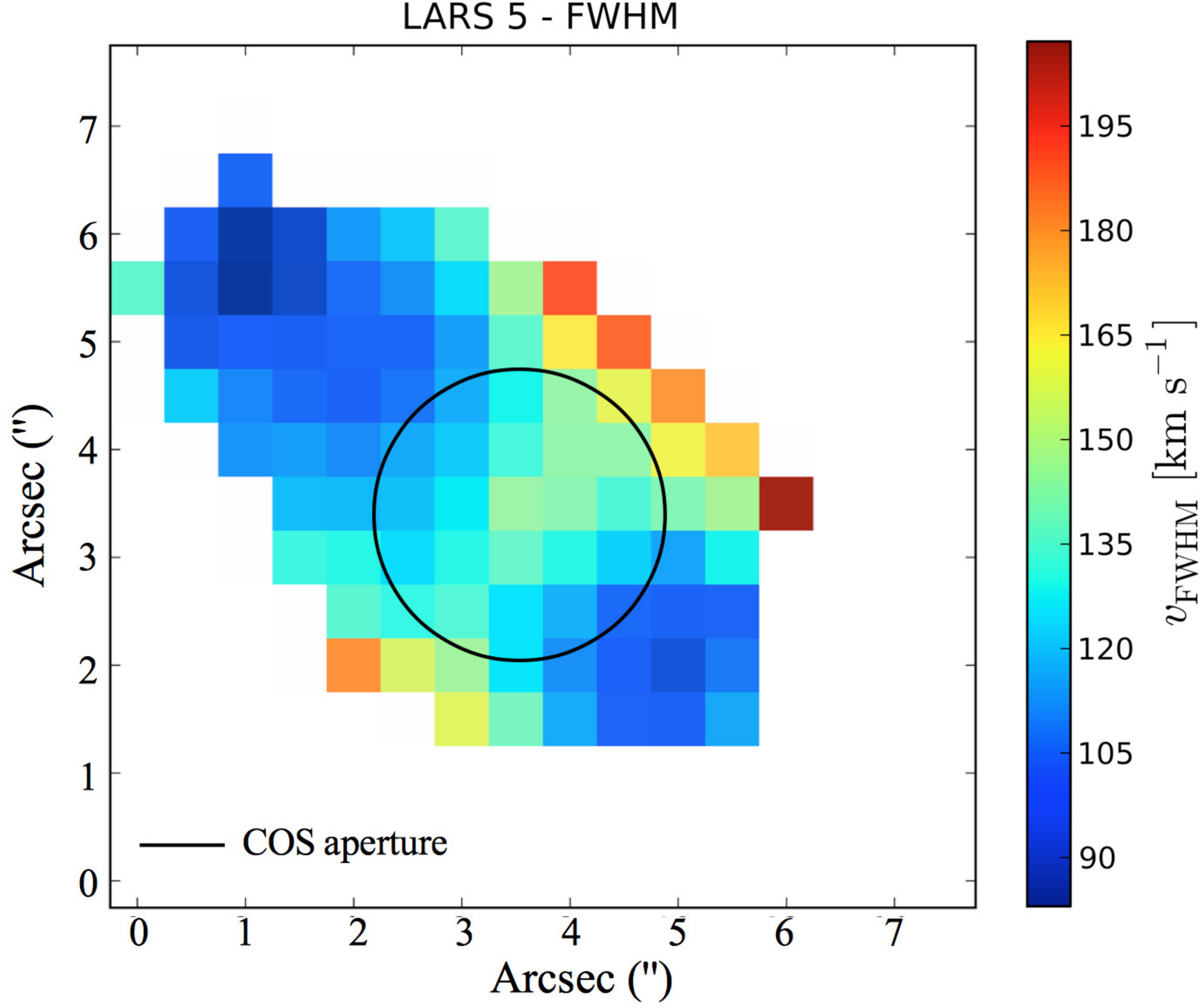}
      \caption{Observed full width at half maximum of the H$\alpha$ line, from the PMAS observations. The spatial resolution of this map is 0.5'' x 0.5''. We also add the exact position of the HST/COS aperture. The errors on the derived FWHM(\ha) are relatively small, from $\pm$ 3 km s$^{-1}$ (in the center of the COS aperture) to $\pm$ 20 km s$^{-1}$ at a large distance from the galaxy disk. 
              }
       \label{Ha_FWHM}
   \end{figure}

\begin{itemize}
\item \textbf{The galaxy disk}: 
The disk represents the most extended structure of Mrk1486 inside the COS aperture. As shown in Fig.\ \ref{ISM_structure}, a multitude of bright UV knots can be identified throughout the galaxy disk, each of them corresponding to an active star-forming region from which a strong ionizing radiation comes from. We used the observed UV image of Mrk1486 to determine the approximate location of the edge of the galaxy disk. The analysis of the UV light curves along the minor axis of the galaxy reveals a sharp decrease in the UV surface brightness SB(UV) above and below the mid-plane of the disk. Using this decrease of SB(UV) as a probe, we estimate the disk of Mrk1486 to be relatively thin, with a vertical scale height of about 245$\pm$15 pc. \\

         \begin{figure*}
   \centering
   \includegraphics[width=92.5mm]{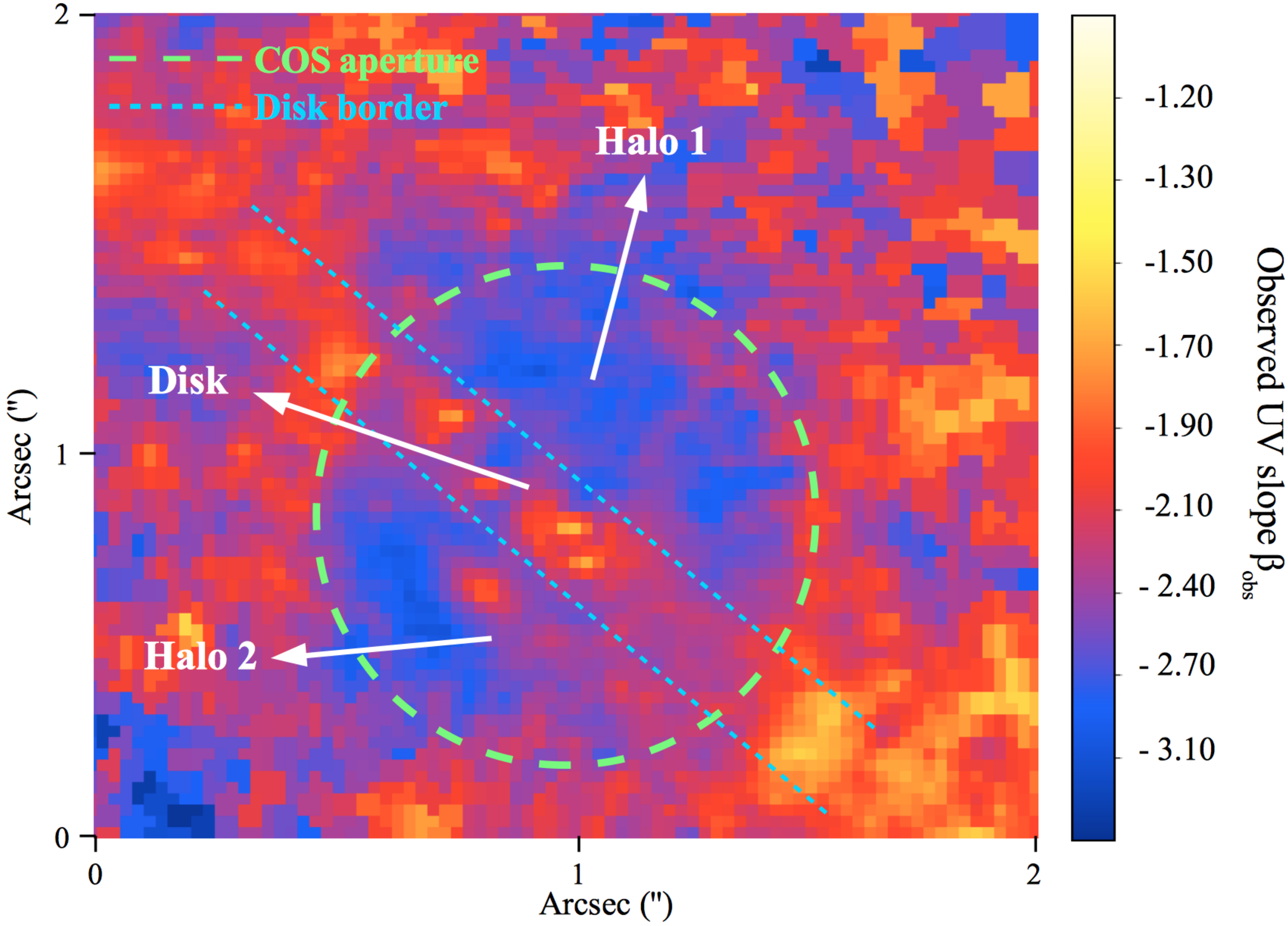}
   \includegraphics[width=90mm]{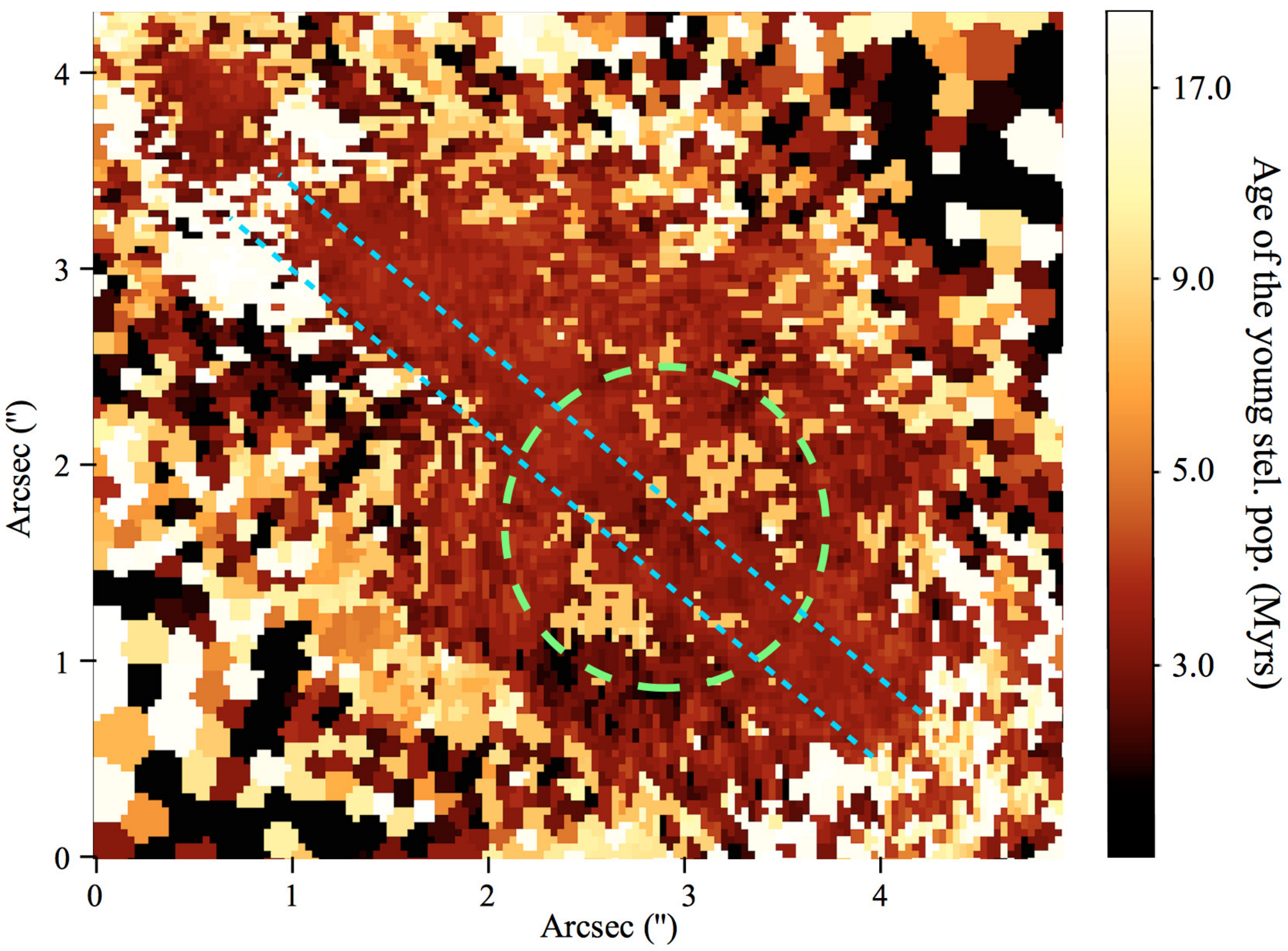}
      \caption{\textit{Left panel}: UV continuum slope $\beta_{obs}$ measured inside the COS aperture between 1540 \AA\ and 3330 \AA. We add to this figure both the location of the COS aperture and the edge of the galaxy disk. While the disk exhibits a reddened UV continuum slope (with $\beta_{obs}$ in the range [-2.60;-1.60]), each outflowing halo is characterized by a blued UV continuum. The $\beta_{obs}$ values of each halo are well below the intrinsic slope $\beta_{int}$ = -2.50 (see text for more details). \textit{Right panel}: Age map of the young stellar population of Mrk1486, as derived by SED fitting of the HST observations on a pixel-by-pixel scale. The field of view is the same as in Figs. \ref{maps} and \ref{Lya_COS_ape}. The green dashed circle corresponds to the region covered by the HST/COS aperture, and the blue dashed lines indicate the edge of the galaxy disk. We represent here the stellar age on a logarithmic scale. We find approximately the same age for the young stars throughout the disk of the galaxy, with a mean age of 3.02$\pm$0.20 Myr. 
              }
       \label{Ebvs_laxs}
   \end{figure*}

\item \textbf{Outflowing halos}: both our HST and PMAS data of Mrk1486 reveal two bipolar outflowing halos along the minor axis of the galaxy disk. First of all, we show in Fig.\ \ref{Ha_FWHM} the observed H$\alpha$ full width at half maximum (FWHM(\ha)) map of Mrk1486, as obtained from our PMAS observations. 
While a relatively broad H$\alpha$ line emerges from the galaxy disk (FWHM(\ha)=150$^{+3}_{-3}$ km s$^{-1}$), FWHM(\ha) clearly increases in a cone shape farther away from the galaxy center (up to FWHM(\ha)=200$^{+10}_{-10}$ km s$^{-1}$ at 3 arcsec from the center of the COS aperture). This increase in the FWHM(\ha) reveals very turbulent and/or fast outflowing materials in which H$\alpha$ photons are either produced or scattered toward the observer. Then, the H$\alpha$ map of Mrk1486 (in Fig.\ \ref{maps}) also reveals large H$\alpha$ filamentary structures that suggest the presence of two bipolar winds above and below the galaxy disk (such H$\alpha$ filaments are mostly observed close to the disk, but also up to a distance of 3 kpc from the galaxy center). In Fig.\ \ref{ISM_structure} we trace the edge of the outflowing halos from the strong Ly$\alpha$ radiation they emit (see Fig. \ref{Lya_COS_ape}). We name them halo 1 and halo 2. 
\end{itemize}
   
\subsection{Ly$\alpha$ and UV properties of each ISM region}
      
In Table\ \ref{obs_brightness} we list the \lya, UV continuum, and dust extinction properties of each region of the ISM. All these components show clearly different Ly$\alpha$ and UV continuum emission properties.

A very bright UV continuum emission is observed from the disk of Mrk1486.  Specifically, the disk contributes more than 75\%\ of the total flux within the COS aperture.  Conversely, the Ly$\alpha$ morphology is very extended throughout the disk, but the gas primarily acts as an absorbing screen that cancels out some of the Ly$\alpha$ emission in the same region.  The net result is Ly$\alpha$ absorption from the disk.


The situation is different for the outflowing halos. As shown in Fig. \ref{Lya_COS_ape}, two very bright regions in Ly$\alpha$ appear slightly above and below the galaxy disk. Moreover, a comparison between the Ly$\alpha$ map of Mrk1486 and the FHWM(H$\alpha$) map (see Figs. \ref{Lya_COS_ape} and \ref{Ha_FWHM}) shows a clear consistency between the location of these two bright Ly$\alpha$ regions and the location of the outflowing halos. This means
that both halos are very bright in Ly$\alpha$ and correspond to the regions from which Ly$\alpha$ photons escape from Mrk1486. 

In addition to a bright Ly$\alpha$ emission, a faint and smooth UV continuum radiation is also detected from the outflowing halos. While this UV continuum emission could be produced by massive stars entrained in the winds, our HST data clearly favor a scenario in which the UV continuum photons are instead produced inside the galaxy disk and scatter on dust grains toward the observer. In theory, this scattering process should lead to a UV continuum that is blued and not reddened by dust in the outflowing halos, as observed in all reflexion nebula around massive stars (such as the open cluster M45). The left panel of Fig. \ref{Ebvs_laxs} shows the observed UV continuum slope\footnote{The UV slope $\beta$ describes the UV continuum flux as ${\rm f_\lambda} \propto {\rm \lambda^{\beta}}$} $\beta_{obs}$ of Mrk1486 between 1540 \AA\ and 3330 \AA\ (rest-frame wavelengths). While the model spectra of \textit{Starburst99} reveal an intrinsic UV slope $\beta_{int} = -2.50$ when assuming a Salpeter initial mass function (IMF) and an instantaneous starburst age of about 3.02$\pm$0.20 Myr (see the right panel of Fig. \ref{Ebvs_laxs}), the UV slope of each outflowing halo remains steeper than this theoretical value. Moreover, the high \ha/H$\beta$ line ratio assigned to the halos in Table \ref{obs_brightness} implies that the amount of dust particles is large in these regions of the galaxy. All these results tend to confirm the option that UV continuum photons are reflected toward the observer in the outflowing halos of the galaxy. Our modeling of the UV radiative transport inside Mrk1486 (see Sect. 6) will allow us to reach a conclusion in this scenario. 

   
\subsection{Possible scenarios for Ly$\alpha$ escape in Mrk1486}

As discussed above, the Ly$\alpha$ photons do not escape from Mrk1486 by moving through the disk of the galaxy. Instead, they do so by taking advantage of the bipolar outflowing halos located sightly above and below the galaxy disk. In this perspective, two different scenarios can be proposed for the path Ly$\alpha$ photons follow to escape from Mrk1486.

The first scenario is that shocks might be at the origin of the bright Ly$\alpha$ emission of Mrk1486. Indeed, given that outflowing halos are very likely produced by strong stellar feedback from supernovae and stellar winds into the neutral gas of the ISM, the Ly$\alpha$ photons might be directly produced by shocks at the base of the outflows. Given the low neutral hydrogen content that these outflowing regions may have, the Ly$\alpha$ photons might easily penetrate the slab of neutral gas and escape toward the observer. 

We can also propose another scenario. Instead of having an efficient production of Ly$\alpha$ photons by shocks at the base of the outflowing halos, most of the Ly$\alpha$ photons might be produced by photoionization near the star clusters (SCs) that are located inside the galaxy disk. In this case, most of the Ly$\alpha$ photons may travel along the outflowing halos where they would escape toward the observer by scattering on cool HI materials.   

In the remaining paper we propose investigating these two different scenarios by identifying the dominant source of Ly$\alpha$ photons within the galaxy disk (Sect. 5) and reproducing the Ly$\alpha$ spectrum of Mrk1486 by modeling the Ly$\alpha$ radiative transport inside the galaxy (Sect. 6). 


\section{Sources and strength of the intrinsic Ly$\alpha$ line inside the COS aperture}

To model the Ly$\alpha$ radiative transfer inside the ISM of Mrk1486 (see Sect. 6), it is important to constrain the sources and strength of the intrinsic Ly$\alpha$ line inside the COS aperture of Mrk1486. 

  \begin{figure}
   \centering
   \includegraphics[width=90mm]{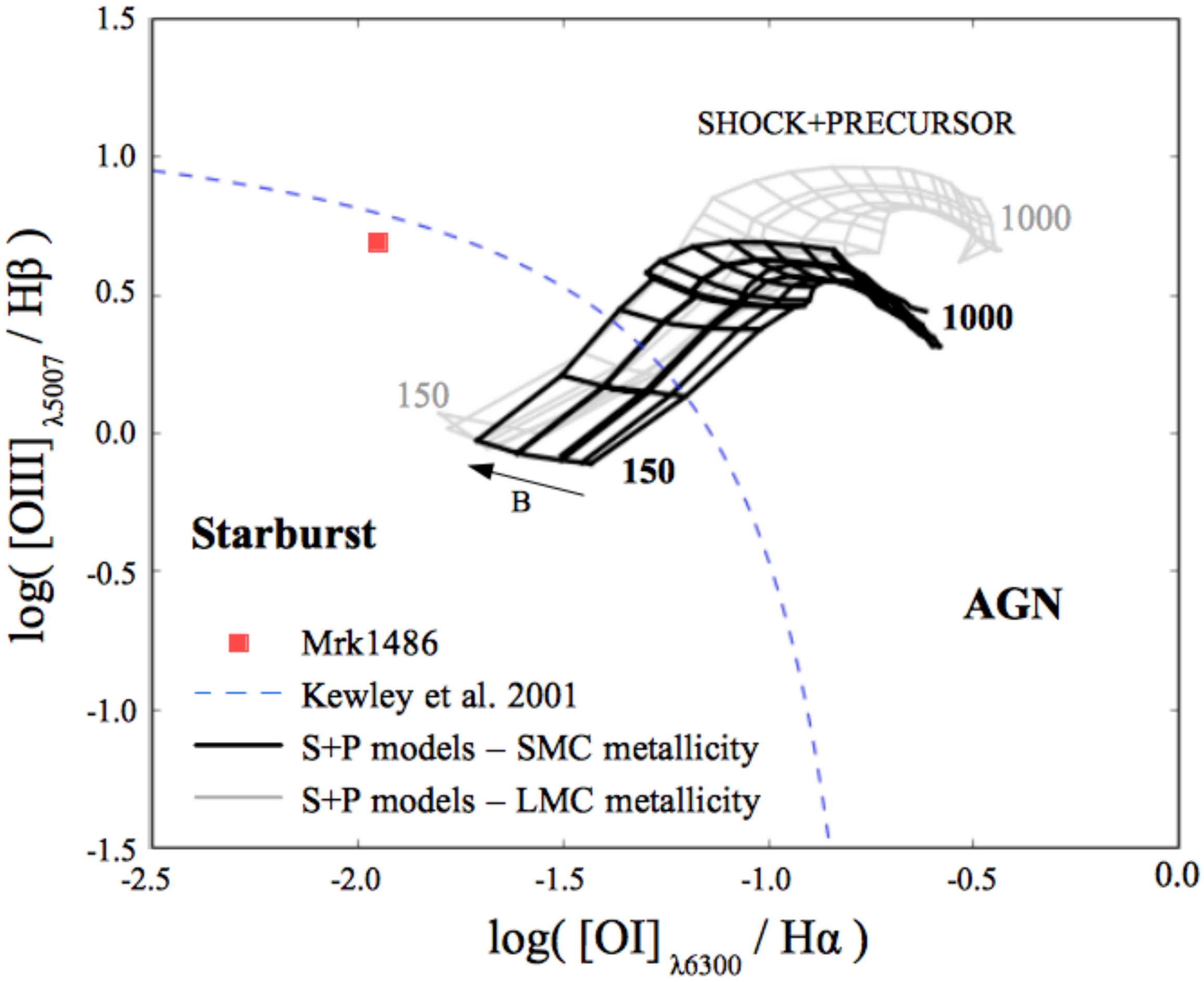}
      \caption{Line diagnostic diagram [OIII]$_{5007}$/H$\beta$ versus [OI]$_{6300}$/H$\alpha$ for Mrk1486 (the error bars of these line ratios are smaller than the red square). In particular, we compare here the line ratios of these objects to the \textit{shock + precursor }(S+P) line ratio models of \cite{allen08}. In all these models, we adopt a very wide range of values for the shock velocity v$_s$ (from 150 km s$^{-1}$ to 1000 km s$^{-1}$, with a step of 50 km s$^{-1}$) and the ISM magnetic field B (from 0.5 to 10 $\nu$G cm$^{3/2}$). We also consider the two atomic abundances available that are most similar to the one of Mrk1486. The thick black zone represents the model grids for the SMC abundance, and the thin gray zone corresponds to the grids for the LMC abundance. The blue dashed line represents the \cite{kewley01} starburst/AGN classification line (i.e. the maximum line ratios possible by pure star-forming photoionization).   
              }
       \label{diagnostic_OIII_OI}
   \end{figure}

\subsection{Dominant source of Ly$\alpha$ photons}

In star-forming galaxies the Ly$\alpha$ line is always produced in dense, warm, ionized and turbulent regions in which the Ly$\alpha$ photons are formed either through recombination of hydrogen atoms or by collisional excitations (i.e., by photoionization or shocks within the ISM). 

We used the line diagnostic diagram of \cite{veilleux87}, [OIII]$_{5007}$/H$\beta$ vs. [OI]$_{6300}$/H$\alpha$ to investigate the dominant excitation mechanism inside the COS aperture of Mrk1486. These optical emission line ratios are particularly sensitive to shocks and can be compared to several photo- and shock-ionization models to reveal the exact origin of the Ly$\alpha$ photons. 
We consider here the [OIII]$_{5007}$/H$\beta$ and [OI]$_{6300}$/H$\alpha$ line ratios of Mrk1486 derived from the integrated Sloan Digital Sky Survey (SDSS) spectrum of the galaxy. We used the SDSS spectrum of Mrk1486 because the center of the SDSS and COS apertures\footnote{Although the SDSS aperture is slightly bigger than the one of the COS spectrograph (with a circular aperture of 3'' for SDSS, instead of 2.5'' for COS), the SDSS spectrum of Mrk1486 can be used to extract some information on the region of the galaxy covered by COS.} match well ($\Delta$RA = 0.02'', $\Delta$DEC = 0.12'').

\begin{table*}
\caption{Optical, UV, and intrinsic Ly$\alpha$ properties of Mrk1486 inside the COS aperture and the galaxy disk. We list here the observed H$\alpha$, H$\beta,$ and UV fluxes (Cols. 2, 3, and 4) and the derived nebular E(B-V)$_{{\rm Neb.}}$ (Col.
6) and stellar extinction E(B-V)$_{{\rm Stel.}}$ (Col. 7). The latter is derived photometrically from the observed stellar continuum slope $\beta_{obs}$ (Col. 5). Finally, we show the derived equivalent width of the intrinsic Ly$\alpha$ line in Col. 8.}       \label{best_param}      
\centering          
\begin{tabular}{c c c c c c c c}     
\hline\hline
Aperture & F$_{obs}$(\ha) & F$_{obs}$(\hb) & F$_{obs}$(UV) & $\beta_{obs}$ & E(B-V)$_{{\rm Neb.}}$ & E(B-V)$_{{\rm Stel.}}$ & EW$_{int}$(\lya) \\ 
& \multicolumn{2}{c}{10$^{-13}$ erg s$^{-1}$ cm$^{-2}$} & 10$^{-15}$ erg s$^{-1}$ cm$^{-2}$ \AA$^{-1}$ & & & \AA \\
(1) & (2) & (3) & (4) & (5) & (6) & (7) & (8) \\
\hline
COS & 1.03$\pm$0.01  & 0.30$\pm$0.01   & 7.50$\pm$1.00   & -2.24$\pm$0.01 & 0.26$\pm$0.07     & 0.03$\pm$0.001      & 130$^{+50}_{-33}$ \\
Galaxy disk & 0.41$\pm$0.04  & 0.11$\pm$0.02  & 5.05$\pm$0.36 & -2.15$\pm$0.01 & 0.34$\pm$0.02  & 0.04$\pm$0.001 & 77$^{+1}_{-3}$\\
\hline  
\hline                  
\label{photometric_properties}
\end{tabular}
\end{table*}  
   
We show in Fig.\ \ref{diagnostic_OIII_OI} the diagnostic diagram [OIII]$_{5007}$/H$\beta$ versus [OI]$_{6300}$/H$\alpha$ and include the line ratios of Mrk1486. 
We also add to this diagram the division of \cite{kewley01} (which represents the theoretical maximum line ratios possible by pure photoionization) and the shock-ionization models of \cite{allen08}. In Fig.\ \ref{diagnostic_OIII_OI}, data points are considered shock-dominated if they lie within the region covered by \textit{shock + precursor} models.
In the models of \cite{allen08}, the emission from shocks consists of two main components: the \textit{shock} layer (the post-shock component, which is the cooling zone of the radiative shock) and the \textit{precursor} (the pre-shock component, which corresponds to the region that is ionized by the upstream photons from the shock zone). Moreover, in these models, different ISM quantities control the line ratios produced by shocks, such as the \textit{metallicity} Z, the \textit{shock velocity} v$_s$, the \textit{preshock zone density} n$_s$ , and the \textit{ISM magnetic field} B. Throughout our analysis, we used the same metallicity Z (in Fig.\ \ref{diagnostic_OIII_OI} we chose the metallicity of the SMC and LMC because the abundance of Mrk1486 is in between the one of these two dwarf galaxies) and the pre-shock gas density ($n_s = 1$ cm$^{-3}$). This means
that the shock velocity v$_s$ and the magnetic field B are the only free parameters that determine the line ratios in the diagnostics. 

As shown in Fig.\ \ref{diagnostic_OIII_OI}, none of the \textit{shock + precursor} models developed with the SMC and LMC metallicities can reproduce the optical-emission line ratios of Mrk1486. 
We therefore conclude that the ionization of hydrogen is very likely dominated by photoionization from the SCs located inside the COS aperture of Mrk1486. We reach the same conclusion by comparing other optical emission line ratios (such as [SII]$_{6710,6730}$/H$\alpha$ and [NII]$_{6583}$/H$\alpha$) to shock-ionization models. 

\subsection{Relative strength of the intrinsic Ly$\alpha$ line}

The young age of the SCs located inside the galaxy disk (of about 3 Myr, see the right panel of Fig. \ref{Ebvs_laxs}) makes it
very likely that a relatively strong Ly$\alpha$ emission line is produced by these star-forming knots. The intrinsic Ly$\alpha$ equivalent width EW$_{int}$(\lya) can be computed inside the COS aperture using the integrated H$\alpha$ and UV continuum fluxes, 

\begin{equation}
{\rm EW}_{\rm int}({\rm Ly}\alpha) = \frac{{\rm F}_{\rm int}({\rm H}\alpha) \times 8.7}{{\rm F}_{\rm int}({\rm UV})} = \frac{{\rm F}_{\rm obs}({\rm H}\alpha) e^{\tau_\alpha} \times 8.7}{{\rm F}_{\rm obs}({\rm UV}) e^{\tau_{\rm UV}}}
,\end{equation}
where F$_{obs}$(\ha) and F$_{obs}$(UV) are the observed H$\alpha$ and UV continuum fluxes inside the COS aperture, $\tau_{\alpha}$ and $\tau_{UV}$ are the dust optical depths for the H$\alpha$ line and the UV continuum radiation, and the factor 8.7 is the most likely intrinsic line ratio \lya/H$\alpha$ in HII regions (case B of recombination theory; Osterbrock 1989). The optical depths $\tau_{\alpha}$ and $\tau_{UV}$ can be derived as 

\begin{equation}
\tau_\alpha = E_{B-V}^{{\rm Neb}}k({\rm H}\alpha) \frac{{\rm ln}(10)}{2.5}
 \text{}\end{equation}

\begin{equation}
\tau_{\rm UV} = E_{B-V}^{{\rm Stel}}k({\rm Ly}\alpha) \frac{{\rm ln}(10)}{2.5}
,\end{equation}
where E$_{B-V}^{{\rm Neb}}$ and E$_{B-V}^{{\rm Stel.}}$ are the nebular and stellar color excess inside the COS aperture. We also consider k(\ha) and k(\lya) as the extinction coefficients at the H$\alpha$ and Ly$\alpha$ wavelengths. Using the SMC extinction law\footnote{The use of the SMC extinction law is motivated here by of the low metallicty of Mrk1486.} of \cite{prevot84} we obtain k(\ha) = 2.36 and k(\lya)= 17.30. The nebular color excess E(B-V)$_{{\rm Neb.}}$ is derived from the Balmer decrement H$\alpha$/H$\beta$ as

\begin{equation}
E(B-V)_{{\rm Neb.}} = \frac{{\rm log}((H\alpha/H\beta)_{obs}/2.86) \times 2.5}{k(H\beta) - k(H\alpha)}
,\end{equation}
where the factor 2.86 is the intrinsic line ratio H$\alpha$/H$\beta$ in HII regions (Case B; Osterbrock 1989) and k(H$\beta$) = 3.13 from the SMC extinction law. The stellar color excess E$_{B-V}^{{\rm Stel.}}$ is also derived photometrically from the observed UV continuum slope $\beta_{obs}$ between 1540 \AA\ and 3330 \AA\ (see Fig. \ref{Ebvs_laxs}),

\begin{equation}
E(B-V)_{{\rm Stel.}} = \frac{\beta_{obs} - \beta{int}}{1.03 \times (k(1540) - k(3330))}
,\end{equation}
where $\beta_{int}$ is the average intrinsic UV slope of the SCs inside the COS aperture. As in Sect. 4.2, we adopt ${\rm \beta_{int}} = {\rm -2.50}$, which is derived from the model spectra of \textit{Starburst99} for an instantaneous burst age of 3.02$\pm$0.20 Myr. Moreover, we use the extinction coefficients k$(1540) = 13.25$ and k$(3330) = 4.83$ from \cite{prevot84}. 

We list in Table\ \ref{photometric_properties} the photometric properties of Mrk1486 inside the COS aperture and the derived EW$_{int}$(Ly$\alpha$). As expected, EW$_{int}$(Ly$\alpha$) is relatively high and reaches a mean value of 130 \AA. Nevertheless, this value might overestimate EW$_{int}$(Ly$\alpha$) in Mrk1486 because of two outflowing halos that exhibit a negative stellar extinction inside the COS aperture (see Fig.    \ref{Ebvs_laxs}). Therefore, we may underestimate both the global stellar extinction E$_{B-V}^{{\rm Stel.}}$ inside this aperture and the intrinsic flux F$_{int}$(UV). 

To avoid the effect of the two outflowing halos in our measurements, we also measured EW$_{int}$(Ly$\alpha$) from the galaxy disk alone. As shown in Table \ref{photometric_properties}, this approach provides a lower value down to 77 \AA. 
This should be taken as a lower limit for EW$_{int}$(Ly$\alpha$) since by restricting the integration area to the disk, we would underestimate the total H$\alpha$ emission (which originates from a more extended region). From our photometric approach we therefore conclude that EW$_{int}$(Ly$\alpha$) is in between 77 \AA\ and 130  \AA, which is roughly consistent with the predictions of \textit{Starburst99} for an instantaneous burst of 3.02$\pm$0.20 Myr \citep[EW$_{int}$(Ly$\alpha$) evolves very rapidly at these ages with a mean value of 90 \AA;][]{schaerer08}.

\section{Ly$\alpha$ line fit of Mrk1486}

  \begin{figure*}
   \centering
   \includegraphics[width=180mm]{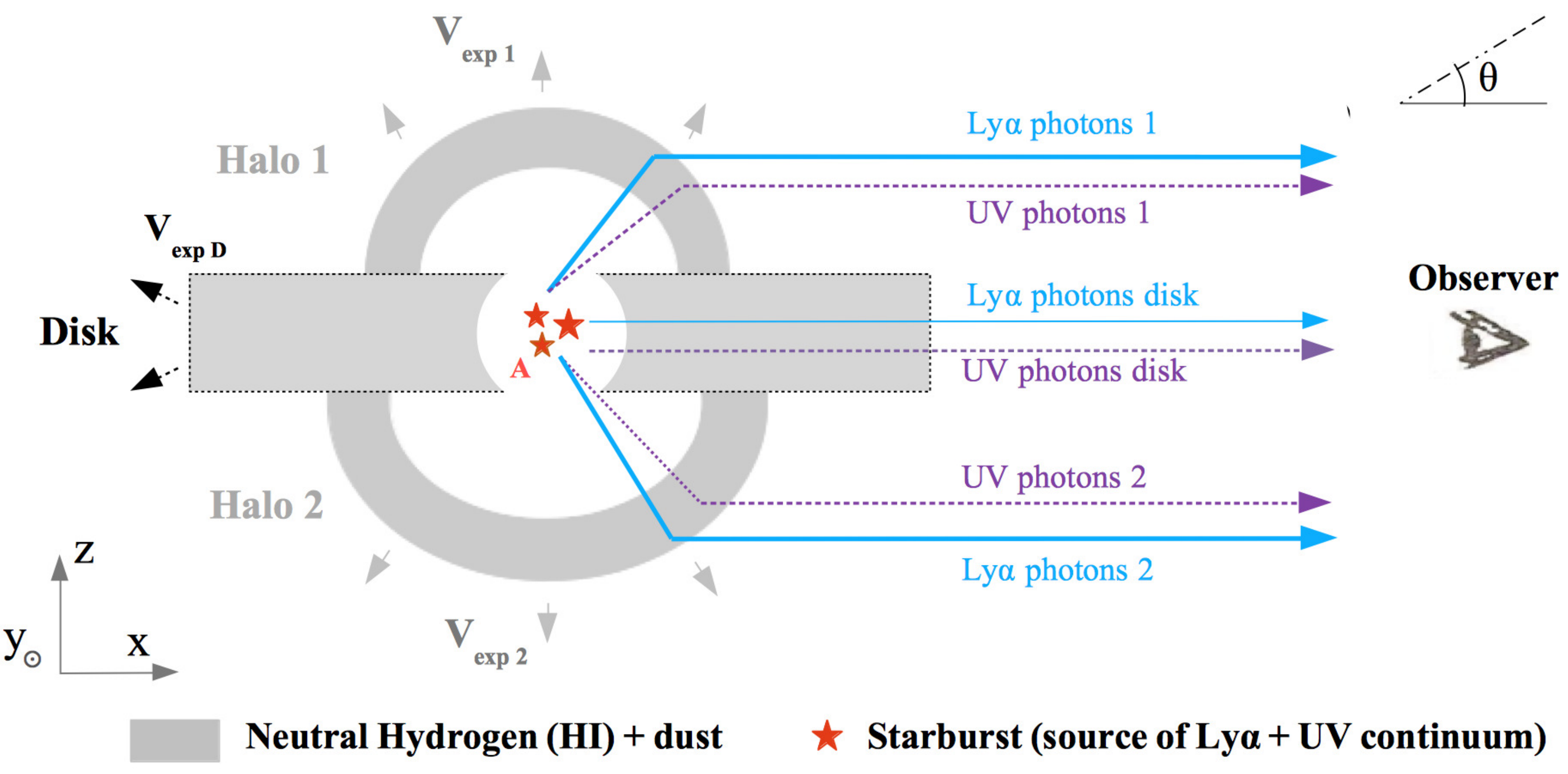}
      \caption{Illustration of the 3D geometry used to fit the Ly$\alpha$ line profile of Mrk1486 and built in a Cartesian grid (x, y, z). It is composed of three different regions that represent the regions seen through the COS aperture of Mrk1486. First of all, we consider a medium named "disk" that represents the disk of the galaxy. This medium is located in the center of the 3D geometry and corresponds to the central part of a radially homogeneously expanding shell of HI and dust. Then, we add two other media above and below the galaxy disk (named halo 1 and halo 2). These two half-shells model the outflowing halos of cool material observed above and below the disk of Mrk1486. We consider a point-like central source of Ly$\alpha$ and (non-ionizing) UV continuum photons at the center of the disk. Finally, the dashed and the thick lines represent the trajectories of UV continuum and Ly$\alpha$ photons that we take into account to fit the Ly$\alpha$ line of Mrk1486 ( $\sqrt{\theta^2 + \phi^2} \le 5$). While Ly$\alpha$ photons are very likely to escape toward the observer after scattering on HI atoms, UV continuum photons do the same either by passing through the galaxy disk or by scattering on dust grains in the outflowing halos.  
              }
       \label{3Dgeo}
   \end{figure*}

In this section we fit the Ly$\alpha$ line profile of Mrk1486 to investigate the ISM physical properties and the Ly$\alpha$ transport inside the COS aperture of the galaxy. In particular, we attempt to reproduce the Ly$\alpha$ line profile of Mrk1486 by assuming a pure scattering of Ly$\alpha$ photons along the bipolar outflowing halos of the galaxy. 

\subsection{Ly$\alpha$ radiative transfer code}

To fit the observed Ly$\alpha$ line of Mrk1486 we used the latest version of the 3D Monte Carlo radiative transfer code MCLya of \cite{verhamme06} and \cite{schaerer11}. For a given geometry of neutral hydrogen and dust, the code computes both the Ly$\alpha$ and adjacent UV continuum radiative transfer, including the physical details of the Ly$\alpha$ scattering, dust scattering, and dust absorption. 

We also supplemented our analysis by computing the H$\alpha$ and H$\beta$ radiative transfer inside the same HI and dust geometry. We did this by assuming scattering and absorption by dust. We list in Table\ \ref{dust_prop} the SMC dust parameters that we adopted for the Ly$\alpha$ line, UV continuum, H$\alpha,$ and H$\beta$ photons throughout this paper \citep{witt00}.

\begin{table}[h]
\caption{Dust parameters (\textit{albedo} A and \textit{scattering phase function asymmetry} g) taken from \cite{witt00} and used throughout this paper. We are interested in four different wavelengths: the Ly$\alpha$ line, its neighboring UV continuum, and the H$\alpha$ and H$\beta$ lines. }
\centering                          
\begin{tabular}{c c c c c c}        
\hline\hline                 
Photons &  & $\lambda$ (\AA) & $\tau_d$/$\tau_V$ & $A$ & $g$\\    
\hline                        
\lya & & 1215.67 & & & \\[-1ex]
 & & & \raisebox{0.5ex}{6.74} & \raisebox{0.5ex}{0.460}&\raisebox{0.5ex}{0.770}\\[-2ex]
UV & & 1235.0 & & & \\
\hline                 
H$\beta$          &  & 4861.1 & 1.191  & 0.493  & 0.621 \\
H$\alpha$           &  & 6562.8 & 0.787 & 0.487  & 0.584 \\
\hline                                   
\end{tabular}
\label{dust_prop}      
\end{table}

\subsection{3D geometry and fit parameters}

\subsubsection{3D geometry}

To fit the apparent ISM structure of Mrk1486 inside the COS aperture (see Fig.\ \ref{ISM_structure}), we adopted a 3D geometry similar to the one shown in Fig. \ref{3Dgeo}. This geometry shows three distinct components as described below. 

\begin{itemize}
\item \textbf{Galaxy disk}: in the center of the 3D geometry, we built a medium of HI and dust that represents the galaxy disk. This medium corresponds to the central part of a radially expanding shell of HI atoms and dust grains (i.e., a shell cut above and below a certain height from the mid-plane). In this section, we always consider a homogeneous distribution of HI and dust inside the shell. 

In the center we placed a source emitting both Ly$\alpha$ line and UV continuum photons. This source is called "source A" in Fig.\ \ref{3Dgeo} (we assume here a point-like central source and isotropic emission). \\ 

\item \textbf{Halo 1}: Above the galaxy disk, we built half of a shell of HI and dust that represents the material that is ejected above the disk of Mrk1486. For simplicity, we chose to distribute both the HI and dust content homogeneously inside the half-shell. Each parcel of this component expands radially from the radiation source A. \\

\item \textbf{Halo 2}: Below the galaxy disk, we placed another half-shell of HI and dust. This component represents the outflowing
gas below the disk of Mrk1486. We also adopted a homogeneous distribution of gas and dust within this structure and considered a radial and constant expansion of this half-shell from source A. 
\end{itemize} 

We built the 3D geometry in a Cartesian grid of ${\rm N} = 128^3$ cells. For our model to be consistent with the real ISM structure of Mrk1486 inside the COS aperture, we chose to constrain the size of each component in our 3D geometry. Figure \ref{ISM_structure} shows that the disk has an apparent surface 1.73 times larger than the one of the halo 1 and 1.05 times larger than the surface
of halo 2. Therefore, all components of our 3D geometry were built as follows: the source of photons is located at the coordinates (64, 64, 64) in the Cartesian grid (x, y, z); the \textbf{disk} is a shell of radius $R_D = 64$ cells cut above ${\rm z} = 79$ and below z = 53; \textbf{halo 1} is a half-shell of radius R$_{H1}$ = R$_D$/2.5 cells that we added on top of the disk; h\textbf{alo 2} is another half-shell of radius R$_{H2}$ = R$_D$/2 cells that we added below the disk. 

Nevertheless, it is important to note that the size we assigned to each component has no noticeable effect on the Ly$\alpha$ radiative transfer inside the 3D geometry. These are instead the physical properties of each component (i.e., velocity, gas temperature, HI, and dust density), which play the most important role in this process. 

  
\subsubsection{Fit parameters}

The physical conditions in each component of our 3D geometry are described by five physical parameters: 

\begin{itemize}
\item the mean expansion velocity \vexp\ 
\item the mean HI column density \nhi
\item the mean dust optical depths $\tau_{Ly\alpha}$ and $\tau_{cont}$ (for the Ly$\alpha$ line and the UV continuum radiation)
\item the Doppler parameter b.
\end{itemize}

The expansion velocity \vexp\ of a shell is always radial and constant from source A (see Fig.\ \ref{3Dgeo}). In practice, the velocity \vexp\ we allotted to each shell of our 3D geometry can be constrained by different observations: 

  \begin{figure}
   \centering
   \includegraphics[width=90mm]{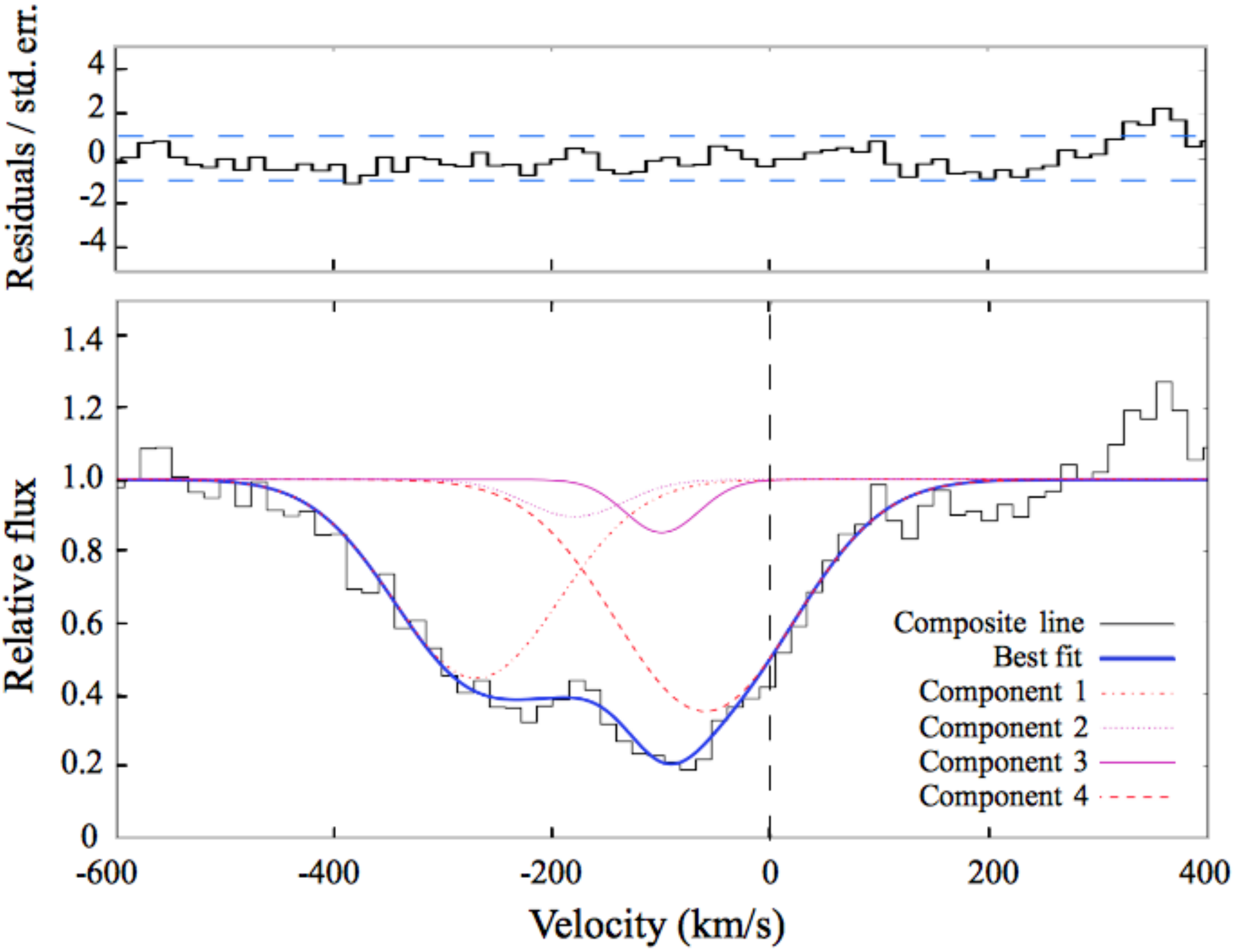}
      \caption{\textit{Bottom panel}: composite line profile of four SiII absorption lines observed in the COS spectrum of Mrk1486 \citep{rivera15}: $\lambda$=1190 \AA, $\lambda$=1193\AA, $\lambda$=1260 \AA,\ and $\lambda$=1304 \AA. These lines all arise from the same ground state. The composite spectrum is transformed here in velocity space using the observed wavelength of stellar lines. Only in this figure, outflows show negative and inflows positive velocities. \textit{Upper panel}: for each velocity bin, we plot the residual over the standard deviation between the observed SiII line and our best fit. The blue dashed lines correspond to 1$\sigma$ deviation.  
              }
       \label{Vexp_shift}
   \end{figure}
   
   \begin{table}
\caption{Best-fit parameters of the SiII composite line in Fig.\ \ref{Vexp_shift}. Each component of the line fit corresponds to a Gaussian that is centered at a velocity v$_{center}$ (Col. 2) and described by a certain width ($\sigma$, Col. 3) and strength (Col. 4).  }       
\label{SiII_composite}      
\centering          
\begin{tabular}{c c c c}     
\hline\hline
Component & v$_{center}$ & width $\sigma$ & strength \\
 & km s$^{-1}$ & km s$^{-1}$ &  \\
(1) & (2) & (3) & (4) \\ 
\hline
1 & -270 & 65 & 0.5566 \\
2 & -180 & 45 & 0.1038 \\
3 & -100 & 34 & 0.1484 \\
4 & -60 & 82 & 0.6494  \\
\hline  
\hline                  
\end{tabular}
\end{table} 

   \begin{table*}
\caption{Grid of models used in this work to fit the Ly$\alpha$ line of Mrk1486. The physical parameters of each component of the 3D geometry are listed in Col. 2. For each parameter in question, we give the range or set of values in Col. 3 and the steps in Col. 4 (the comment "\textit{fixed}" means that the parameters is not considered as a free-parameter).}             
\label{table:free_param}      
\centering          
\begin{tabular}{c l c l c l c }     
\hline\hline
Region & Physical parameters & Range of values & Step \\
\hline
         & v$_{expD}$ [km s$^{-1}$] &  100 &  \textit{fixed} \\
         & \nhi$_{D}$ [10$^{20}$ cm$^{-2}$] & [1 , 10] & 1.0 \\
\textbf{Disk}   & $\tau_{Ly\alpha D}$ &  3.5  &  \textit{fixed}\\
         & $\tau_{cont D}$ & 0.3 & \textit{fixed}\\
         & $b_{D}$ [km s$^{-1}$] & 12.8 & \textit{fixed}\\
\hline                
            & v$_{exp1}$[km s$^{-1}$] &  100 or 180 & -\\
\textbf{Halo 1}  & N$_{HI1}$ [10$^{19}$ cm$^{-2}$] & [1 , 10] & 1.0 \\
            & $\tau_{Ly\alpha 1}$ = $\tau_{cont1}$ &  [0.6 , 2.2] & 0.1  \\
            & $b_1$ [km s$^{-1}$] & [12.8 , 42.8] & 5.0 \\
\hline
            & v$_{exp2}$ [km s$^{-1}$] &  100 or 180 & -\\
\textbf{Halo 2}  & N$_{HI2}$ [10$^{19}$ cm$^{-2}$] & [1 , 10] & 1.0  \\
            & $\tau_{Ly\alpha 2}$ = $\tau_{cont2}$ & [1.2 , 3.0] & 0.1 \\
            & $b_2$ [km s$^{-1}$] &  [12.8 , 42.8] & 5.0 \\
\hline  
\hline  
 \textbf{Intrinsic Ly$\alpha$ line}    & EW$_{int}$(\lya) [\AA]& [77 , 130]  & 1.0 \\
   &  FWHM$_{int}${\lya} [km s$^{-1}$] & [50 , 150] & 20.0 \\
\hline
\hline    
\end{tabular}
\end{table*} 
   
\begin{itemize}
\item \textbf{From the Ly$\alpha$ line profile of Mrk1486}. As shown in \cite{verhamme06} and \cite{duval14}, the mean expansion velocity \vexp\ of an ISM plays the most important role in the formation of the observed Ly$\alpha$ line profile of a galaxy. In particular, the observed wavelength of some redshifted bumps on top of an asymmetric Ly$\alpha$ line prove to be good tracers of the mean ISM expansion velocity \vexp, such as the backscattering redshifted peak, as already pointed out by \cite{verhamme06} for expanding shells\footnote{Comparing the observed wavelength of the backscattering bump $\lambda_{em}$ to the Ly$\alpha$ line center $\lambda_{Ly\alpha}$, this leads to either $\Delta$v(em - Ly$\alpha$) $\approx$ \vexp\ for low HI column densities \nhi\ ($<$ 10$^{19}$ cm$^{-2}$) or $\Delta$v(em - Ly$\alpha$) $\approx$ 2$\times$\vexp\ for high \nhi\ ($>$ 10$^{19}$ cm$^{-2}$).}. As shown in Fig. \ref{Lya_COS_ape} (right panel), the Ly$\alpha$ line of Mrk1486 exhibits a P-cygni profile in which three significant peaks can be identified. These are located at the rest-frame wavelengths $\lambda_1$ = 1216.37 \AA, $\lambda_2$ = 1216.55 \AA,\ and $\lambda_3$ = 1217.15 \AA. 
If we assume that these bumps correspond to the backscattering redshifted features produced by different outflowing shells inside the ISM of Mrk1486, we can estimate, very roughly, the mean expansion velocities \vexp\ we should consider in our analysis. For the bump seen at $\lambda_3$ = 1217.15 \AA, a mean expansion velocity \vexp = c($\lambda_3$ - $\lambda_{Ly\alpha}$)/(2$\times$$\lambda_{Ly\alpha}$) $\approx$ 182 km s$^{-1}$ would be needed. For the formation of peaks 1 and 2 the same calculation leads to a mean \vexp\ $\approx$ 100 km s$^{-1}$. Therefore, the components of our 3D geometry should exhibit a mean expansion velocity of about 180 km s$^{-1}$ or 100 km s$^{-1}$  to reproduce the bumps observed on top of the Ly$\alpha$ line profile.
\item \textbf{From the shift between the low ionization states (LIS) and the stellar lines of the galaxy}. 
LIS metal absorption lines are established as good tracers of neutral gas \citep{pettini02a,erb12,sandberg13} and have been employed in studies of relations between neutral ISM kinematics and Ly$\alpha$ radiative transfer \citep{kunth98,shapley03}. \cite{rivera15} measured the shift of four different UV SiII absorption lines in the COS spectrum of Mrk1486. We show in Fig.\ \ref{Vexp_shift} the composite line profile of these SiII absorption lines. Qualitatively, we note that most of the SiII line flux is blueshifted with respect to the stellar lines of the galaxy, suggesting that most of the neutral gas of Mrk1486 is outflowing with respect to the stars. Quantitatively, we can also note that the velocity range of the neutral gas is very broad (from -400 to 100 km s$^{-1}$), indicating that distinct subsystems (clouds of neutral gas) move outward at slightly different velocities inside the ISM. In Fig.\ \ref{Vexp_shift} we provide our best line profile fit, and we list in Table\ \ref{SiII_composite} the best-fit parameters of each component. Overall, four velocity components are detected along the whole length of the SiII composite line. They are all described by a mean expansion velocity \vexp\ and a velocity dispersion $\Delta$\vexp. Among these four components, two very bright ones seem to trace the gas kinematics inside the galaxy disk (components 1 and 4), and two faint ones may imprint the gas kinematics inside the two outflowing halos (components 2 and 3). Interestingly, the mean \vexp\ of these two faint components and the ones derived from the bumps on top of the Ly$\alpha$ line in Fig.  \ref{Lya_COS_ape}
are highly consistent. 
\end{itemize}

To fit the Ly$\alpha$ line profile of Mrk1486, we treated the mean expansion velocity of the shells of our 3D geometry in the following way. In Table\ \ref{table:free_param} we adopt a mean expansion velocity \vexp\ $\approx$ 100 km s$^{-1}$ for the disk in our 3D geometry 
(as the average \vexp\ of components 1 and 4 in Fig.\ \ref{Vexp_shift}). For the outflowing halos, we considered two possible velocities \vexp: 100 km s$^{-1}$ and 180 km s$^{-1}$. The best expansion velocity ascribed to each component is derived from our best fit of the Ly$\alpha$ line profile. The velocity dispersion $\Delta$\vexp\ listed in Table\ \ref{SiII_composite} was not applied in our 3D geometry. On the one hand, it would be very challenging to apply such velocity dispersions inside pure homogeneous shells of HI and dust (in reality, $\Delta$\vexp\ indicates that the ISM of Mrk1486 is clumpy, each clump moving outward at a slightly different velocity). On the other hand, our work presented in \cite{duval14} showed that the effect produced by $\Delta$\vexp\ on the emerging Ly$\alpha$ line profile is very weak and not dominant compared to the one produced by the mean ISM expansion velocity \vexp. Throughout our analysis we therefore only considered the mean expansion velocity \vexp\ at which most of the gas is outflowing in each component of our 3D geometry. 

The mean HI column density \nhi\ of a shell is simply related to the HI density (n$_H$) and the thickness (L) by 
\begin{equation}
\nhi = n_H \times L    \ \ \     (cm^{-2})
.\end{equation}
We always adopted an arbitrary HI density n$_{HI}$ = 0.05 cm$^{-3}$ throughout this work. We therefore varied \nhi\ by changing the thickness L of the shells. As we did not have any observational constraint on \nhi, we considered \nhi\ a free parameter in each shell of our 3D geometry (see Table\ \ref{table:free_param}). 

The mean dust optical depths $\tau_{Ly\alpha}$ and $\tau_{cont}$ measure the dust attenuation that affects the nebular Ly$\alpha$ line and the UV continuum emission in each shell of our 3D geometry. We list in Table\ \ref{table:free_param} the values of $\tau_{Ly\alpha}$ and $\tau_{cont}$ we considered throughout our analysis. For each shell we adopted a narrow range of $\tau_{Ly\alpha}$  ,
which provides a color excess E(B-V)\footnote{In our numerical simulations we derive the color excess ${\rm E(B-V)}$ of a shell from the escape fraction of the H$\alpha$ and H$\beta$ lines.} that is consistent with the nebular extinction E$_{B-V}^{{\rm Neb}}$ shown in Table \ref{table:obs_proper}. We treated $\tau_{cont}$  in the following way. For the galaxy disk we adopted $\tau_{cont} = 0.7$ because this provides a color excess consistent with the stellar one measured in this region of the galaxy E$_{B-V}^{{\rm Stel}} = 0.04$ (see Table  \ref{photometric_properties}). For the two outflowing halos we adopted $\tau_{cont}$ = $\tau_{Ly\alpha}$. This is because most of the radiation detected from the bipolar winds is produced inside the galaxy disk and scatter on dust grains toward the observer (see Sect. 4.2). Therefore, both the ionized gas and the UV continuum radiation should suffer the same dust attenuation inside the outflowing halos. 

The Doppler parameter $b = \sqrt{v_{\rm th}^2 + v_{\rm turb}^2}$ here measures the random (thermal + turbulent) motions of the HI gas in our 3D geometry. It is considered as a free parameter. We list in Table\ \ref{table:free_param} the range of values we considered throughout our analysis (from ${\rm b} = 12.8       {\rm km s}^{-1}$, corresponding to the HI dispersion observed within T=10$^4$ K clouds, to b = 42.8 km s$^{-1}$).

  \begin{figure*}
   \centering
   \includegraphics[width=140mm]{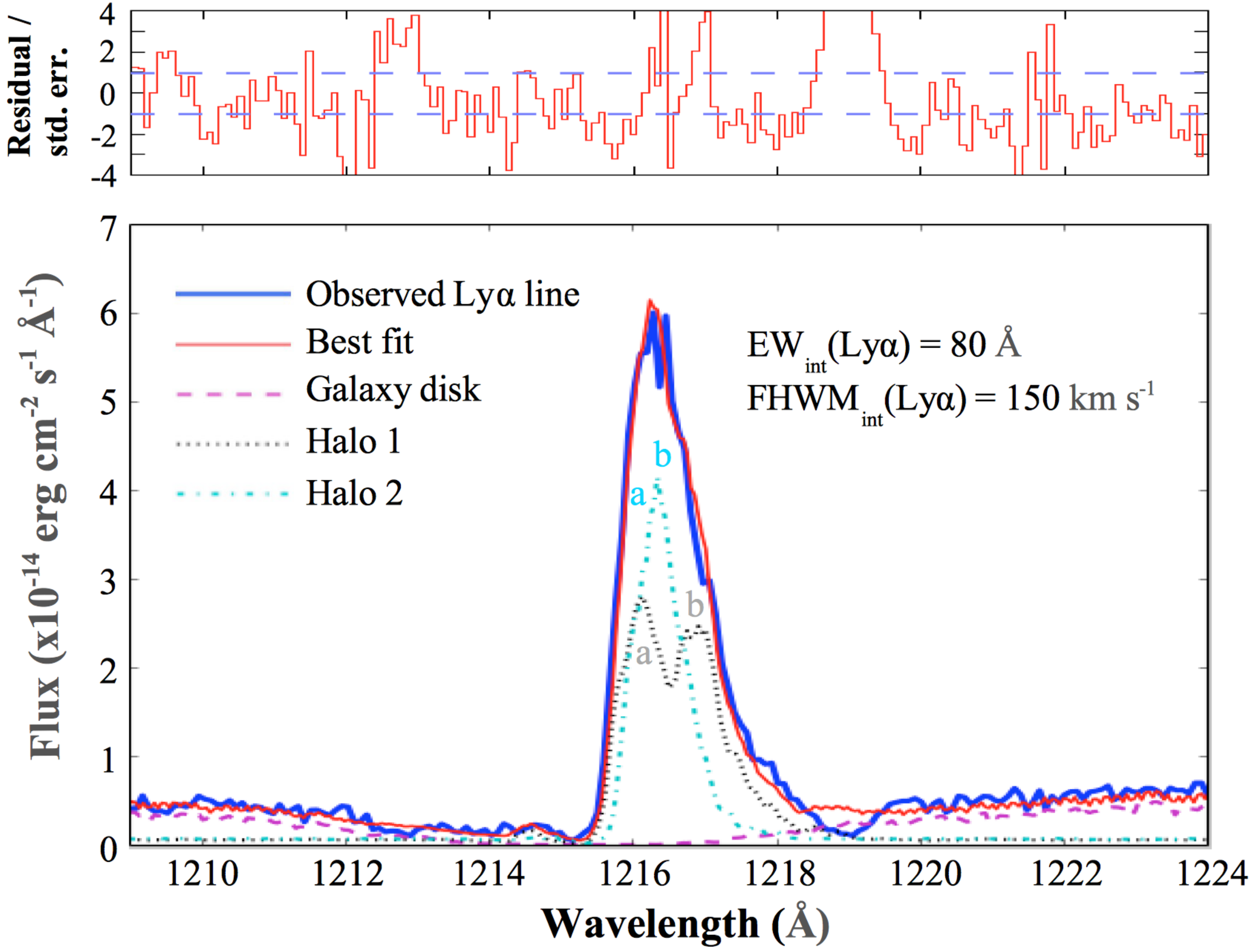}
      \caption{Best Ly$\alpha$ line fit of Mrk1486. \textit{Lower panel}: the blue lines correspond to the observed Ly$\alpha$ line of Mrk1486, the red line represents our best fit. Three components are included in this composite line. The magenta dashed
line shows the spectrum that escapes from the galaxy disk. The components of halo 1 and halo 2 are shown as a black dotted and a blue dashed-dotted line, respectively. The letters "a" and "b" refer to Fig.\ \ref{3Dgeo_clumpy}: bump "a" is formed by Ly$\alpha$ photons that escape without experiencing any backscattering, while bump "b" is formed by Ly$\alpha$ photons that undergo one backscattering event. \textit{Upper panel}: for each wavelength bin, we plot the residual over the standard deviation between the observed Ly$\alpha$ line and our best fit. The blue dashed lines correspond to 1$\sigma$ deviation. Inside the wavelength range of the Ly$\alpha$ line, 70\% of the data points lie inside the 1$\sigma$ deviation region.     
              }
       \label{best_fit}
   \end{figure*}
   
   \begin{table*}
\caption{Best-fit parameters of the 3D geometry. For each component of our 3D geometry, we list the photometric properties in Cols. 2 and 3 and the derived physical parameters (\vexp, \nhi, $\tau_{Ly\alpha}$, b) in Cols. 4-7.}        
\label{best_param}      
\centering          
\begin{tabular}{c c c c c c c}     
\hline\hline
Region & Ly$\alpha$ contribution & UV contribution & \vexp & N$_{HI}$ & $\tau_{Ly\alpha}$ (E(B-V)) & b (temperature)  \\
 & & & km s$^{-1}$ & cm$^{-2}$ &  & km s$^{-1}$ (K)  \\
(1) & (2) & (3) & (4) & (5) & (6) & (7) \\ 
\hline
Galaxy disk & 0\% & 80\% & 100 & 4x10$^{20}$ & 3.50 (0.34) & 12.8 (10 000)  \\
Halo 1 & 55\% & 9\% & 180 & 4x10$^{19}$ & 1.10 (0.09) & 22.8 (31 000) \\
Halo 2 & 45\% & 11\% & 100 & 3x10$^{19}$ & 1.20 (0.11) & 22.8 (31 000) \\
\hline  
\hline                  
\end{tabular}
\end{table*}

\subsection{Intrinsic spectra}

In Fig.\ \ref{3Dgeo} we consider a single source of Ly$\alpha$  and UV continuum photons, named source A. The spectrum emitted by this source is built in the following way: in the region of Ly$\alpha$ we always assumed that the spectrum consists of a flat UV continuum (constant in the number of photons per frequency interval) plus the Ly$\alpha$ line, characterized by a Gaussian with an equivalent width EW$_{int}$(\lya) and a FWHM$_{int}$(\lya). 

Throughout this work we adopted FWHM$_{int}$(\lya) in the range [50 , 150] km s$^{-1}$. While FWHM$_{int}$(\lya) $<$ 100 km s$^{-1}$ would correspond to a realistic intrinsic width from ionized nebulae in star-forming galaxies \citep{teplitz00, baker04, erb03, swinbank11}, we considered a higher FWHM$_{int}$(\lya) to mimic the effective broadening of the integrated Ly$\alpha$ line caused by large-scale motions of nebulae inside the HST/COS aperture. We adopted FWHM$_{int}$(\lya) up to 150 km s$^{-1}$ to make the width of the Ly$\alpha$ line consistent with the width of the integrated H$\alpha$ line observed at the galaxy center (see Fig.\ \ref{Ha_FWHM}).  

For the intrinsic Ly$\alpha$ equivalent width EW$_{int}$(\lya), we considered a range of values from 77 \AA\ to 130 \AA. This corresponds to the intrinsic Ly$\alpha$ equivalent width measured inside the COS aperture (see Table\ \ref{photometric_properties}).

\subsection{Fitting method}

To fit the Ly$\alpha$ line of Mrk1486, 
we used the Monte Carlo code MCLya to model the Ly$\alpha$ and UV continuum radiative transfer inside the 3D geometry shown in Fig.\ \ref{3Dgeo}. 

In the output of the Monte Carlo simulations, MCLya provides useful information on the escaping Ly$\alpha$ and UV photons: the \textit{emission direction} of the photons (described by the two angles $\theta$ and $\phi$), the Ly$\alpha$ and UV continuum \textit{escape fraction,} and the output \textit{Ly$\alpha$ line profile}. Our simulations were performed for typically 1000-2000 frequency points around the Ly$\alpha$ line center with a spacing of 20-10 km s$^{-1}$. 

As illustrated in Fig.\ \ref{3Dgeo}, the global Ly$\alpha$ spectrum was built using the Ly$\alpha$ and UV continuum photons that escape from the 3D geometry toward the observer. More precisely, we took photons within 5 degrees of the line of sight ($\sqrt{\theta^2 + \phi^2} \le 5$ from the axis galaxy-observer) into account. As shown in Fig. \ref{3Dgeo}, the Ly$\alpha$ photons can reach the observer after scattering on HI atoms inside the disk, halo 1, and halo 2. The UV continuum photons do the same either by penetrating the disk or by scattering on dust grains within the two outflowing halos.

The best-fit parameters we fixed for each component of our 3D geometry (\vexp, \nhi, $\tau_{Ly\alpha}$, $\tau_{cont}$, b) were derived afterward on the basis of two different criteria.

\begin{enumerate}
\item We determined the set of parameters that provides the best fit of the Ly$\alpha$ line profile. We did this by minimizing the chi-square $\chi$$^2$ of the line fitting. This parameter was computed in the wavelength range [1212 ; 1218] \AA\ to avoid the effect of any strong LIS and HIS absorption lines in our line fitting (see Fig.\ \ref{Lya_COS_ape}, right panel). 
Nonetheless, we also checked the quality of the line fit by eye (as the lowest $\chi$$^2$ may not necessarily correspond to the best Ly$\alpha$ line fit because of the large number of data points involved in the calculation).
\item From the sets of parameters that provide the lowest $\chi$$^2$, we selected the best from a photometrical method. From each set, we computed the amount that each component of our 3D geometry (disk, halo 1, and halo 2) contributes to the total amount of \lya\ and UV continuum photons received by the observer in Fig. \ref{3Dgeo}. We thus selected the set of parameters from which these \lya\ and UV contributions are consistent with the parameters measured from our HST imaging (see Table\ \ref{table:obs_proper}).
\end{enumerate}

\subsection{Best fits of the Ly$\alpha$ line}

\subsubsection{Best-fit parameters of Mrk1486}

Figure\ \ref{best_fit} shows our best fit of the Ly$\alpha$ line. We list in Table\ \ref{best_param} the best-fit parameters for each component of our 3D geometry. 

We fit the Ly$\alpha$ line of Mrk1486 with ${\rm EW_{int}(Ly\alpha)} = 80    {\rm \AA}$. This value is consistent with the intrinsic Ly$\alpha$ equivalent width derived inside the COS aperture of Mrk1486. Furthermore, we obtained our best fit with FWHM$_{int}$(\lya) = 150 km s$^{-1}$. As shown in Fig.\ \ref{Ha_FWHM}, this large FWHM$_{int}$(\lya) is consistent with the width of the integrated H$\alpha$ emission line that escapes from the galaxy disk.

The Ly$\alpha$ dust optical depth $\tau_{Ly\alpha}$ of each component ($\tau_{Ly\alpha D}$ = 3.50, $\tau_{Ly\alpha 1}$ = 1.10 and $\tau_{Ly\alpha2}$ = 1.20) provides a color excess E(B-V) that is globally consistent with the nebular component derived photometrically from our HST data (see Table\ \ref{table:obs_proper}). Nonetheless, the color excess derived for halo 2 remains slightly lower than the observed one (although it still lies within the error bars assigned to this parameter). 

For the continuum optical depth $\tau_{cont}$, the value assigned to each component ($\tau_{cont D}$ = 0.7, $\tau_{cont 1}$ = $\tau_{Ly\alpha 1}$ and $\tau_{cont 2}$ = $\tau_{Ly\alpha 1}$) proves to be exactly the one we need to fit the Ly$\alpha$ of Mrk1486 properly. 

We also find a large diversity of HI column densities (\nhi$_{D}$ = 4$\times$$10^{20}$ cm$^{-2}$, \nhi$_1$ = 4$\times$$10^{19}$ cm$^{-2}$ and \nhi$_2$ = 3$\times$$10^{19}$ cm$^{-2}$). As expected, the galaxy disk exhibits the highest \nhi, while those derived for the two outflowing halos are in the same order of magnitude.  

The mean expansion velocities \vexp\ we assigned to the shells in Table\ \ref{table:free_param} seem to be the ones we need to fit the Ly$\alpha$ line profile of Mrk1496. In other words, we find v$_{expD}$ = 100 km s$^{-1}$ for the galaxy disk, v$_{exp1}$ = 180 km s$^{-1}$ for halo 1 and v$_{exp2}$ = 100 km s$^{-1}$ for halo 2. This result confirms that fast outflows occur inside the COS aperture of Mrk1486. 

Finally, a relatively warm neutral gas is entrained in the outflowing halos (b = 22.8 km s$^{-1}$, or T $\approx$ 30 000 K). We were able to constrain the Doppler parameter $b$ from its effects on the output Ly$\alpha$ line profile. While higher b leads to the formation of a too prominent blue bump in the emerging Ly$\alpha$ line profile, lower values lead to a too narrow and faint Ly$\alpha$ line. The derived temperature T $\approx$ 30 000 K is well below the thermal ionization of hydrogen (T$_{ioni.}$ $<$ 10$^5$ K), which is consistent with the presence of neutral hydrogen inside the outflowing halos.

\subsubsection{Uncertainties on each parameter}

The characteristic uncertainties of most of the parameters are relatively low. We derived these uncertainties from the quality of the Ly$\alpha$ line fit and the sampling we adopted in our grid of models (see Table\ \ref{table:free_param}).

\begin{figure}
  \centering
   \includegraphics[width=90mm]{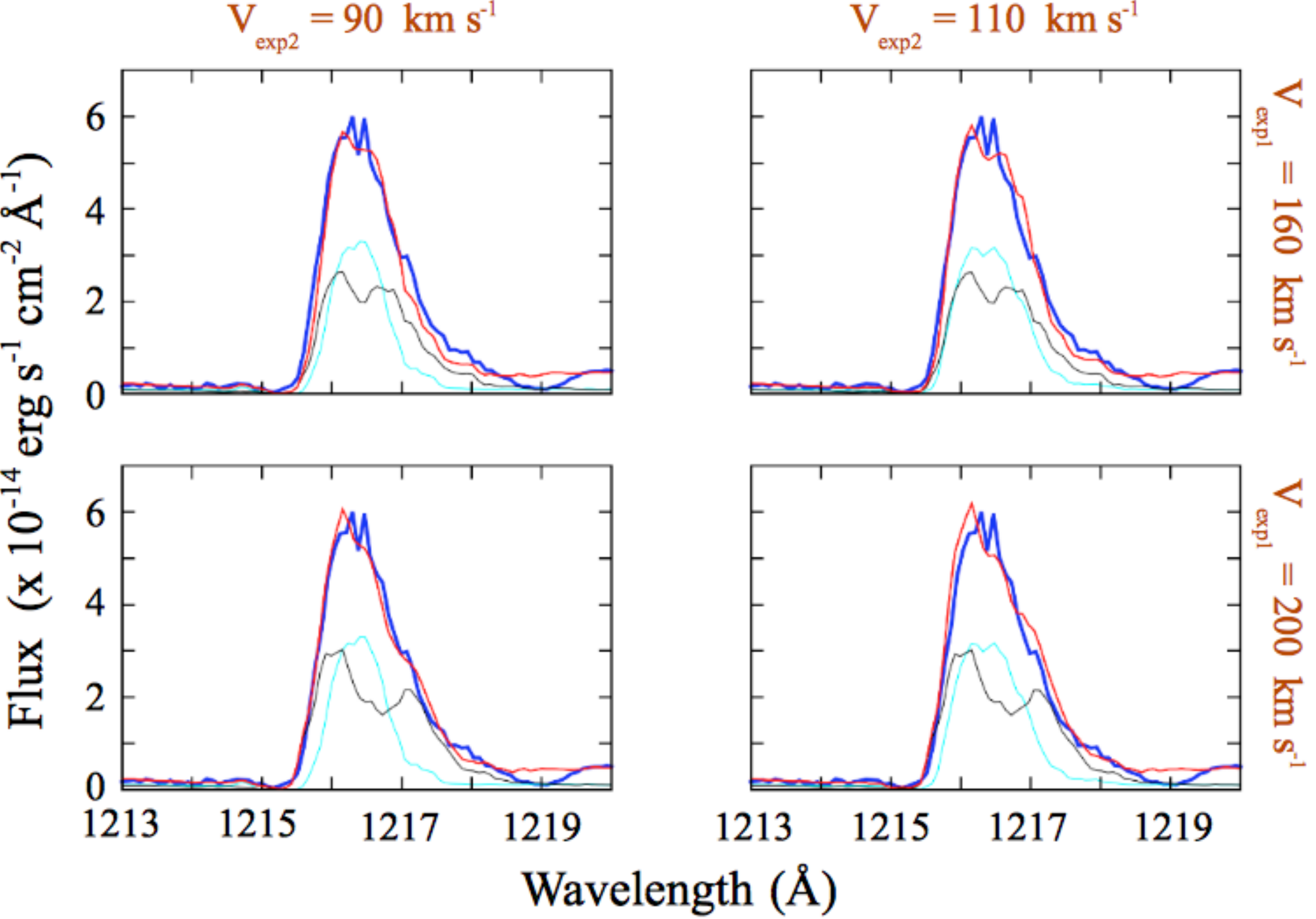}
      \caption{Evolution of the composite Ly$\alpha$ line profile (red line) as a function of the expansion velocity of halo 1 (V$_{exp1}$) and of halo 2 (V$_{exp2}$). The blue line shows the observed Ly$\alpha$ line of Mrk1486 through the COS aperture, the gray line shows the line emerging from halo 1, and the clear blue line shows the one escaping from halo 2. In the four captions, we only vary V$_{exp1}$ for halo 1 and V$_{exp2}$ for halo 2. The other physical parameters do not change. We use the parameters listed in Table\ \ref{best_param}. We always assume an intrinsic Ly$\alpha$ line with EW$_{int}$(\lya) = 80 \AA\ and FWHM$_{int}$(\lya) = 150 km s$^{-1}$.
              }
       \label{Effect_vexp}
   \end{figure}

Although not treated as a free parameter in our analysis, we estimate the characteristic uncertainty on the expansion velocity of the outflowing halos to be $\sim$ 10\% (halos 1 and 2). This low uncertainty on \vexp\ is explained by the relatively high sensitivity of the output line profile to the mean expansion velocity \vexp\ of the shells \citep{verhamme06, duval14}. 
We illustrate in Fig.\ \ref{Effect_vexp} the sensitivity of the emerging Ly$\alpha$ line profile on the expansion velocity of each outflowing halo. Here, we only adopted the most extreme velocities \vexp\ because of the uncertainty of 10\%\ assigned to this parameter. Although we always reproduce the overall Ly$\alpha$ line profile of Mrk1486, we note that a change in \vexp\ does not provide a good fit of some bumps either, nor a Ly$\alpha$/UV contribution for each halo that is consistent with those observed in our HST imaging (see Table\ \ref{obs_brightness}). 

The characteristic uncertainty on the dust optical depths $\tau_{Ly\alpha}$ and $\tau_{cont}$ reaches 10\%\ for each component of the 3D geometry. Figure \ref{chi_square} shows that this uncertainty has been derived from the quality of the line fit and the constraints on the strength of each line component. Figure 13 shows for the quality of the Ly$\alpha$ line fit that the output Ly$\alpha$ profile is very sensitive to dust parameters. In other words, a very narrow range for $\tau_{Ly\alpha 1}$ or $\tau_{Ly\alpha 2}$ can provide a good fit of the Ly$\alpha$ line of Mrk1486 (in particular, a small decrease in $\tau_{Ly\alpha 1}$ or $\tau_{Ly\alpha 2}$ provides a too prominent backscattering red peak; peak $b$ in Figs.\ \ref{best_fit} and\ \ref{3Dgeo_clumpy}). 
Figure 15 shows for the strength of each component that only a very narrow range of dust optical depths can provide a Ly$\alpha$ flux from each component that is consistent with those derived from the HST imaging (see Table\ \ref{obs_brightness}).

  \begin{figure}
   \centering
   \includegraphics[width=90mm]{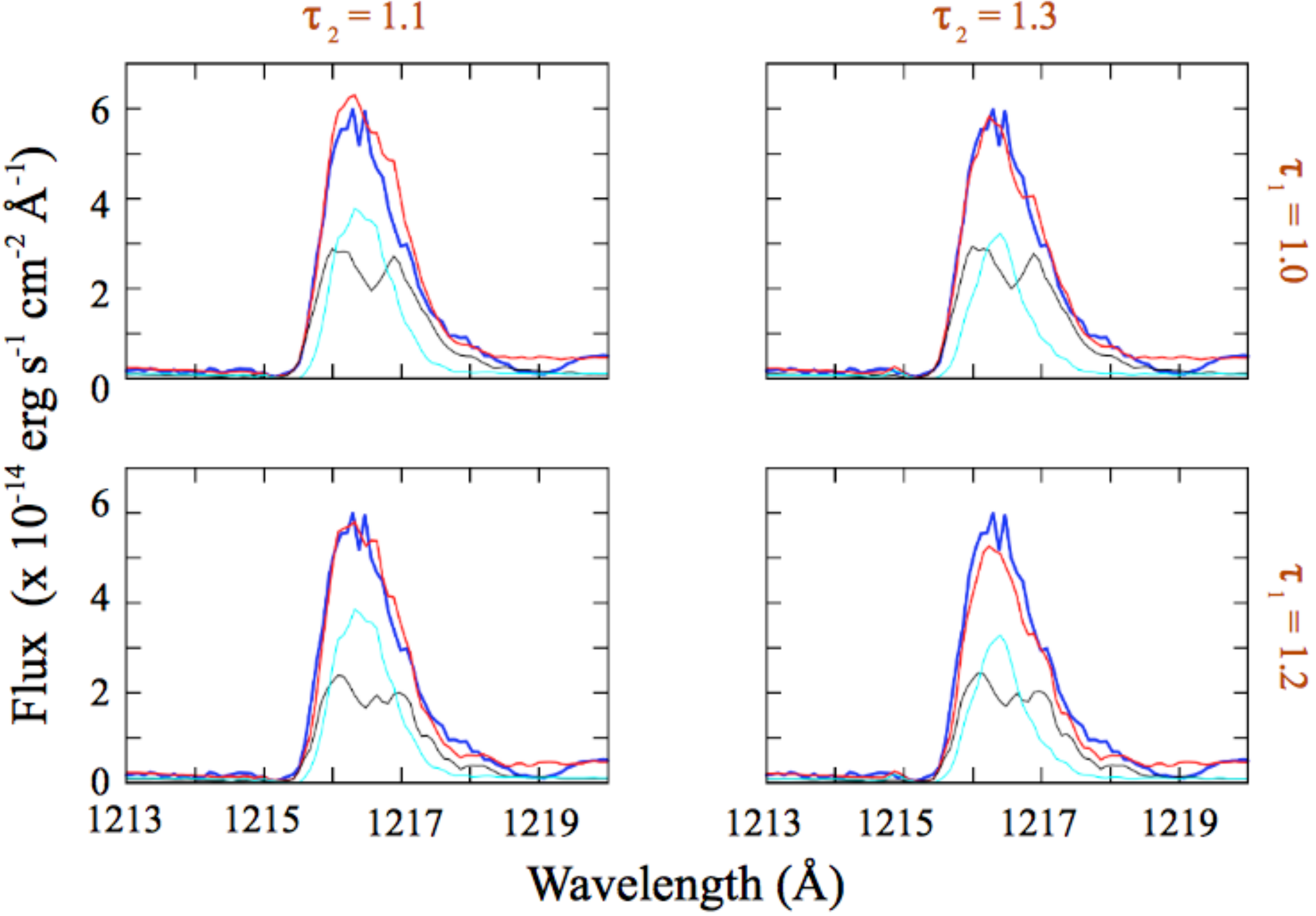}
      \caption{Same as Fig.\ \ref{Effect_vexp}, but only the dust optical depth of halo 1 ($\tau_{Ly\alpha 1}$) and halo 2 ($\tau_{Ly\alpha 2}$) varies. The other physical parameters do not change, we adopt those listed in table\ \ref{best_param}. 
              }
       \label{Effect_Tau}
   \end{figure}

  \begin{figure}
   \centering
   \includegraphics[width=90mm]{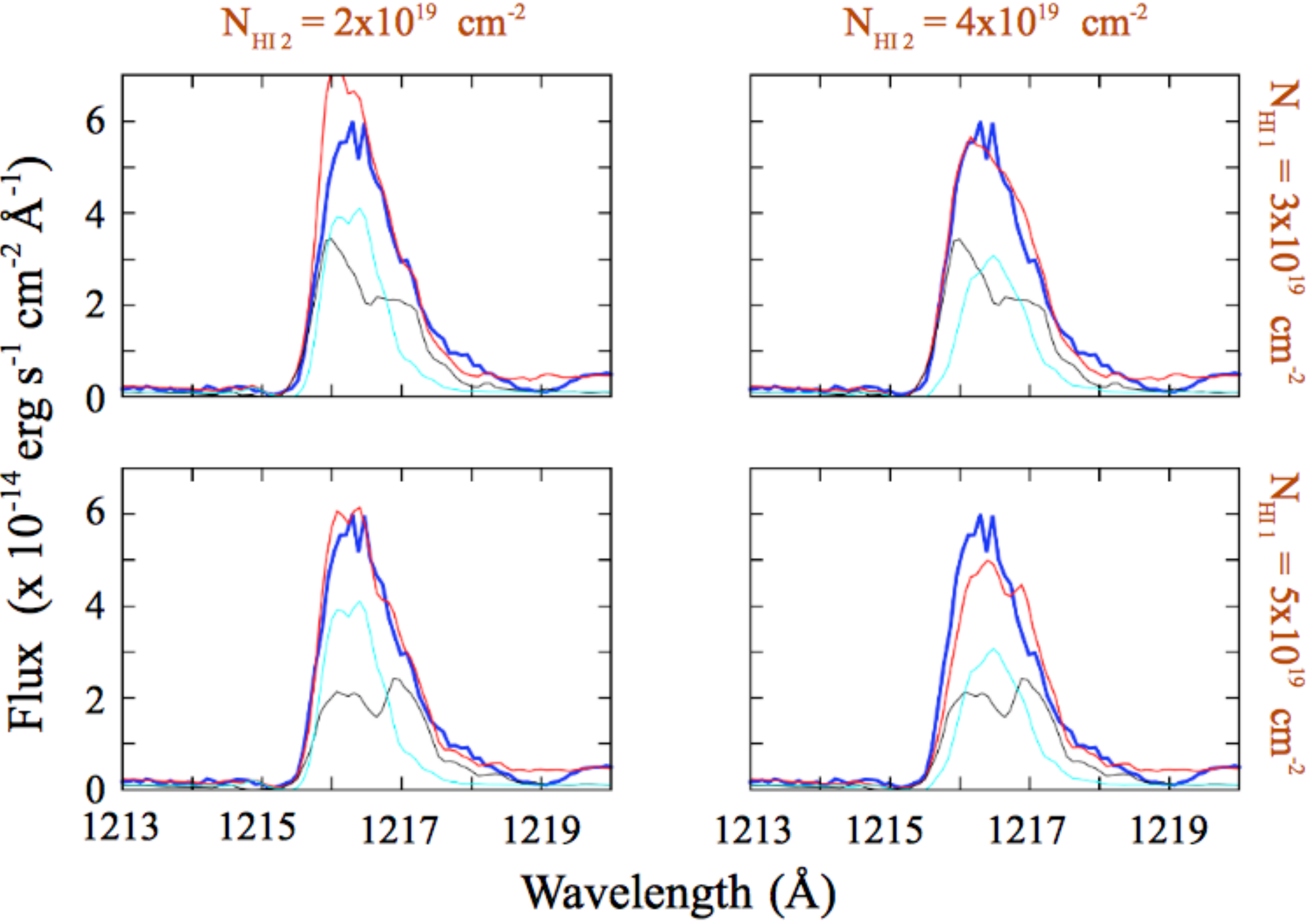}
      \caption{Same as Fig.\ \ref{Effect_vexp}, but only the HI column density of halo 1 (N$_{HI 1}$) and halo 2 (N$_{HI 2}$)
varies. The other physical parameters do not change, we adopt those listed in Table\ \ref{best_param}. 
              }
       \label{Effect_NHI}
   \end{figure}




For the HI column density \nhi, we obtain an uncertainty slightly lower than 10 \% for the outflowing halos ($\pm$ 10$^{19}$ cm$^{-2}$) and up to 20 \%\ for the galaxy disk ($\pm$ 10$^{20}$ cm$^{-2}$). \nhi\ has a complex impact on the Ly$\alpha$ line profile that escapes from each component. Figure\ \ref{Effect_NHI} shows that an increase in \nhi\ for halo 1 causes the two red peaks to less
pronounced (the first peak, noted 'a' in Fig. \ref{3Dgeo}, decreases with respect to the second, noted 'b', because backscattering becomes stronger). For halo 2, an increase in \nhi\ leads to a broader Ly$\alpha$ line and moves the redshifted peak of the Ly$\alpha$ away from the line center. 

The Doppler parameter $b$ shows an uncertainty of about 20\% from our grid of models. Overall, we note that a Doppler parameter b higher than 27 km s$^{-1}$ provides a blue bump that is too strong compared to the observed one. We could correct that by using a very low intrinsic equivalent width EW$_{int}$(\lya) in our 3D geometry, but such a strategy requires adopting an EW$_{int}$(\lya) twice as low as those we measured inside the COS apperture of Mrk1486 (see Sect. 5). All of this prevents us from obtaining a good line fit with a Doppler parameter b higher than 27 km s$^{-1}$ in each halo.

Finally, we obtain an uncertainty of about 15\% for the relative strength EW$_{int}$(\lya). In our fitting model, the best EW$_{int}$(\lya) mostly depends on the contribution of the galaxy disk to the total UV surface brightness (about 80$\%$ in Fig. \ref{best_fit}). While most of the Ly$\alpha$ emission escapes from the outflowing halos and the UV continuum radiation emerges from the disk, a small increase (decrease) in the UV contribution from the galaxy disk leads to a strong increase (decrease) in EW$_{int}$(\lya) to be able to fit the Ly$\alpha$ line of Mrk1486. 

\subsubsection{Degeneracies on the solution}

\begin{figure}
  \centering
   \includegraphics[width=90mm]{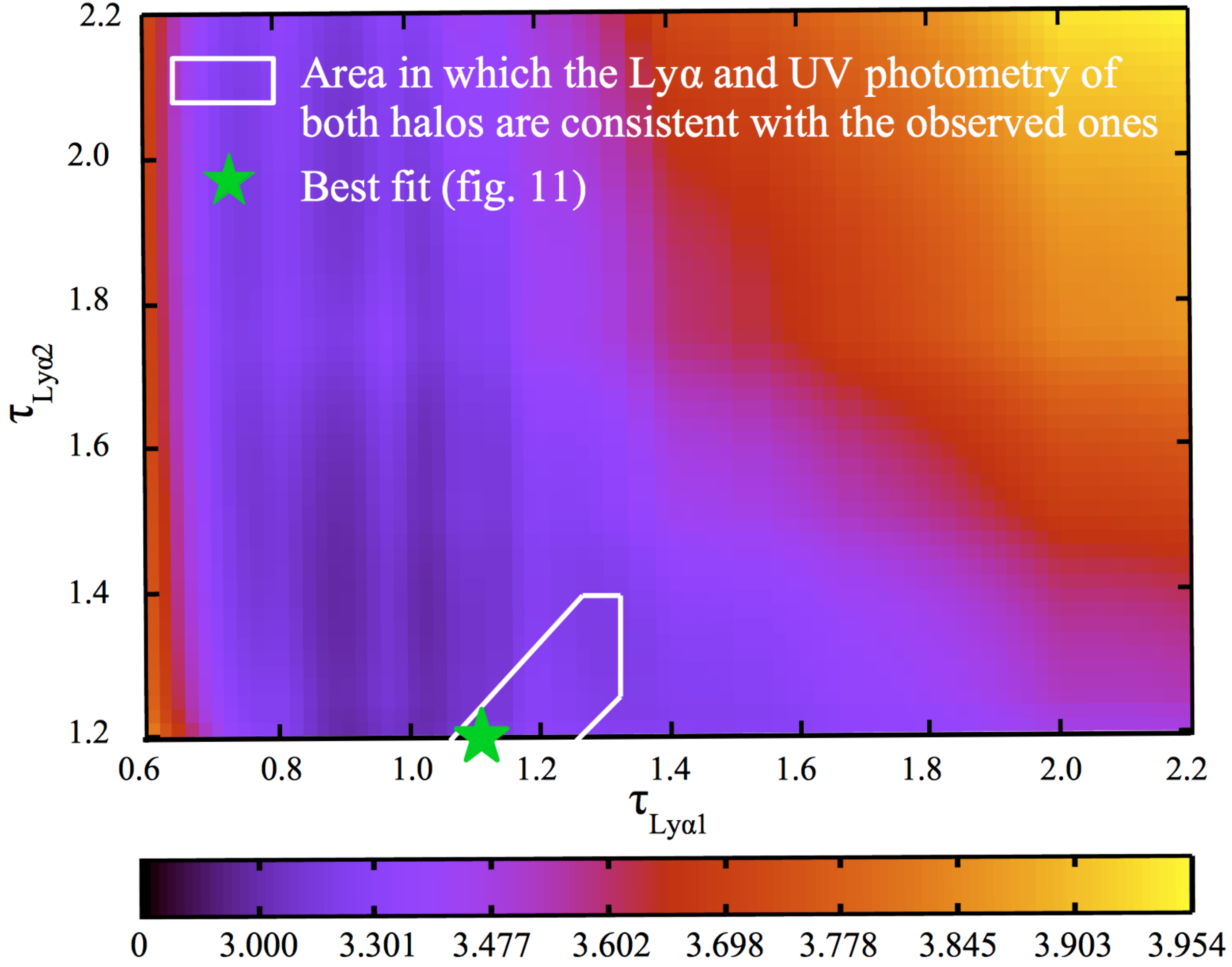}
      \caption{log($\chi^2$) map for the Ly$\alpha$ line fit with three components (disk, halo 1, and halo 2). Here we vary only the dust optical depth of each outflowing halo: $\tau_{Ly\alpha 1}$ and $\tau_{Ly\alpha 2}$. While the intrinsic Ly$\alpha$ equivalent width is allowed to vary from 77 \AA\ to 130 \AA, all other physical parameters do not change (we use those listed in Table \ref{best_param}). A degeneracy is clearly seen (darkest zones), running diagonally for the lowest $\tau_{Ly\alpha 1}$ values. The white area shows the range for $\tau_{Ly\alpha 1}$ and $\tau_{Ly\alpha 2}$ from which the strength of each line component (halos 1 and 2) are consistent with the one observed from the HST imaging (see Table \ref{obs_brightness}): halo 1 and halo 2 contribute up to 56$^{+2}_{-3}$ \% and 44$^{+3}_{-2}$ \% of the total Ly$\alpha$ flux and up to 11$^{+3}_{-2}$\% and 12$^{+3}_{-1}$ \% of the total UV continuum flux, respectively. 
              }
       \label{chi_square}
   \end{figure}

As shown in Fig. \ref{chi_square}, a degeneracy exists on our solution if we focus only on the quality of the Ly$\alpha$ line fit (by minimizing $\chi^2$). This was clearly expected when fitting the Ly$\alpha$ line profile of a star-forming galaxy with three components. However, by using our robust Ly$\alpha$ and UV photometrical constraints on each line component (see Table \ref{obs_brightness}), we can considerably improve this degeneracy. As shown by the white area in Fig. \ref{chi_square}, all those photometrical constraints are collected in a small area. Therefore, the best fit of the Ly$\alpha$ line must be found within this small region, as indicated by the green star in Fig. \ref{chi_square}. 


\section{Discussion}

\subsection{Comparison with other \lya\ line fits of Mrk1486}

Orlitova et al. (in prep.) attempt to fit the \lya\ line profile of Mrk1486 using a single expanding shell of HI and dust that surrounds a central source of \lya\ and UV continuum photons. The use of this simplified geometry is motivated by the work of \cite{verhamme08}, \cite{schaerer08} and \cite{dessauges10}, in which a spherically symmetric expanding shell of HI and dust is able to reproduce the wide variety of observed \lya\ line profiles in LBGs and LAEs at various redshifts. With four physical parameters to describe the single shell (\vexp, \nhi, $\tau_{dust}$ , and b),  Orlitova et al. (in prep.) find a high degeneracy on the solutions when fitting of the Ly$\alpha$ line of Mrk1486. Moreover, when a good fit is obtained, the set of parameters related to the fit is not necessarily consistent with the observational constraints (\vexp\ and $\tau_{dust}$).  

The model we used here proceeds slightly further than this because we now take into account constraints from the geometry of the galaxy. As seen in Fig.  \ref{3Dgeo}, this requires using three different shells to fit the Ly$\alpha$ line of Mrk1486. This model solves some problems encountered by Orlitova et al. (in prep.), such as obtaining a reasonable fit of the Ly$\alpha$ line profile with a set of physical parameters that are consistent with those measured photometrically and spectroscopically from our available HST data (see Tables \ref{obs_brightness} and \ref{best_param}).    
  

\subsection{Escape of Ly$\alpha$ photons from Mrk1486}

In Sect. 4.3 we proposed two different scenarios to explain how Ly$\alpha$ photons might escape from Mrk1486 inside the COS aperture. On the one hand, we proposed that the Ly$\alpha$ photons might be produced by shocks at the base of the outflowing halos before escaping straight toward the observer. On the other hand, we proposed that Ly$\alpha$ photons might be produced by photoionization in the environment of bright SCs inside the galaxy disk. In this way, most of the Ly$\alpha$ photons may travel along the axis of the bipolar outflowing halos in which they would scatter on neutral hydrogen toward the observer. 

Based on the dominant source of ionizing radiation inside Mrk1486 (Sect. 5) and the good fit of the Ly$\alpha$ line profile (Sect. 6), we conclude that the second scenario fits better than the first. With our best fit of the Ly$\alpha$ line profile, our model reproduces the shape and strength of the Ly$\alpha$ line very well by assuming a pure scattering of Ly$\alpha$ photons on the cool HI material entrained in the outflowing halos (see Fig. \ref{3Dgeo}). Moreover, we obtain a very good consistency between the physical quantities of our 3D geometry and those measured photometrically and spectroscopically from our available data (\vexp\ and $\tau_{Ly\alpha}$, see Tables \ref{obs_brightness} and \ref{best_param}). This therefore leads to the scenario in which the Ly$\alpha$ photons of Mrk1486 are produced inside the galaxy disk before traveling along the outflowing halos and scattering on cool HI materials toward the observer.


Nonetheless, shocks may be at work at the base of the outflowing halos of Mrk1486. When we assume only a pure scattering of Ly$\alpha$ photons on the outflowing HI material, we note that the derived E(B-V) of halo 2 in Table\ \ref{best_param} remains slightly lower than the one derived photometrically from our HST data (see Table\ \ref{obs_brightness}). A way to correct this weak discrepancy would be to assume a brighter intrinsic Ly$\alpha$ radiation inside halo 2. In this way, a higher E(B-V) would be needed to obtain an emerging Ly$\alpha$ line similar to the one shown in Fig. \ref{best_fit}.  
This additional intrinsic Ly$\alpha$         radiation may come from shocks in this region of the galaxy. However, shocks always are a subdominant mechanism of Ly$\alpha$ production inside Mrk1486.


\subsection{HI material entrained in the outflowing halos}

In Table\ \ref{best_param} we derive a relatively high HI column density \nhi\ for the outflowing halos of Mrk1486. This indicates that a substantial HI mass is outflowing along the minor axis of the galaxy disk. Such a phenomenon is commonly observed from a large portion of local star-forming galaxies (Veilleux et al. 1995; Heckman et al. 2000; Rupke et al. 2002, 2005b, 2005c; Schwartz \&\ Martin 2004; Martin 2005). 

The HI column density \nhi\ derived in table   \ref{best_param} is the HI quantity that affects the Ly$\alpha$ photons in each outflowing halo.  
From the apparent size of each halo inside the COS aperture (see Fig.\ \ref{Lya_COS_ape}), we can convert this HI column density in terms of HI mass M$_{HI}$. For this purpose we simply consider that each bright Ly$\alpha$ outflowing halo exhibits a half-sphere shape of radius R and volume V from the galaxy center. Therefore, we can derive M$_{HI}$ as 

\begin{equation}
M_{HI} = V \times \rho_{HI} = \frac{4 \pi R^2}{6} \times N_{HI} \times m_{HI}   
,\end{equation}
where $\rho_{HI}$ is the average HI density in the outflowing halo, R is its apparent radius, \nhi\ is the (radial) HI column density, and m$_{HI}$ is the mass of the hydrogen atom. In Fig. \ref{Lya_COS_ape} we obtain an apparent radius of R $\sim$ 390 pc for the halo located above the galaxy disk and R $\sim$ 450 pc for the one located below the disk. This leads to a substantial mass M$_{HI}$ $\approx$ 10$^5$ M$_{\odot}$ for the cool HI material that affects the Ly$\alpha$     line in each halo. Concerning the total HI mass that is entrained in the outflowing halos, the derived M$_{HI}$ can only be seen as a lower limit of this global quantity.

There is no active galactic nucleus (AGN) at the galaxy center, which means that the two outflowing halos could only be driven by the kinetic energy and momentum from stellar winds and supernovae. However, of these two possible sources of energy, our available data clearly favor the role played by the stellar winds in the formation of the outflowing halos inside Mrk1486. First of all, Mrk1486 is classified as a Wolf-Rayet star-forming galaxy because
of its strong and broad He[II] $\lambda$4686 emission line \citep{schaerer99,izotov97c}. This suggests that some of the gas mass motions are driven by winds from Wolf-Rayet stars. Moreover, given the young age of 3 Myr for the starburst of Mrk1486, the models of \textit{Starburst99} provide mass-loss ($M_*$) rates and energy returned to the ISM ($E_*$) that lead to an integrated momentum injection from the brightest SCs (\textit{p$_*$} = $\int{M_*V_*dt}$ = $\int{\sqrt{{2M_*E_*}}dt}$) that is high enough to entrain a HI mass of 10$^5$ M$_{\odot}$ at \vexp\ = 100 km s$^{-1}$ and \vexp\ = 180 km s$^{-1}$ in each halo. 

As stellar winds dominate the mechanical energy deposition in the ISM for t $<$ 3 Myr \citep{veilleux05}, supernovas seem to play a subdominant role in the formation of the outflowing halos in Mrk1486.

  \begin{figure}
   \centering
   \includegraphics[width=90mm]{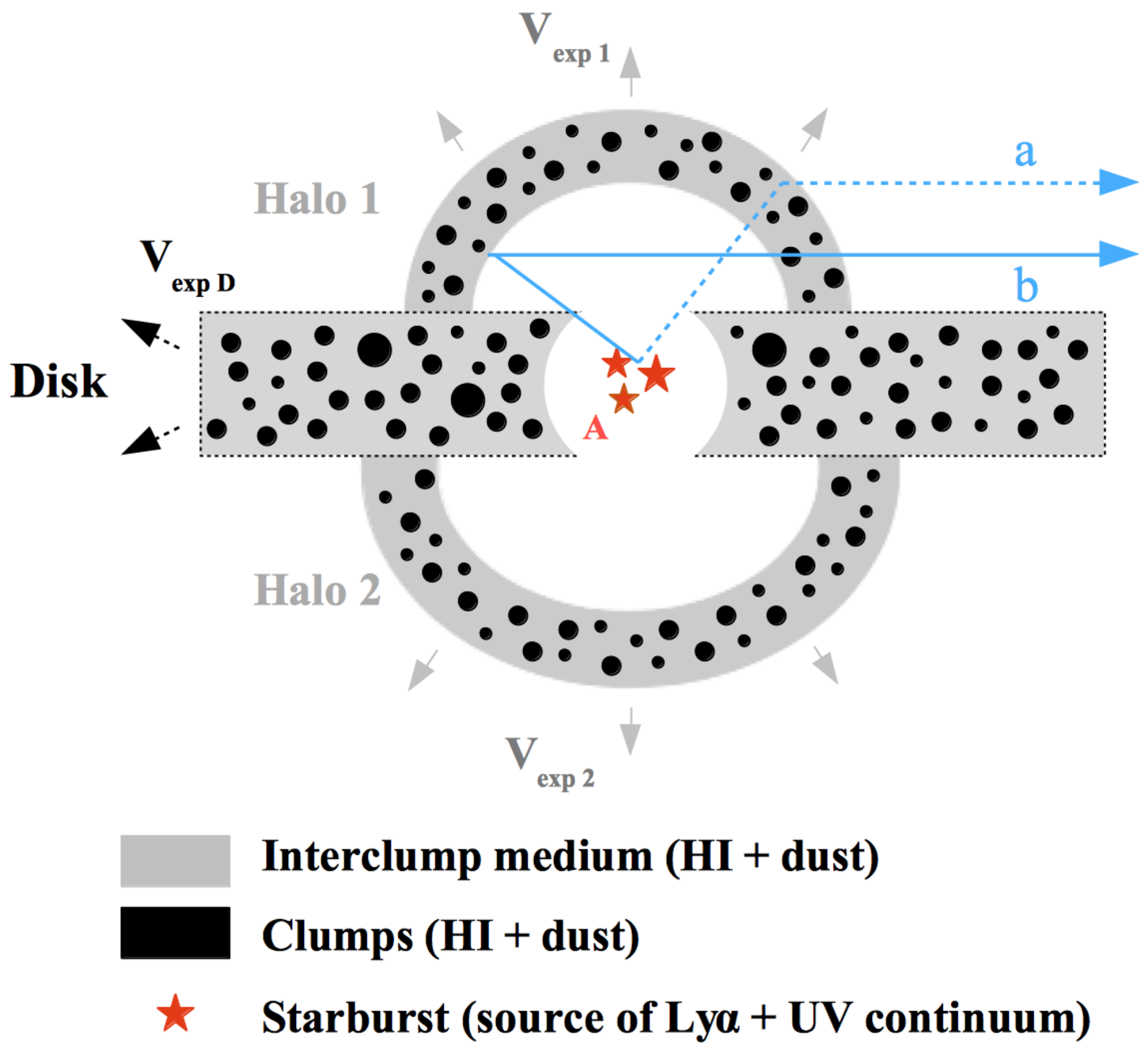}
      \caption{Scheme of our 3D geometry in which the distribution of HI and dust is clumpy within the galaxy disk and the outflowing halos. The observer is located on the right side, at infinity. We also add to this figure the most likely path followed by each group of Ly$\alpha$ photons in Fig.\ \ref{best_fit}: while the Ly$\alpha$ photons "a" escape from the outflowing halo without undergoing any backscattering (they simply scatter on HI atoms toward the observer), the Ly$\alpha$ photons "b" experience one backscattering before escaping from the outflowing halo.  
              }
       \label{3Dgeo_clumpy}
   \end{figure}

\subsection{HI and dust clumpiness inside the disk and the outflowing halos}

     \begin{figure}
   \centering
   \includegraphics[width=90mm]{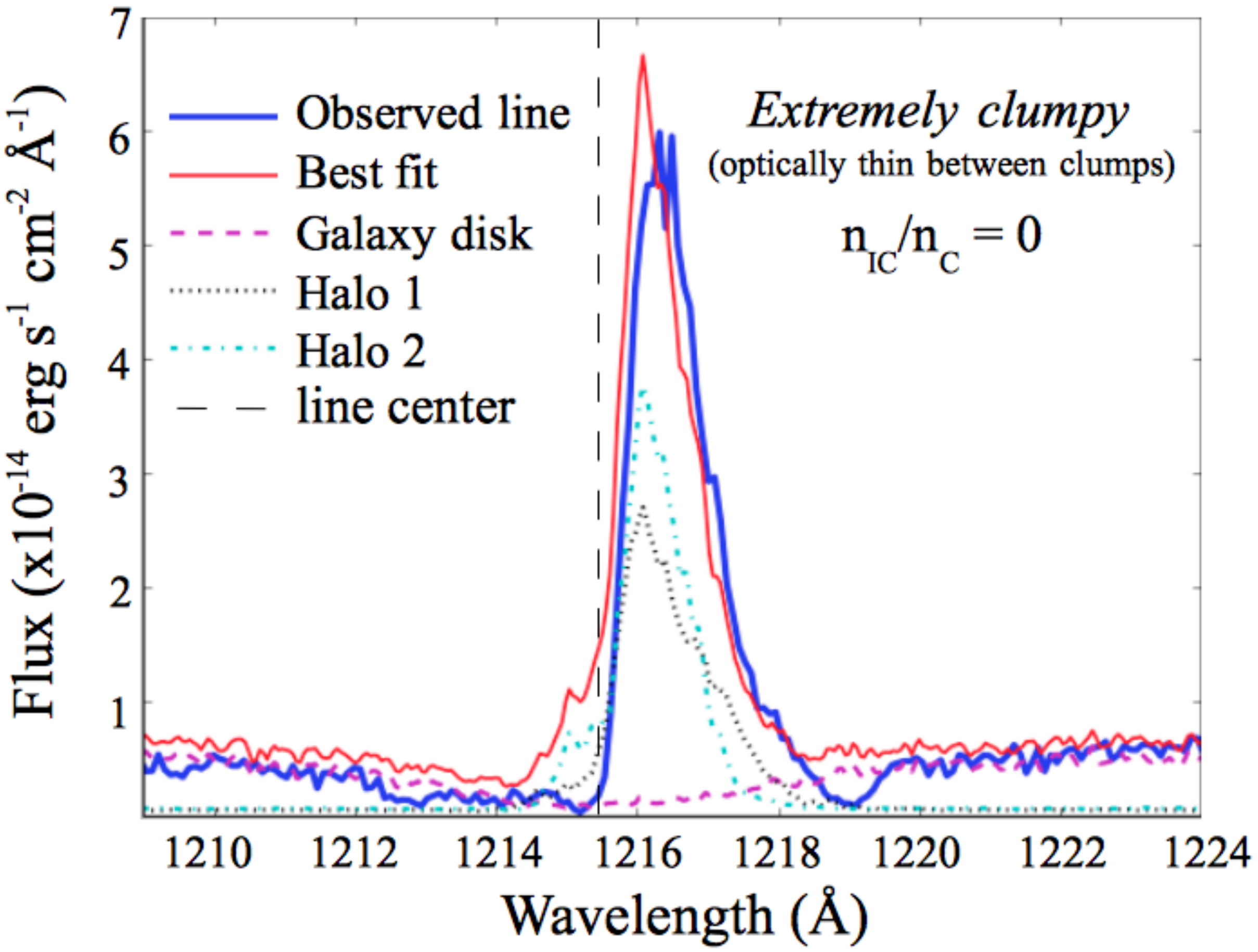}
      \caption{Comparison between the observed Ly$\alpha$ line profile of Mrk1486 (blue thick line) and the one escaping from our 3D geometry when assuming an extremely clumpy distribution of HI and dust inside the galaxy disk and the outflowing halos (i.e., with an optically thin interclump medium, n$_{IC}$/n$_C$ = 0). The magenta dashed line shows the spectrum that escapes from the galaxy disk. The spectra of wind 1 and wind 2 are shown as a black dotted and a blue dashed-dotted line, respectively. When assuming n$_{IC}$/n$_C$ = 0 in each component, we always obtain a strong emission at the Ly$\alpha$ line center.
              }
       \label{extremely_clumpy}
    \end{figure}
    
       \begin{figure*}
   \centering
   \includegraphics[width=90mm]{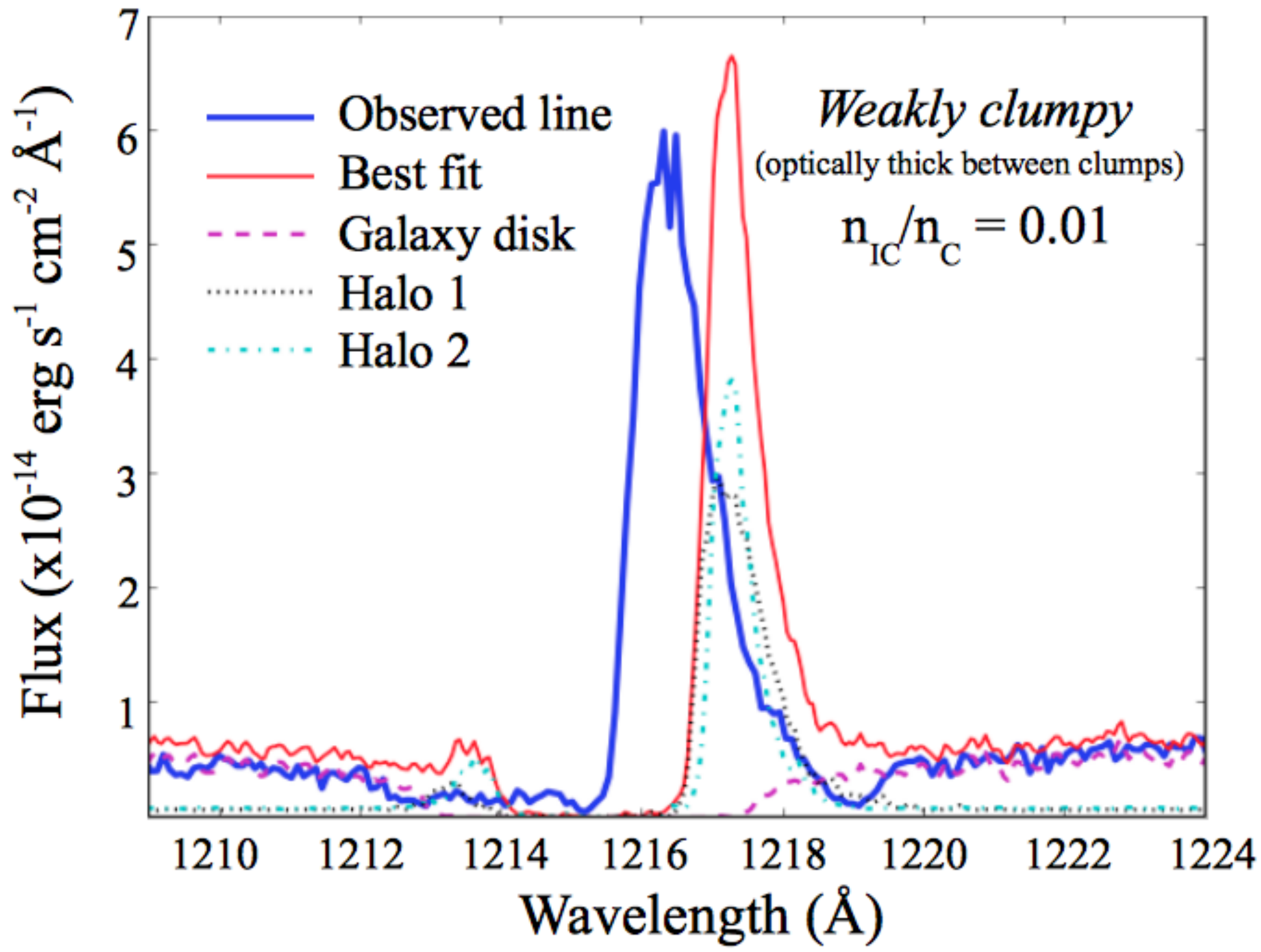}
   \includegraphics[width=90mm]{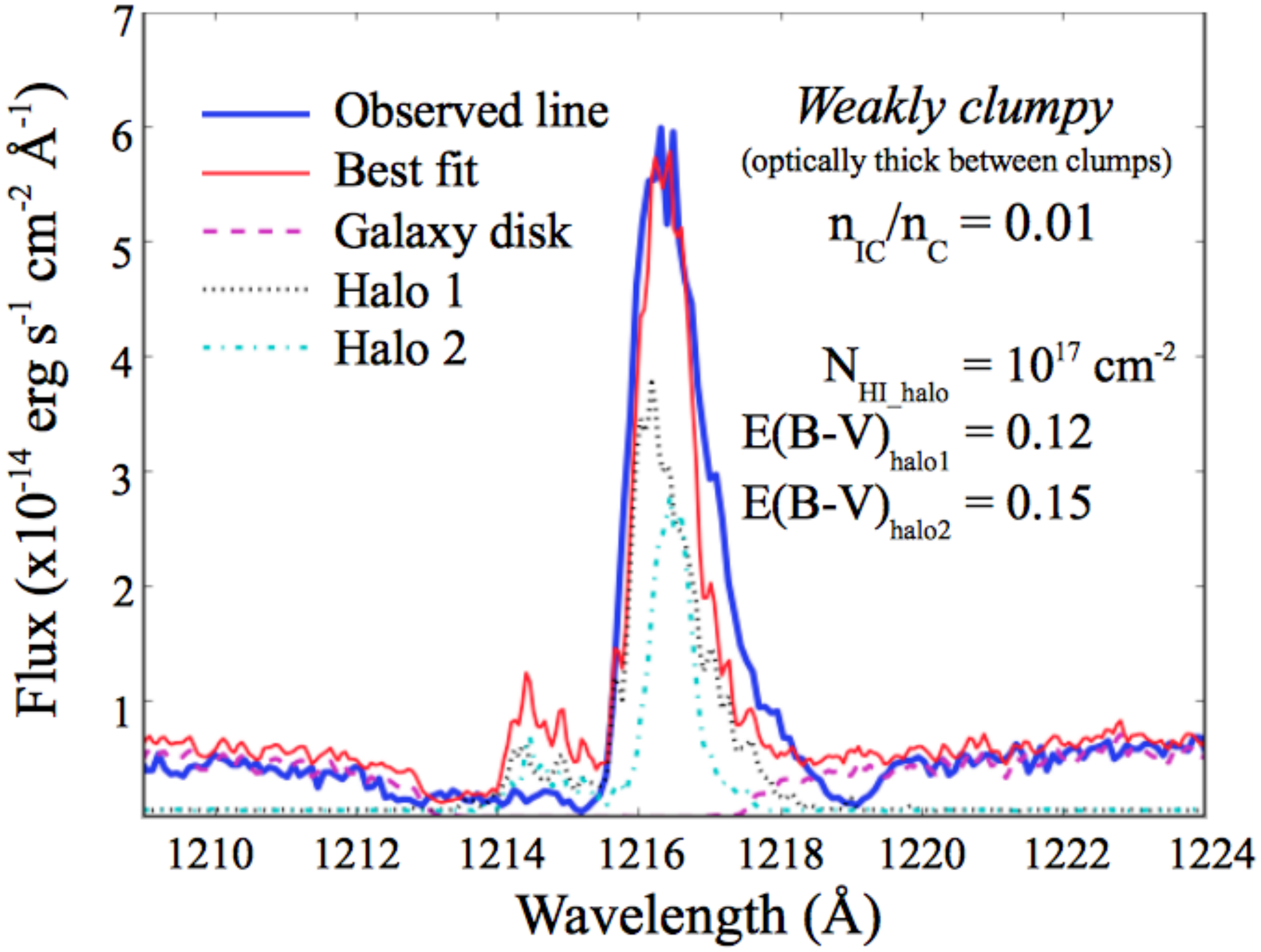}
      \caption{Comparison between the observed Ly$\alpha$ line profile of Mrk1486 (blue thick line) and the one escaping from our 3D geometry when assuming a weakly clumpy distribution of HI and dust inside the galaxy disk and the outflowing halos (i.e., with an optically thick interclump medium, n$_{IC}$/n$_C$ = 0.01). The magenta dashed line shows the spectrum that escapes from the galaxy disk. The spectra of halo 1 and halo 2 are shown as
a black dotted line and blue dashed-dotted line, respectively. \textit{Left panel}: the red thin line shows the Ly$\alpha$ line profile escaping from our 3D geometry when adopting the fit parameters listed in Table\ \ref{best_param}. Under these conditions, we note a clear shift of the blue and red bump peaks that is due to the transfer of Ly$\alpha$ photons in the very warm interclump medium (T = 10$^6$ K, which leads to a broader Voigt absorption profile for \lya). \textit{Right panel}: the red thin line shows our best fit of the observed Ly$\alpha$ line profile when adopting new \nhi\ and \taua\ for the two outflowing halos of our 3D geometry. We find here an HI column density \nhi\ = 10$^{17}$ cm$^{-2}$ for each wind and a color excess E(B-V) = 0.12 and 0.15 for halo 1 and halo 2, respectively. 
              }
       \label{clumpy}
   \end{figure*}

Throughout the paper we succeed in reproducing the Ly$\alpha$ line profile of Mrk1486 by computing the radiative transfer of Ly$\alpha$ and UV continuum photons inside a 3D geometry showing a homogeneous distribution of HI and dust particles around the source of photons (see Fig.\ \ref{3Dgeo}). Although our simplified model provides a very good result, a homogeneous distribution of gas and dust seems clearly inconsistent with observations of real ISMs. There is indeed ample evidence for a multiphase ISM in star-forming galaxies, galactic fountains, and outflowing halos \citep{geach14, veilleux09}.  \citep{cox74,mckee77,mckee95,spitzer90,maclow04,geach14, veilleux09}. In the most frequently employed picture of \cite{mckee77}, most of the volume of the ISM is occupied by hot (HIM, T $\approx$ 10$^6$ K), warm (WNM, T $\approx$ 10$^4$ K), and cold (molecular phase, T $\approx$ 10$^2$ K) media in thermal equilibrium. 

The disk and outflowing halos of
MRK1486 should exhibit an inhomogeneous distribution of HI and dust in reality. In this perspective, the neutral gas (HI) and dust should be distributed in clumps, leaving a large part of the volume of the galaxy occupied by an interclump medium that
is extremely warm and very poor in neutral hydrogen (see illustration in Fig.\ \ref{3Dgeo_clumpy}). 

The sensitivity of the Ly$\alpha$ line profile to the gas distribution in the ISM of galaxies \citep{giavalisco96,hansen06,laursen13,duval14,verhamme14}
enables us to use the Ly$\alpha$ line of Mrk1486 as a probe of the clumpiness of the HI distribution inside the disk and outflowing halos. As shown in \cite{duval14}, a clumpy distribution of HI and dust can affect the Ly$\alpha$ line profile in three different ways:

\begin{itemize}
\item A strong emission at the Ly$\alpha$ line center ($\lambda_{Ly\alpha}$ = 1215.67 \AA, in rest-frame) must emerge from an extremely clumpy ISM characterized by an optically thin interclump medium for Ly$\alpha$ photons\footnote{This emission at line center is due to either a direct escape of Ly$\alpha$ photons through holes that may appear between clumps, or by scattering of Ly$\alpha$ photons off of the surface of clumps (as initially suggested by Neufeld 1991).}.
\item Expanding clumpy ISMs tend to form a stronger blue bump at the expense of the red one (see Fig. 13 in Duval et al. 2014). 
\item In clumpy ISMs, the effect of the dust on the Ly$\alpha$ line profile is less efficient than in homogeneous media. This can lead to an intense Ly$\alpha$ emission escaping from a very dusty clumpy ISM (while an absorption line profile should emerge from a homogeneous medium with the same dust content).
\end{itemize}

Throughout this subsection we have adopted different clumpy distributions of HI and dust inside our 3D geometry and tested whether we can reproduce the Ly$\alpha$ spectrum of Mrk1486 with clumpy models. We show in Figs.\ \ref{extremely_clumpy} and \ref{clumpy} the evolution of the Ly$\alpha$ line profile as a function of the gas clumpiness in our 3D geometry. A detailed explanation of the construction of our clumpy media is given in Appendix A. We only mention here that we distributed all the HI and dust content of a shell in clumps (with a high HI density n$_{C}$) in between clumps (with a low HI density n$_{IC}$). We also assumed different temperatures inside and between the clumps, mostly to ensure the stability of the clumps by thermal equilibrium (T = 10$^4$ K in clumps by analogy to the WNM, and T = 10$^6$ K in the interclump medium as measured in the HIM of the MW). Finally, following a similar approach as \cite{duval14}, the clump distribution inside the shells is characterized by a volume filling factor FF= 0.23 and a certain covering fraction (CF = 1) around the photon sources. Unless otherwise specified, we always adopted the physical conditions listed in Table\ \ref{best_param} in Figs.\ \ref{extremely_clumpy} and \ref{clumpy}. Our main results are summarized below. \\
\\
\textbf{Clumpiness of the outflowing halos}: by fitting the Ly$\alpha$ line profile of Mrk1486 from different clumpy models,        \textit{we conclude that the Ly$\alpha$ line can only be reproduced if the bipolar outflowing halos of the galaxy are composed of a homogeneous medium of HI and dust}. This is exactly a homogeneous medium of the kind that allows us to reproduce the Ly$\alpha$ line profile of Mrk1486.


  \begin{figure*}
   \centering
   \includegraphics[width=175mm]{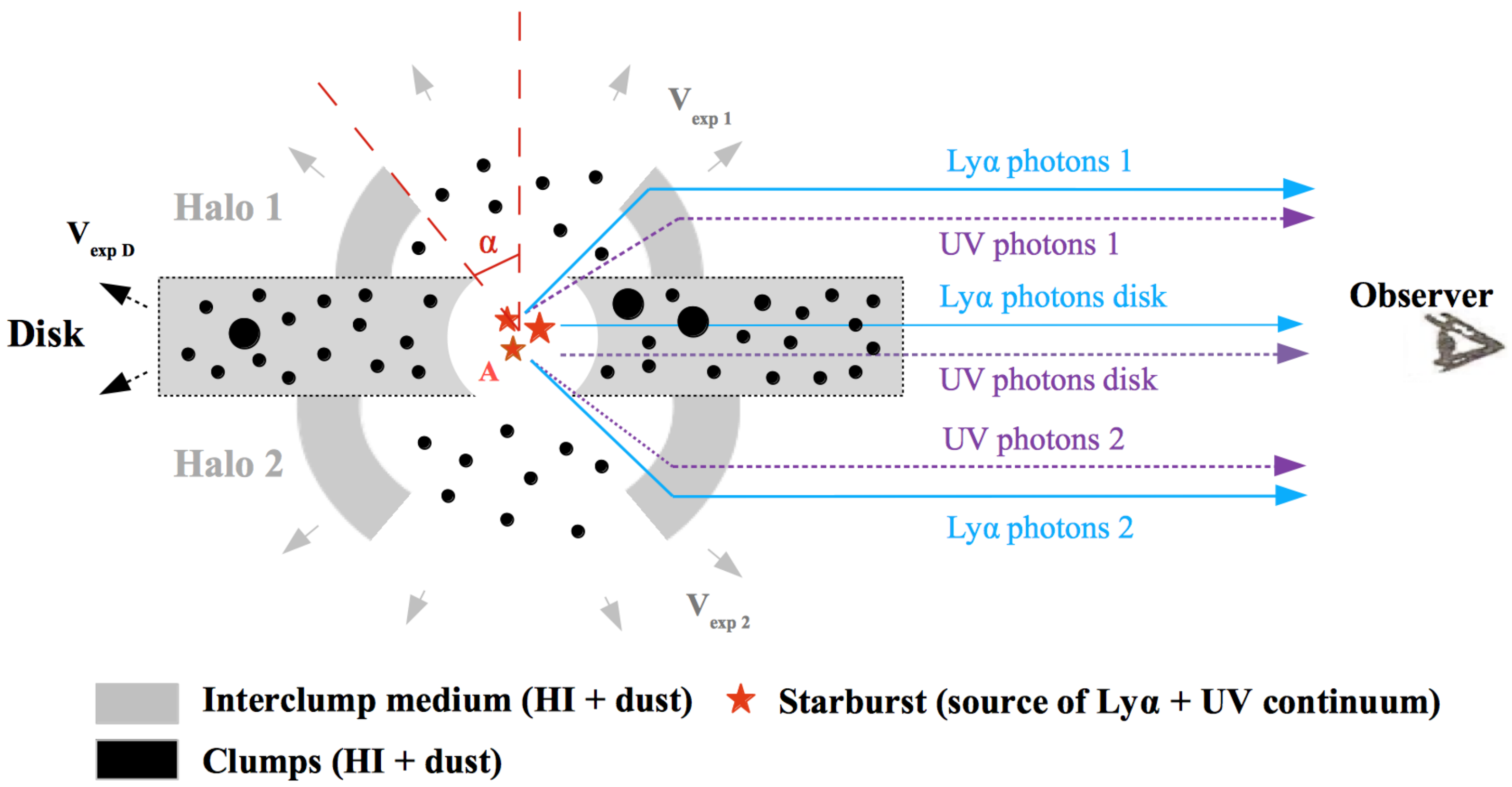}
      \caption{Final scheme that represents the possible gas distribution inside the disk and the outflowing halos of Mrk1486. This scheme is built from the conclusions of our modeling of the Ly$\alpha$ radiative transfer inside the galaxy (Sects. 6 and 7). This figure is similar to the one shown in Fig.\ \ref{3Dgeo}, but the outflowing halos can be now opened in a conical shape that is defined with a cone angle $\alpha$. The composite SiII line profile obtained from our COS spectrum indicates a clumpy distribution of gas and dust inside the medium of the \textbf{galaxy disk}. Moreover, the clumps are very likely distributed around the source of photons such as the total covering fraction ${\rm CF} = 1$. The external layer of the \textbf{outflowing halos} corresponds to a homogeneous medium of HI and dust. This medium plays the most important role in the Ly$\alpha$ transport (all Ly$\alpha$ photons scatter toward the observer from this region). Clumps of HI and dust may composed the internal part of the halos, but their effect on the Ly$\alpha$ radiative transport is negligible.           
          }
       \label{final_geometry}
   \end{figure*}

Before demonstrating this result, we recall here that the two outflowing halos are the only regions of Mrk1486 that show Ly$\alpha$ in emission inside the COS aperture (see Table \ref{obs_brightness}). Therefore, the Ly$\alpha$ line profile of Mrk1486 is only formed within these two outflowing regions. 

In Fig.  \ref{extremely_clumpy} we have tested the most extreme scenario in which both outflowing halos are composed of cold neutral clumps (of HI and dust) embedded in a completely ionized interclump medium (a total density ratio n$_{IC}$/n$_{C}$ = 0). As seen in this figure, we cannot fit the observed Ly$\alpha$     line profile of Mrk1486 (the observed and the modeled Ly$\alpha$ lines show a radically different flux at the line center). This clearly dismisses the scenario of an extreme distribution of HI and dust in the two outflowing halos of the galaxy. 

In Fig.\ \ref{clumpy} we therefore tested another scenario in which the outflowing halos show a weakly clumpy distribution of HI and dust (characterized by an interclump medium that is optically thick for \lya). Once again, this scenario seems to be dismissed because of the poor fits of the Ly$\alpha$ line profile. In the left panel of Fig.\ \ref{clumpy}, we first note that the difference of temperature inside and between the clumps forces us to find new physical parameters to eventually fit the Ly$\alpha$ line profile of Mrk1486 ($\tau_{Ly\alpha}$ and \nhi\ in particular). We show the new fit parameters in the right panel of Fig.\ \ref{clumpy}. Nevertheless, two main problems rule out the scenario of a weakly clumpy gas distribution inside the outflowing halos. On the one hand, the Ly$\alpha$ line profile that escapes from the weakly clumpy media always exhibits a too strong blue bump compared to the observed one. On the other hand, the new best-fit parameters reveal a very low \nhi = 10$^{17}$ cm$^{-2}$ that leads to a very narrow Ly$\alpha$ line. Such a narrow line cannot fit the observed line profile of Mrk1486. 

From these negative results, our last option would be to assume a homogeneous distribution of HI and dust within the outflowing halos of Mrk1486. This is exactly what we tested in Sect. 6. Given the very good line fit shown in Fig. \ref{best_fit}, we conclude that the gas distribution is most likely homogeneous
within each outflowing halo of Mrk1486. This homogeneous medium proves to be the one that affects the Ly$\alpha$ photons the most within those galactic winds. 

\begin{figure}
   \centering
   \includegraphics[width=90mm]{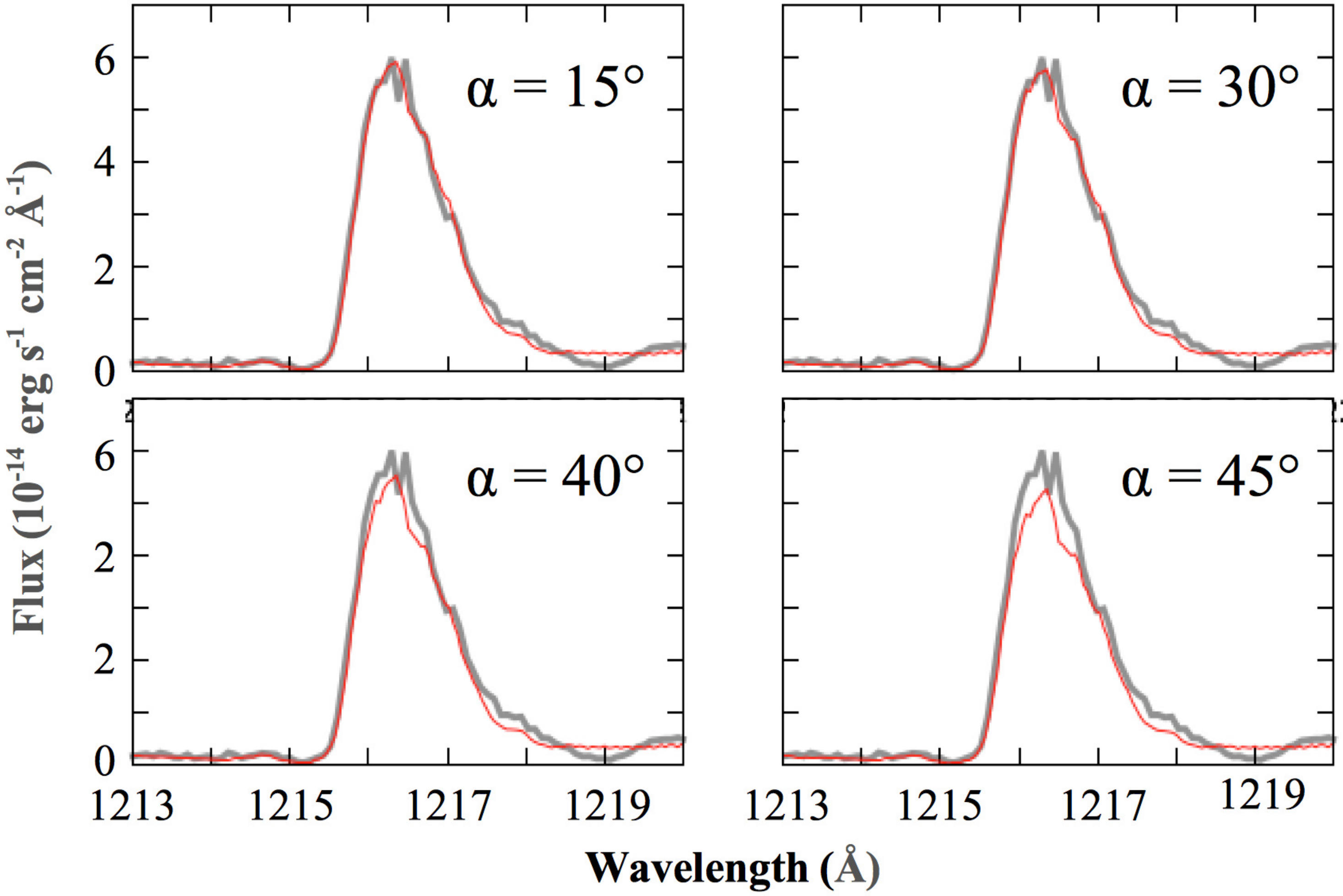}
      \caption{Evolution of the composite Ly$\alpha$ line profile (thin red line) as a function of the cone angle $\alpha$ in Fig. \ref{final_geometry} ($\alpha$ = 15$^o$, 30$^o$, 40$^o$ , and 45$o$). This figure has been obtained by calculating the Ly$\alpha$ and UV radiative transfer inside the geometry shown in Fig.  \ref{final_geometry}. The thick gray line shows the observed Ly$\alpha$ line of Mrk1486 through the COS aperture.  In the four captions we adopt the same physical parameters as in Table \ref{best_param} and the intrinsic Ly$\alpha$ line is characterized by EW$_{int}$(\lya) = 80 \AA\ and FWHM$_{int}$(\lya) = 150 km s$^{-1}$.
              }
       \label{fit_final_geo}
   \end{figure}

We therefore propose in Fig.\ \ref{final_geometry} a structure that may describe the gas distribution inside the outflowing halos of Mrk1486. In this figure each halo is a shell that is opened in a conical shape (with a cone angle $\alpha$). 
The envelopes of these outflowing halos are characterized by a homogeneous distribution of HI and dust. These media scatter most of the Ly$\alpha$ photons toward the observer and play the most important role in the formation of the observed Ly$\alpha$ line profile of Mrk1486.
We show in Fig. \ref{fit_final_geo} that this gas distribution provides a very good fit of the Ly$\alpha$ line profile of Mrk1486 for even a very large cone angle $\alpha$ up to 40$^o$ for the outflowing halos (for higher angles the flux of the output Ly$\alpha$ line decreases as the homogeneous media become too weak to scatter the Ly$\alpha$ photons efficiently toward the observer). This good fit supports the idea that the gas distribution shown in Fig. \ref{final_geometry} may be close to real one inside Mrk1486. \\
\\
\textbf{Clumpiness of the galaxy disk}: from the observed Ly$\alpha$ line profile of Mrk1486, no clear conclusion can be formulated regarding the clumpiness of the gas distribution within the galaxy disk. As a result of the high \nhi\ = 4x10$^{20}$ cm$^{-2}$ and E(B-V) = 0.34 of the galaxy disk, a pronounced Ly$\alpha$ absorption line always emerges from the disk (see Figs.\ \ref{best_fit}, \ref{extremely_clumpy} and \ref{clumpy}). Nevertheless, two conclusions can be formulated from previous results. First, the composite SiII line profile of Mrk1486 shown in Fig.\ \ref{Vexp_shift} \textit{clearly indicates a clumpy distribution of gas and dust within the galaxy disk}. The low covering fraction of gas per velocity bin (f$_c$ $<$ 0.8) and the broad velocity dispersion of the SiII line profile indicate that distinct subsystems (clouds of neutral gas) move outward at slightly different velocities inside the ISM of Mrk1486.  
Second, although clumpy, we support that the total covering fraction of neutral clumps CF is unity around the sources of photons to ensure no escape of Ly$\alpha$ photon from the disk (all lines of sight from the sources of Ly$\alpha$ photons must be intersected by at least one clump). 



\subsection{Comparison with simulated Ly$\alpha$ transport in edge-on disk galaxies}

As mentioned above, recent numerical simulations have highlighted the strong dependence of the Ly$\alpha$ properties of disk galaxies on the inclination at which they are observed \citep{verhamme12, behrens14}. These results are based on radiative transfer calculations of Ly$\alpha$ and UV continuum photons through high-resolution hydrodynamical simulations of isolated low-mass disk galaxies showing similar physical properties as Mrk1486 (very inclined disk-like galaxies, same sources of Ly$\alpha$ photons, presence of large-scale galactic winds in the minor axis of the elongated galaxy, a stellar mass M$^*$ for the simulated galaxies that is similar or twice as low as the one of Mrk1486). 

Overall, two important effects produced by the inclination of a disk galaxy on Ly$\alpha$  have been demonstrated in these works:

\begin{itemize}

\item The observed Ly$\alpha$ equivalent width EW(\lya) must vary reasonably with the inclination at which the galaxy disk is observed \citep[from -5 \AA\ when the disk is seen edge-on to 80 \AA\ when it is seen face-on;][]{verhamme12}. 

\item The Ly$\alpha$ line profile is expected to vary with the inclination of the disk because of the kinematic structure of the ISM, which is different from face-on (strong outflows) to edge-on (static velocity fields).

\end{itemize}

While the first effect suggests possible strong observational biases in surveys of \lya-emitting galaxies that rely on Ly$\alpha$ luminosity and EW selections (mostly face-on disk galaxies might satisfy the narrowband excess criterion that is commonly used to identify LAEs; EW(\lya) $>$ 20 \AA), the second effect leads to the formation of either a double-peaked Ly$\alpha$ line profile from edge-on disk galaxies or to an asymmetric redshifted line profile from face-on disk galaxies. 

Interestingly, the Ly$\alpha$ properties of Mrk1486 seem to contradict all these predictions. 
On the one hand, Mrk1486 shows a total EW(\lya) = 45 \AA\ that indicates a very easy escape of Ly$\alpha$ photons from this object \citep[Mrk1486 may be selected in deep Ly$\alpha$ narrowband surveys;][]{hayes14}. On the other hand, a clear P-Cygni profile emerges from the edge-on disk galaxy of Mrk1486 (see Fig.\ \ref{Lya_COS_ape}). 

Without calling into question the accuracy of the work of \cite{verhamme12} and \cite{behrens14}, the discrepancy observed between the Ly$\alpha$ properties of Mrk1486 and the simulated galaxies seems to originate in a small difference in the galactic wind properties of these objects. Although the simulated galaxies of \cite{verhamme12} and \cite{behrens14} exhibit very extended tenuous and ionized outflowing materials from the disk, the galactic winds of Mrk1486 differ from the simulated ones in that \textit{they entrain a higher mass of neutral gas (HI) at large distances from the galaxy disk}. These very extended HI outflowing materials from the core of Mrk1486 may explain why the scattering of Ly$\alpha$ photons toward the observer is more efficient than in the simulated galaxies. This difference perfectly explains the higher escape of Ly$\alpha$ photons and the formation of a P-cygni line profile in Mrk1486.  


The presence of fast bipolar galactic winds seems to be a common feature of star-forming disk galaxies (Heckman 1990, Martin 1999, Veilleux et al. 2005 for a recent review). 
Given the clear advantage Ly$\alpha$ photons take of the outflowing halos of Mrk1486 to easily escape toward the observer, we may generalize this mechanism to a large number of edge-on disk galaxies.
In particular, this may explain the origin of very bright Ly$\alpha$ emission lines observed from elongated disk objects at high redshift \citep{Shibuya14}. In conclusion, this mechanism may challenge the strong observational biases in surveys of \lya-emitting galaxies (as edge-on disk galaxies might also be detected as LAEs) and the robustness of the viewing angle effect on the Ly$\alpha$ properties of disk galaxies \citep{,laursen07,verhamme12,behrens14}. 

\section{Summary}

We carried out a detailed study of Ly$\alpha$ radiative transfer in Mrk1486, a highly inclined (edge-on) disk star-forming galaxy located at redshift z$\sim$0.0338. This galaxy belongs to the \textit{Lyman Alpha Reference Sample} (LARS), a sample of 14 star-forming galaxies at redshift between 0.028 and 0.18 that were imaged and spectroscopically observed with the Hubble Space Telescope (HST) from far-UV to optical wavelengths (up to $\sim 8000$ \AA). 

The global Ly$\alpha$ properties of Mrk1486 that include its high luminosity (L$_{Ly\alpha}$ = 1.11$\times$10$^{42}$ erg s$^{-1}$), equivalent width (EW(\lya) = 45 \AA), and escape fraction (\flya\ = 0.174; Hayes et al. 2014) make this object stand apart from the other galaxies of the LARS sample and suggest an easy escape of Ly$\alpha$ photons from the galaxy. This result seems surprising in view of the strong viewing-angle effect an edge-on disk galaxy is expected to produce on the Ly$\alpha$ line properties, as revealed by recent numerical simulations \citep{verhamme12,behrens14}. In other words, it seems that a particular mechanism is at work in Mrk1486 and helps Ly$\alpha$ photons to easily escape from the core of the galaxy. 

Throughout this work we aimed to determine which path the Ly$\alpha$ photons followed to easily escape from the ISM of Mrk1486. For this purpose, we modeled the Ly$\alpha$ radiative transport inside the ISM of Mrk1486 from a detailed study of our HST imaging and radiation transfer calculations performed using the 3D Ly$\alpha$ Monte Carlo code MCLy$\alpha$ of \cite{verhamme06} and \cite{schaerer11}. 

Our main results are summarized as follows.

\begin{itemize}
\item Morphologically, the ISM of Mrk1486 exhibits a complex structure. Three distinct regions were identified.  On the one hand, a galaxy disk 4 kpc long and 245$\pm$15 pc thick represents the most extended structure of Mrk1486. On the other hand, the analysis of IFU H$\alpha$ spectroscopic PMAS data of Mrk1486 has revealed two extended outflowing halos above and below the galaxy disk. 
\item The disk and the outflowing halos of Mrk1486 show radically different \lya, H$\alpha,$ and UV continuum emission properties. While the UV continuum and H$\alpha$ radiation are mostly concentrated in the disk, this region of the galaxy exhibits a pronounced Ly$\alpha$ absorption line. Conversely, outside the disk bright and extended Ly$\alpha$ structures are visible. In particular, we identified two very bright Ly$\alpha$ regions above and below the galaxy disk. The location of these two regions is perfectly consistent with the one of the outflowing halos.     
\item Because Ly$\alpha$ photons escape from Mrk1486 through the bipolar outflowing halos, we proposed two different scenarios to explain the path followed by Ly$\alpha$ photons to escape from Mrk1486. The first scenario proposed that Ly$\alpha$ photons may be produced by shocks at the base of the outflowing halos before escaping straight toward the observer. The second scenario proposed that Ly$\alpha$ photons might be produced by photoionization inside the galaxy disk. In this way, most of the Ly$\alpha$ photons may travel along the bipolar outflowing halos where they would scatter on neutral hydrogen toward the observer. 
\item Using specific optical emission lines ratios that are particularly sensitive to shocks ([OIII]$_{5007}$/H$\beta$ over [OI]$_{6300}$/H$\alpha$, [NII]/H$\alpha$ or [SII]$_{6710,6730}$/H$\alpha$), we investigated the dominant source of Ly$\alpha$ photons at the galaxy center. For this purpose, we compared the optical line ratios to the photo- and shock-ionization models of \cite{allen08}. We concluded that the ionization of hydrogen is very likely dominated by photoionization from star clusters (SCs) inside the disk of Mrk1486. This is consistent with the second scenario we described above for the path Ly$\alpha$ photons may follow to escape from Mrk1486. 
\item The derived age of the starburst of Mrk1486 is 3.02$\pm$0.20 Mryr (as obtained from our best SED fit of the HST data). Moreover, we found the intrinsic Ly$\alpha$ equivalent width EW$_{int}$(\lya) to be in the range [77, 130] \AA\ inside the COS aperture of the galaxy (this aperture was placed on the brightest UV knot). 
\item Using the 3D Ly$\alpha$ Monte Carlo code MCLya, we studied Ly$\alpha$ and UV continuum radiative transfer inside an expanding 3D geometry of HI and dust that models the ISM structure of Mrk1486 at the galaxy center (see Fig.\ \ref{3Dgeo}). From these numerical simulations, and assuming an intrinsic Ly$\alpha$ line showing EW$_{int}$(\lya) = 80 \AA, we successfully reproduced the strength and shape of the Ly$\alpha$ line of Mrk1486 (observed through the HST/COS spectrum of the galaxy). 
\item Out best fit of the Ly$\alpha$ line profile confirms the scenario in which Ly$\alpha$            photons are produced inside the galaxy disk and escape toward the observer by scattering on the HI material of the outflowing halos. 
\item The Ly$\alpha$ profile of Mrk1486 also provides useful information on the physical properties of the outflowing halos (\vexp, \nhi, E(B-V) and temperature). In particular, each halo shows a high HI column density (up to \nhi\ = 3-4$\times$10$^{19}$ cm$^{-2}$) that indicates that a HI mass of 10$^5$ M$_{\odot}$ is entrained in the outflows. We found that this HI mass is entrained at \vexp\ = 180 km s$^{-1}$ and \vexp\ = 100 km s$^{-1}$ above and below the galaxy disk. Finally, we estimated the average temperature of this HI material to be T $\approx$ 30 000 K.
\item The Ly$\alpha$ line profile also provides robust information on the clumpiness of the gas distribution inside the galaxy disk and the bipolar outflowing halos of Mrk1486. We showed in Fig.\ \ref{final_geometry} the most likely structure of the galaxy as revealed by our modeling of the Ly$\alpha$ radiative transport. Regarding the outflowing halos, the envelope of the winds must be characterized by a homogeneous medium of HI and dust. This medium plays the most important role in the Ly$\alpha$ transport and allows us to reproduce the Ly$\alpha$ line profile of Mrk1486. Concerning the galaxy disk, the LIS SiII line profile reveals a clumpy distribution of gas inside the ISM (with a total covering fraction ${\rm CF} = 1$).   
\end{itemize}

Fast and large-scale outflowing halos or galactic-winds are frequently visible from edge-on disks of star-forming galaxies because of their favorable geometries. The clear advantage that \lya\ photons take of the bipolar outflowing halos of Mrk1486 to easily escape from this object suggests that the same mechanism may be at work in a large number of highly inclined disk galaxies that are observed at low-z and high-z. 
This may require a critical review of the strong observational biases in surveys of high-z \lya-emitting galaxies \citep[as edge-on disk galaxies are also detected as LAEs;][]{Shibuya14} and the robustness of the viewing angle effect on the Ly$\alpha$ properties of disk galaxies.

\phantom{\citep{dickey89,veilleux05,lehnert99,leitherer95, leitherer99,neufeld91,hayes_b15}}

\begin{acknowledgements}

The PMAS data were based on observations collected at the Centro Astronómico Hispano Alemán (CAHA) at Calar Alto, operated jointly by the Max-Planck Institut fuer Astronomie and the Instituto de Astrofísica de Andalucía (CSIC).
FD is grateful for financial support from the Japan Society for the Promotion of Science (JSPS) fund. MH acknowledges the support of the Swedish Research Council (Vetenskapsr{\aa}det) and the Swedish National Space Board (SNSB), and is Fellow of the Knut and Alice Wallenberg Foundation. DK is funded by the Centre National d\'\ Etudes Spatiales (CNES). E.Z. acknowledges research funding from the Swedish Research Council (project 2011-5349). IO was supported by the grant GACR 14-20666P of Czech Science Foundation and the long-term institutional grant RVO:67985815 and PL acknowledges support from the ERC-StG grant EGGS-278202. 

\end{acknowledgements}

\begin{figure*}
   \centering
   \includegraphics[width=150mm]{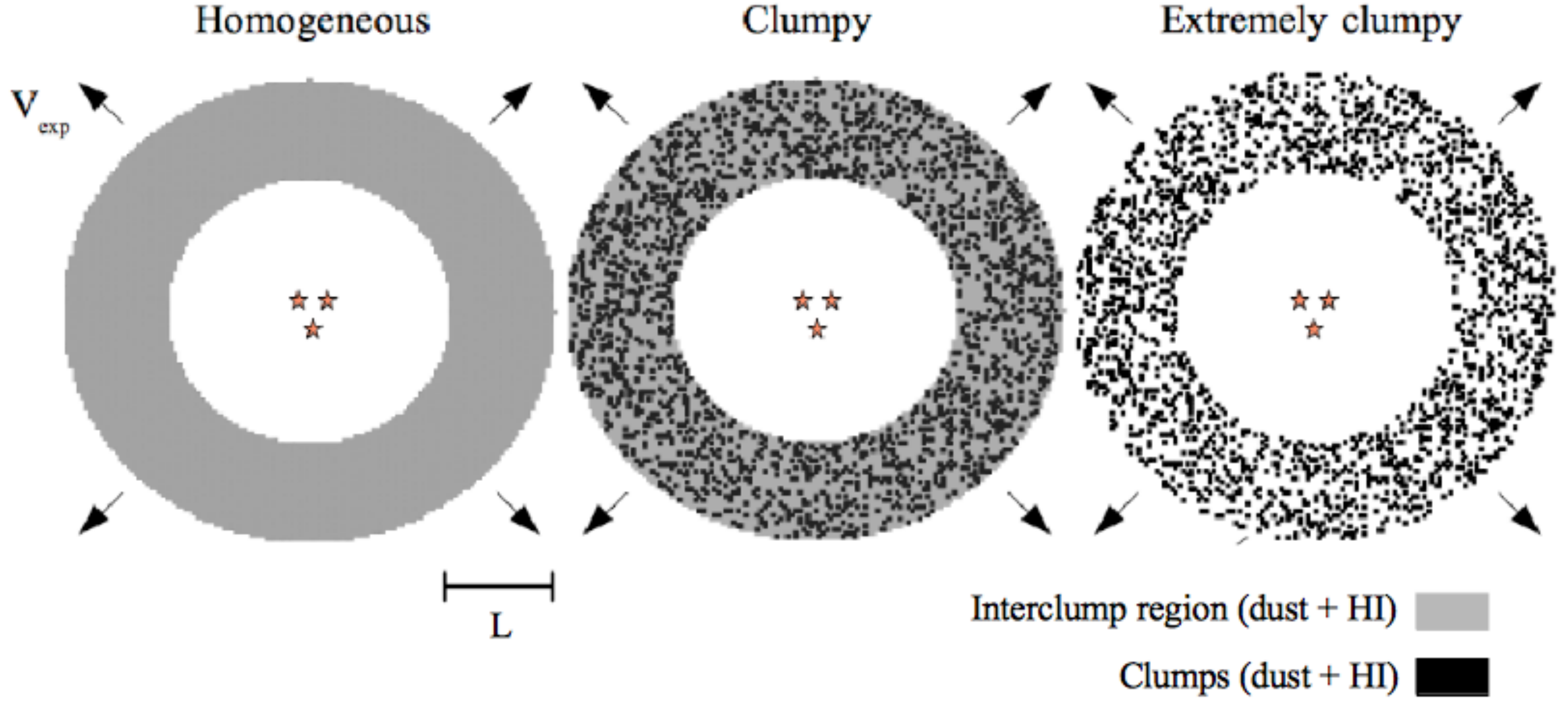}
      \caption{Representation of some 3D homogeneous and clumpy geometries. The stellar distribution is always localized at the center of the shell, while the dust and the Hi content are distributed around it. The dust and Hi distribution can be homogeneous (left), clumpy (middle), or extremely clumpy (right). In a clumpy distribution, the clumps and the interclump medium receive high and low densities of dust and HI, respectively. Nonetheless, the interclump medium remains optically thick for Ly$\alpha$ photons. For an extremely clumpy distribution, all the dust and the HI content are distributed in clumps, leaving the interclump medium optically thin for Ly$\alpha$ photons. 
              }
       \label{shells}
   \end{figure*}

\bibliographystyle{aa}
\bibliography{article}
\phantom{espace}
\textbf{APPENDIX A} \\
\\
\textbf{- Construction of clumpy media in the 3D geometry} \\
\\
In Sect. 7.3 we distributed the HI and dust content in a clumpy way in all components of our 3D geometry (see Fig.\ \ref{3Dgeo_clumpy}). As explained below, we follow the same approach as \cite{duval14} to build all clumpy media (see Fig.\ \ref{shells}). 

Three geometrical parameters characterize the clumpy distribution of HI and dust in a shell:

\begin{itemize}
\item the density contrast n$_{IC}$/n$_{C,}$
\item the volume filling factor $FF, \text{and}$
\item the covering factor $CF.$
\end{itemize} 

  \begin{figure}
   \centering
   \includegraphics[width=90mm]{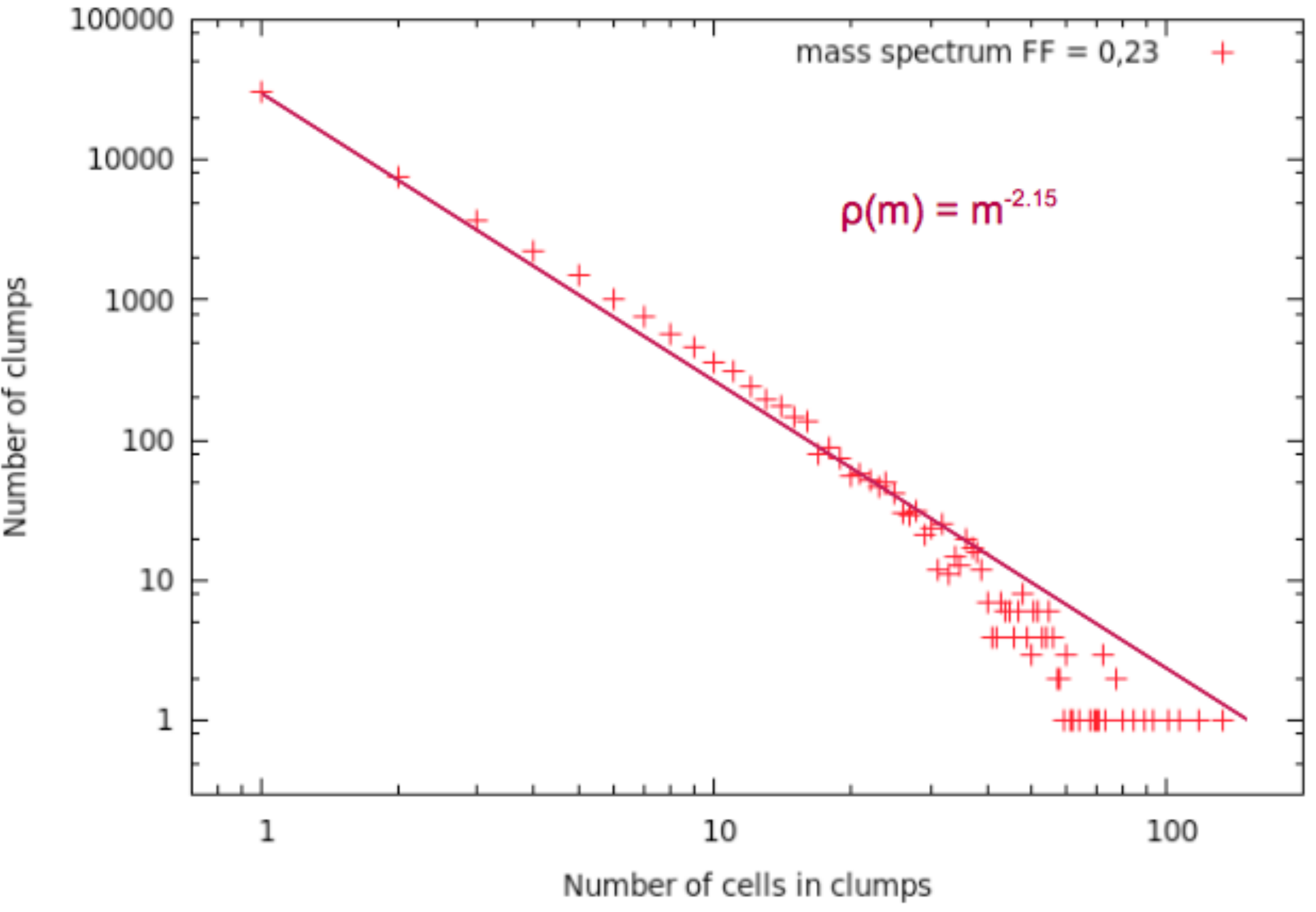}
      \caption{Variation in clump size by adopting different filling factors FF in a shell geometry defined with R$_{min}$ = 49 and R$_{max}$ = 64 cells. The filling factors used here are FF = 0.07 (circles), FF = 0.13 (diamonds), and FF = 0.23 (triangles). We obtain a mass spectrum $\rho(m) \propto m^{-2.04}$ when adopting FF = 0.23. This is the most consistent with observations of diffuse interstellar clouds. We therefore adopt FF = 0.23 throughout this study.
              }
       \label{mass_spectrum}
   \end{figure}

The density contrast n$_{IC}$/n$_{C}$ corresponds to the ratio between the HI density in clumps (n$_{C}$) and in between clumps (n$_{IC}$), the volume filling factor (FF) measures the fraction of the volume occupied by the clumps inside the shell, and the covering factor (CF) is defined as the fraction of the solid angle that is covered by the clumps as seen from the center of the shell.

All clumps are randomly distributed inside a shell. As clumps we considered, as did \cite{witt96, witt00}, all cells directly connected with each other by at least one face. We then determined their mass spectrum, which approximately follows a power law $\rho$(m) $\propto$ m, where m is the mass of a clump. Throughout the paper we adopt a filling factor FF = 0.23. This choice allows us to reproduce the interstellar mass spectrum of HI clouds of the Milky Way and nearby galaxies \citep[$\rho$(m) $\propto$ m$^{-2.04}$;][]{dickey89}. We illustrate in Fig.\ \ref{mass_spectrum} the mass spectrum obtained with FF= 0:23 (red curve) in a shell geometry built in a Cartesian grid of N=128$^3$ cells and with a thickness L = R$_{max}$ - R$_{min}$, where R$_{min}$ = 49 and R$_{max}$ = 64 cells. This mass spectrum does not change when
the grid resolution is increased.


Given a choice of the clump volume filling factor FF, we can control the covering factor CF by changing the thickness of the shell. More precisely, models with different covering factors are constructed by varying R$_{min}$. 

For the physical conditions in clumpy media, we assumed different HI densities and temperatures inside and between the clumps. First of all, starting from an homogeneous shell of HI density n$_{HI}$, we distributed the same HI content in high density in clumps (n$_{C}$) and low density in the interclump medium (n$_{C}$). More precisely, we always have the relation  
\begin{equation}
  n_{HI} = {\rm FF} \nc + (1-{\rm FF}) \nic.        
\end{equation} 
Moreover, when studying a clumpy shell, we always considered a higher temperature in the interclump medium (T$_{IC}$) than in clumps (T$_{C}$). This is mostly motivated by the stability of clumps, which is due to a pressure equilibrium between the two phases of our clumpy media. Throughout our work, we considered for each phase a temperature that is commonly measured in real ISMs. In particular, we
made calculations assuming a temperature of T = 10$^6$ K in the interclump medium (like those measured in the HIM), but a lower temperature in all clumps (T = 10$^4$ K, by analogy to the WNM).


\end{document}